\documentclass[10pt,final,journal,a4paper,oneside,twocolumn]{IEEEtran}

\usepackage{cite}

\usepackage{amsmath,amsfonts,amssymb}
\usepackage{xcolor}
\usepackage{mathdots}
\usepackage{bbm}
\usepackage{mathtools, cuted}
\usepackage{amssymb,bm}
\usepackage{tabularx}
\usepackage{booktabs}
\usepackage{multirow}
\usepackage{tikz}
\usetikzlibrary{calc,positioning}
\usepackage{overpic}
\usepackage{float}

\makeatletter
\let\MYcaption\@makecaption
\makeatother
\usepackage[font=footnotesize]{subcaption}
\makeatletter
\let\@makecaption\MYcaption
\makeatother

\usepackage{bbm}
\usepackage{mathtools, cuted}
\usepackage{amssymb,bm}
\usepackage{hyperref}
\usepackage{ragged2e}
\usepackage{overpic}
\usepackage{color}
\usepackage[super]{nth}
\usepackage{textcomp}

\usepackage{tikz}
\usetikzlibrary{calc,positioning}
\usepackage{array}
\newcolumntype{R}{>{$}r<{$}} 
\newcolumntype{C}{>{$}c<{$}} 
\usepackage[crop=pdfcrop]{pstool}
\usepackage{cancel}
\usepackage{dblfloatfix} 
\usepackage{tabularx}
\usepackage[usestackEOL]{stackengine}
\usepackage{blindtext}
\usepackage{siunitx}
\usepackage{tikz}
\usetikzlibrary{shapes.geometric, arrows,automata,positioning,shapes.multipart}
\usepackage{tabularx,colortbl}

\usepackage[super]{nth}

\usetikzlibrary{fit,backgrounds} 

\newcommand{\htext}[1]{%
	\makebox[0pt]{\Centerstack{#1}}
}

\newcommand{\vtext}[1]{%
	\makebox[0pt]{\rotatebox[origin=c]{90}{\Centerstack{#1}}}
}

\newcommand{\vtextf}[1]{%
	\makebox[0pt]{\rotatebox[origin=c]{270}{\Centerstack{#1}}}
}

\newcommand{\Et}{\tilde{E}}
\newcommand{\Ht}{\tilde{H}}
\newcommand{\Pt}{\tilde{P}}

\newcommand{\Jt}{\tilde{J}}
\newcommand{\Kt}{\tilde{K}}

\newcommand{\Etf}{\tilde{\mathcal{E}}}
\newcommand{\Htf}{\tilde{\mathcal{H}}}
\newcommand{\Ptf}{\tilde{\mathcal{P}}}

\newcommand{\E}{\mathbf{E}}
\newcommand{\R}{\mathbf{R}}

\newcommand{\C}{\mathbf{C}}
\renewcommand{\H}{\mathbf{H}}

\newcommand{\nh}{\mathbf{\hat n}}

\newcommand{\J}{\mathbf{J}}
\newcommand{\K}{\mathbf{K}}
\newcommand{\X}{\mathbf{X}}

\renewcommand{\r}{\mathbf{r}}

\renewcommand{\L}{\mathbf{L}}

\newcommand{\Nv}{\mathbb{N}}
\newcommand{\bbmatrix}{\begin{bmatrix}}
\newcommand{\ebmatrix}{\end{bmatrix}}
\newcommand{\I}{\mathbb{I}}
\newcommand{\U}{\mathbb{U}}

\newcommand{\chia}[2]{\chi_\text{#1}^{#2}}

\newcommand{\chib}[2]{\bar{\chi}_\text{#1}^{#2}}

\newcommand{\kv}{\mathbf{k}}

\newcommand{\Efd}{\tilde{\mathbf{{E}}}}
\newcommand{\Cfd}{\tilde{\mathbf{{C}}}}

\newcommand{\Lf}{\mathbf{{L}}}
\newcommand{\Hfd}{\tilde{\mathbf{{H}}}}
\newcommand{\Pfd}{\tilde{\mathbf{{P}}}}
\newcommand{\Mfd}{\tilde{\mathbf{{M}}}}
\newcommand{\Jfd}{\tilde{\mathbf{{J}}}}
\newcommand{\Kfd}{\tilde{\mathbf{{K}}}}

\newcommand{\RT}{\mathbf{{R}}_\text{T}}
\newcommand{\NT}{\mathbf{{N}}_\text{T}}

\newcommand{\Esfd}{\tilde{\boldsymbol{\mathcal{E}}}}

\newcommand{\Lo}{\mathbf{\mathcal{L}}}
\newcommand{\Ro}{\mathbf{\mathcal{R} }}
\newcommand{\st}{\text{s}}
\newcommand{\Nh}{\mathbf{\hat n}}

\newcommand{\0}{\varnothing}

\newcommand{\Cv}{\mathbb{C}}

\newcommand{\Pv}{\overline{\mathbb{P}}}

\newcommand{\Sv}{\mathbb{S}}

\newcommand{\Rv}{\mathbb{R}}
\newcommand{\Xv}{\mathbb{X}}

\newcommand{\Lv}{\mathbb{L}}
\newcommand{\Ev}{\mathbb{E}}

\newcommand{\Dv}{\mathbb{D}}

\newcommand{\Jv}{\mathbb{J}}

\newcommand{\Hv}{\mathbb{H}}

\newcommand{\Kv}{\mathbb{K}}

\newcommand{\chit}{\overline{\overline{\chi}}}

\newcommand{\ptt}{\text{p}}
\newcommand{\rbb}{\mathbbm{r}}
\newcommand{\St}{\text{S}}

\newcommand{\Lff}{\boldsymbol{\mathfrak{L}}}
\newcommand{\Ef}{\boldsymbol{\mathfrak{E}}}
\newcommand{\Hf}{\boldsymbol{\mathfrak{H}}}
\newcommand{\Xf}{\boldsymbol{\mathfrak{X}}}

\newcommand{\Df}{\boldsymbol{\mathfrak{D}}}

\newcommand{\Ft}{\text{F}}

\newcommand{\ee}{\text{ee}}
\newcommand{\ab}{\text{ab}}
\newcommand{\zz}{{zz}}
\newcommand{\zx}{{zx}}
\newcommand{\zy}{{zy}}

\newcommand{\nn}{{xx}}  
\newcommand{\yy}{{yy}}

\renewcommand{\tt}{{yy}}

\newcommand{\mm}{\text{mm}}
\newcommand{\me}{\text{me}}
\newcommand{\emm}{\text{em}}

\usepackage{empheq}

\definecolor{burntorange}{rgb}{0.8, 0.28, 0.0}
\definecolor{myGreen}{rgb}{0.0, 0.5, 0.0}
\definecolor{amber}{rgb}{0.8, 0.28, 0.0}
\definecolor{ceruleanblue}{rgb}{0.16, 0.28, 0.75}

\definecolor{ao}{rgb}{0.0, 0.5, 0.0}
\definecolor{cobalt}{rgb}{0.0, 0.28, 0.67}
\definecolor{amber}{rgb}{0.8, 0.36, 0.27}
\definecolor{ltblu}{rgb}{0.23, 0.27, 0.29}

\usepackage{balance}

\begin{document}
\title{Part 2: Spatially Dispersive Metasurfaces - IE-GSTC-SD Field Solver with Extended GSTCs}

\author{Tom J. Smy, João G. Nizer Rahmeier, \IEEEmembership{Student Member, IEEE}, Jordan Dugan\\ and Shulabh Gupta \IEEEmembership{Senior Member, IEEE}
\thanks{ Tom J. Smy, João G. Nizer Rahmeier, Jordan Dugan, and Shulabh Gupta are with Carleton University, Ottawa, Canada (e-mail: tjs@doe.carleton.ca). }}

\maketitle

\begin{abstract}
An Integral Equation (IE) based field solver to compute the scattered fields from spatially dispersive metasurfaces is proposed and numerically confirmed using various examples involving physical unit cells. The work is a continuation of Part-1~\cite{Part_1_Nizer_SD}, which proposed the basic methodology of representing spatially dispersive metasurface structure in the spatial frequency domain, $\boldsymbol{k}$. By representing the angular dependence of the surface susceptibilities in $\boldsymbol{k}$ as a ratio of two polynomials, the standard Generalized Sheet Transition Conditions (GSTCs) have been extended to include the spatial derivatives of both the difference and average fields around the metasurface. These extended boundary conditions are successfully integrated here into a standard IE-GSTC solver, which leads to the new IE-GSTC-SD simulation framework presented here. The proposed IE-GSTC-SD platform is applied to various uniform metasurfaces, including a practical short conducting wire unit cell, as a representative practical example, for various cases of finite-sized flat and curvilinear surfaces. In all cases, computed field distributions are successfully validated, either against the semi-analytical Fourier decomposition method or the brute-force full-wave simulation of volumetric metasurfaces in the commercial Ansys FEM-HFSS simulator.  
\end{abstract}



\section{Introduction}

Electromagnetic Metasurfaces are constructed using sub-wavelength resonators of diverse geometrical shapes and material characteristics, which give rise to their tailored macroscopic responses. These microscopic resonators act as engineered electric and magnetic scatterers operating on the incoming incident fields to shape the electromagnetic fields in either space or time or both \cite{Metasurface_Review}. This has led to a powerful wave engineering paradigm, leading to a variety of applications across the electromagnetic spectrum, ranging from cloaking and illusions to holograms, and real-time reconfiguration of wireless environments, to name a few~\cite{Optical_MS_Reconfig, Reconfigr_MS, CloakingReview, MS_review_Yu, Smy_Close_ILL, smy2020IllOpen, Fink_AI_Metasurface, dugan_smy_gupta_2021}.

Since the metasurfaces are constructed from sub-wavelength resonating particles but are typically electrically large, they are naturally multi-scale in nature. As a result, the determination of electromagnetic fields scattered off them for a given incident field is typically a computationally expensive task if performed using standard brute force commercial full-wave simulators. Consequently, surface susceptibilities, $\bar{\bar{\chi}}(\omega)$ have recently been used as compact full-wave simulation models of wide-variety of practical metasurfaces with good success \cite{smy2021iegstc, Compact_Chi_Tiukuvaara}. They represent zero thickness sheet models involving electric and magnetic surface polarization densities, which when coupled with the Generalized Sheet Transition Conditions (GSTCs)~\cite{KuesterGSTC,IdemenDiscont}, can describe the average macroscopic fields around the metasurface and their angular scattering properties for arbitrary incident fields~\cite{Karim_Angular_MS,Karim_Bianiso_MS, GenBCEM, Xiao:2019aa}.

The surface susceptibilities relate the average fields with the induced surface polarizations, and so far in the literature, they have been typically restricted to modeling local interactions with the surface only. Such a point-by-point interaction consequently results in angular independent surface susceptibilities, and the metasurfaces can be referred to as \emph{spatially non-dispersive}. This has been found to be sufficiently adequate for modeling deeply sub-wavelength resonant structures. However, typical practical metasurfaces are not always deeply sub-wavelength, and their unit cell periodicities may easily reach close to a wavelength in size, as, for instance, in all-dielectric Huygen's structures~\cite{Cylindrical_DMS, West_DMS_Lens, AllDieelctricMTMS, Soichi_Huygen_mmWave}. In such cases, the field interaction with the metasurfaces is typically non-local, and the metasurfaces become \emph{spatially dispersive}. Such structures, in general, cannot be modeled using constant dipolar surface susceptibility models (even if the normal surface polarizations are included) to describe their angular scattering behavior.

Very little work has been done to model spatially dispersive metasurfaces and has been typically limited to weak spatial dispersion (SD) with limited success~\cite{Asadchy2017, Albooyeh_SD,achouri2021multipolar, achouri2021extension}. To address this problem, in Part-1 of this work \cite{Part_1_Nizer_SD}, we have proposed a simple method to model a general spatially dispersive metasurface, whose angle dependence of the surface susceptibilities have been expressed as a ratio of two polynomials of the spatial frequency, $\mathbf{k}_{||}$ (i.e., the transverse wave vector). This, when transformed into the spatial domain via inverse spatial Fourier transform, leads to an extended form of GSTCs, which involves spatial derivatives of both the difference and the average fields around the metasurface. To the best of our knowledge, this has been the most general treatment of spatially dispersive metasurfaces in the literature.

As mentioned above, surface susceptibilities are compact zero thickness models of otherwise finite thickness volumetric metasurfaces, which are ideal alternatives to rigorously and efficiently compute scattered macroscopic fields. While several works have been reported in the literature integrating standard GSTCs into a variety of numerical platforms such as Finite Difference (FD) \cite{Caloz_MS_Siijm, Vaheb_FDTD_GSTC, Smy_Metasurface_Space_Time} and Integral Equation (IE) based methods \cite{smy2021iegstc, Smy_Close_ILL, dugan_smy_gupta_2021, Caloz_MS_IE}, for instance, they naturally have been limited to modeling spatially non-dispersive metasurfaces with angle independent surface susceptibilities. This work (Part-2) continues the general treatment of spatially non-dispersive metasurfaces from Part-1 and integrates the general spatial boundary conditions via the extended GSTCs into the IE simulation framework. We develop the integrated matrix formulation of extended GSTCs and the IE-based field propagation, rigorously and efficiently describing field scattering in response to arbitrarily specified incident fields.

This work is structured as follows. Sec.~II presents the extended GSTCs and develops the fields equations that describe a general spatially dispersive metasurface and present an important specialization based on a physically motivated Lorentz oscillator. Sec.~III presents the integrated IE-GSTC formulation equipped with spatial dispersion, i.e., IE-GSTC-SD. This section transforms the desired fields equations into a matrix form and integrates with the IE field propagators, resulting in a system matrix formulation that must be self-consistently solved to compute the unknown surface currents and the subsequent scattered fields. Sec.~IV presents the verification of IE-GSTC-SD by comparing it with a semi-analytical Fourier decomposition method applied to uniform structures excited with 2D Gaussian beams. Next, the method is used to model an example practical structure composed of an electric dipole formed using a finite-length conducting wire, which exhibits an elementary Lorentzian form of spatial dispersion (Sec.~V). Fields scattered from finite-sized flat and curvilinear structures are presented and compared with brute-force 3D full-wave simulations performed in Ansys FEM-HFSS to successfully retrieve intricate interference field patterns that are otherwise missed using spatially non-dispersive models. Finally, conclusions are presented in Sec.~VI.

\section{Spatially Dispersive Metasurfaces}

\subsection{Generalized Sheet Transition Conditions (GSTCs)}

The electromagnetic field interaction with an equivalent zero thickness metasurface model is governed by the GSTCs as given by
\begin{subequations}\label{Eq:ConvGSTC}
\begin{align}
	\Delta \Efd &= j\omega (\Nh \times  \Mfd) - \nabla_{T}\left(\frac{\Pfd_{n}}{\epsilon}\right)\\
	\Delta \Hfd  &= -j\omega (\Nh \times\Pfd) - \nabla_{T}\left(\frac{\Mfd_{n}}{\mu}\right),
\end{align}
\end{subequations}
where $\Pfd$ and $\Mfd$ are the temporal frequency domain electric and magnetic surface polarizations. These surface polarizations, for a \emph{spatially non-dispersive metasurfaces}, following local field interactions are given by
\begin{subequations}\label{Eq:PM_NSD}
\begin{align}
    \Pfd &= \epsilon_0 \chit_\ee \Efd_\text{av} + \frac{1}{\eta_0} \chit_\text{em} \Efd_\text{av}\\
    \Mfd &= \mu_0 \chit_\mm \Hfd_\text{av} + \frac{1}{\eta_0} \chit_\text{me} \Efd_\text{av}
\end{align}
\end{subequations}
where the susceptibility tensors are $3\times 3$ in size and defined by:
\begin{align*}
    \chit_\ab  = \bbmatrix \chi^{xx}_\ab&\chi^{xy}_\ab&\chi^{xz}_\ab\\
                                 \chi^{yx}_\ab&\chi^{yy}_\ab&\chi^{yz}_\ab\\
                                 \chi^{zx}_\ab&\chi^{zy}_\ab&\chi^\zz_\ab\ebmatrix
\end{align*}
where $\ab \in [\ee,\mm,\me,\emm]$. For later simplicity, but without loss of generality, let us ignore the bi-anisotropic terms and the normal polarization, so that we have,
\begin{align} \label{Eq:GSTCsimple}
    \Pfd &= \epsilon_0 \chit_\ee \Efd_\text{av}, \quad 
    \Mfd = \mu_0 \chit_\mm \Hfd_\text{av}
\end{align}
Each of the susceptibility terms present in these equations creates a contribution to the polarizations in \eqref{Eq:PM_NSD}; for example, $\chi_\ee^\zz$ creates a component (e.g., for a metasurface lying along $y-$axis and located at $x=0$ excited with a TE mode, with $E_z$, $H_y$ and $H_x$ components), 
\begin{align} \label{Eq:Pzz}
    \Pt_z^z = \epsilon_0 \chi_\ee^\zz \Et_{z,av}
\end{align}
where the superscript on $P_z^z$ indicates this is the portion of $P_z$ generated by $E_{z,\text{av}}$. The total $P_z$ would thus be,
\begin{align}
    \Pt_z &= \Pt_z^x + \Pt_z^y + \Pt_z^z \notag\\
    &= \epsilon_0 \chi_\ee^\zx \Et_{x,\text{av}} + \epsilon_0 \chi_\ee^\zy \Et_{y,\text{av}} + \epsilon_0 \chi_\ee^\zz \Et_{z,\text{av}}.
\end{align}
This equation represents that the electric (and analogously magnetic) surface polarization at any position of the metasurface is given by the spatial product of the susceptibility and the average fields at that position only, i.e., a local response. Alternatively, it can be seen as assuming a set of angle independent susceptibilities that only captures the field scattering behavior of \emph{spatially non-dispersive} metasurface unit cells.

\subsection{Spatial Dispersion and the Extended GSTCs}

For a spatially dispersive metasurface, the induced electric (and magnetic) surface polarizations on the surface not only depend on the local average fields but also the fields across the metasurface~\cite{Part_1_Nizer_SD}. Assuming a vertical surface at $x=0$ we can write,
\begin{align}\label{Eq:PzS}
    \Pt_z^z(y) = \epsilon_0 \chi_\ee^\zz(y) \ast \Et_{z,\text{av}}(y)
\end{align}
where $\ast$ represents a spatial convolution. This relationship can be equivalently expressed in the \emph{spatial frequency domain, $k_y$} (each $k_y$ represents an obliquely propagating plane-wave), using the Fourier transform $\mathcal{F}_y\{\cdot\}$, so that\footnote{The various fields and field components are represented as $\E(\r,t)$ for the space and time domains, $\Efd(\r, \omega)$ for the temporal frequency domain, $\boldsymbol{\mathcal{E}}(\kv,t)$ for the spatial frequency domain, and $\Esfd(\kv,\omega)$ for the spatial and temporal frequency domains. Time convention used is $e^{j\omega t}$.} 
\begin{align}\label{Eq:PzSF}
    \Ptf_z^z(k_y) = \epsilon_0 \chi_\ee^\zz(k_y)\cdot \Etf_{z, \text{av}}(k_y)
\end{align}
which is a simple product of the average fields and the surface susceptibilities, as opposed to convolution in space, making it a suitable choice for compact unit cell description and later numerical computation.

In order to describe a general unit cell, we can represent $\chi_\ee^\zz(k_y)$ as a ratio of two polynomials to capture the angular dependence of the metasurface. Such a form incorporates possible zeros and poles of the surface susceptibilities, which effectively captures the metasurface response for the sweeping angle of incident plane-waves~\cite{Part_1_Nizer_SD}. Such a form reads,
\begin{align}\label{Eq:Ratio}
\Ptf_z^z(k_y) = \epsilon_0\left(\frac{\sum_m a_m k_y^m}{\sum_n b_n k_y^n}\right)\cdot\Etf_{z,\text{av}}(k_y)
\end{align}
where $a_m$ and $b_n$ are known complex coefficients, which can be extracted from unit cell simulations of a given metasurface structure. 

\begin{figure}[tbp]
\begin{center}
   	 \begin{overpic}[width=0.7\linewidth,grid=false,trim={0cm 0cm 0cm 0cm},clip]{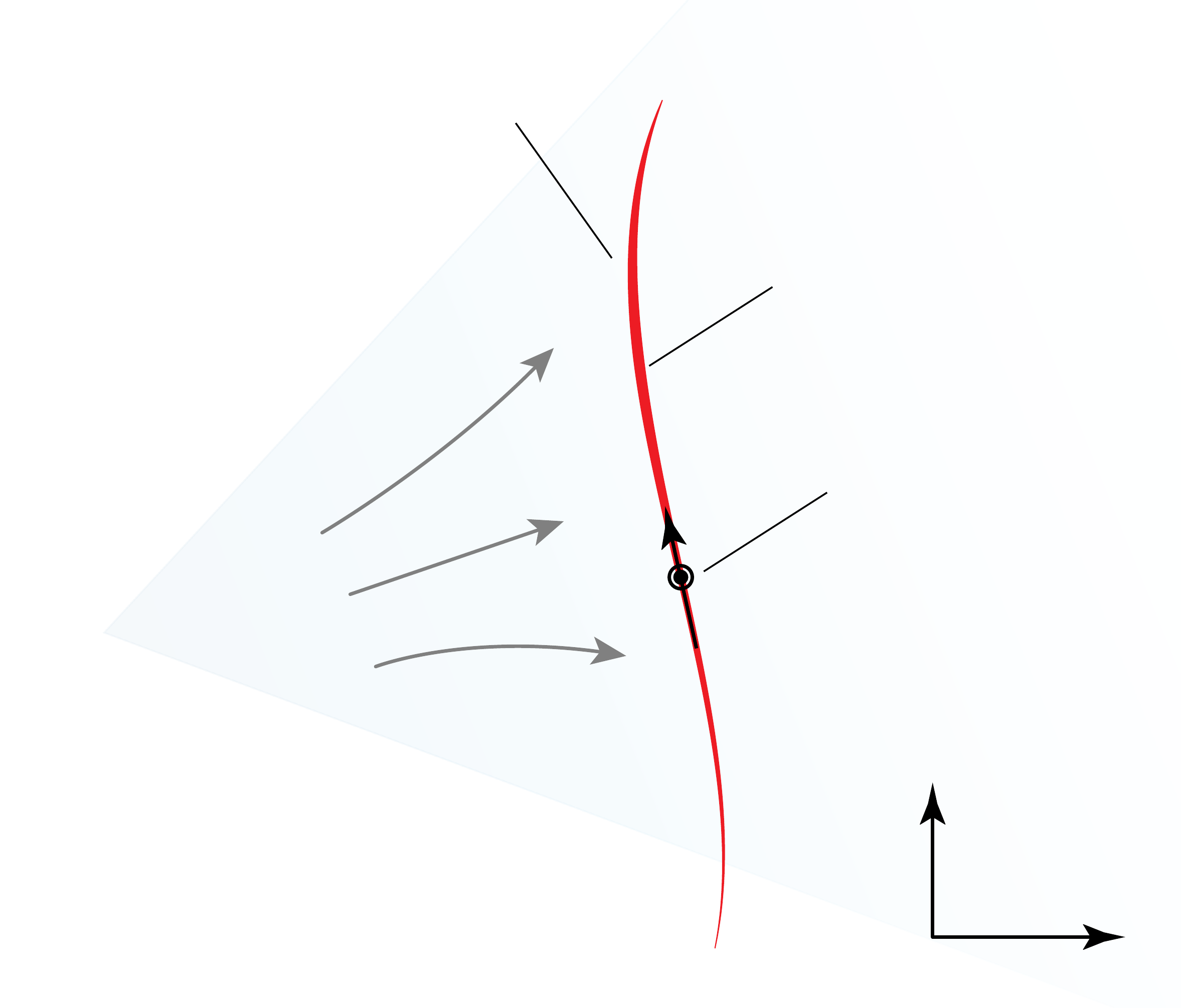}
        \put(21,  78){\htext{\scriptsize \color{amber}\shortstack{Zero-Thickness \\ \textbf{\textsc{Spatially Dispersive Metasurface}} \\ $\bar{\bar{\chi}}_\text{ee}(\r,\theta)$,~$\bar{\bar{\chi}}_\text{mm}(\r, \theta)$ }}}
         \put(15,  15){\htext{\scriptsize \shortstack{Specified Incident Fields,\\ $\psi^\text{inc.}(\r)$: TE Mode, $\{E_z, H_x,H_y\}$}}}
         \put(94,  45){\htext{\scriptsize \shortstack{Tangential Surface Polarization,\\ $\{P_z,M_y\}$ [e.g. \eqref{Eq:PzS} or \eqref{Eq:PzSF}]}}}
          \put(84,  65){\htext{\scriptsize \shortstack{Extended GSTCs,\\  \eqref{Eq:GenGSTC1} \& \eqref{Eq:GenGSTC2}}}}
            \put(98,  5.5){\htext{\scriptsize $x$}}
            \put(79.5, 21){\htext{\scriptsize $y$}}
    \end{overpic}
\end{center}
\caption{The field scattering problem considered in this work, where a spatially dispersive metasurface characterized by angle-dependent tangential surface susceptibilities is illuminated with an incident wave. For simplicity but without loss of generality, spatially symmetric surfaces with no bi-anisotropic or normal susceptibility terms are assumed throughout this work, along with TE mode excitation. }\label{fig:Problem}
\end{figure}
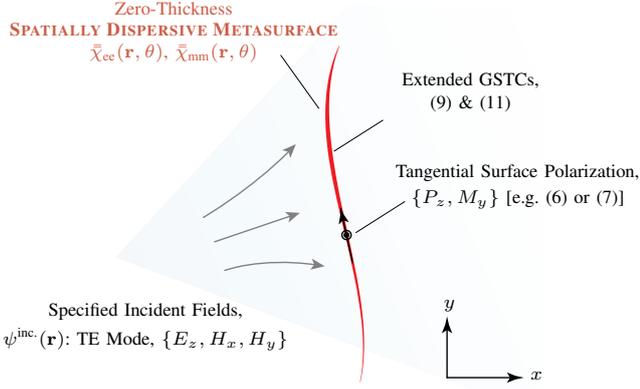

To express this relationship in the spatial domain and represent a spatial boundary condition across the metasurface, we utilize the GSTCs of \eqref{Eq:ConvGSTC}. For a vertical surface with surface normal $\nh = \{1 \; 0 \; 0 \}$, we note that $\Delta H_y = j\omega \Pt_z$, and thus we can associate a portion of $\Delta H_y$ with each portion of $\Pt_z$ and transform back to the spatial domain. This would give for the $\Ptf_z^z(k_y)$ portion,
\begin{align}\label{Eq:GenGSTC1}
\sum_n b_n j^n  \frac{\partial^n \Delta\Ht_y^z}{\partial y^n} = j\omega \sum_m a_m j^n \frac{\partial^m \Et_{z,\text{av}}}{\partial y^m}
\end{align}
resulting in general form of \eqref{Eq:Pzz} that incorporates spatial derivatives of both $\Delta \Ht_y^z$ and $\Et_{z,\text{av}}$. The total field difference is then given by,
\begin{align} \label{Eq:DeltaHy}
    \Delta \Ht_y = \Delta \Ht_y^x + \Delta \Ht_y^y + \Delta \Ht_y^z
\end{align}
with an equation of the form of \eqref{Eq:GenGSTC1} being defined for all three terms. Likewise, we would define similar relationships for the other components ($\Delta \Et_z$), i.e.
\begin{align}\label{Eq:GenGSTC2}
\sum_n d_n j^n  \frac{\partial^n \Delta \tilde{E}_z}{\partial y^n} &= j\omega \sum_m c_m j^m  \frac{\partial^m \tilde{H}_{y,\text{av}}}{\partial y^m}.
\end{align}
The two boundary equations \eqref{Eq:GenGSTC1} and \eqref{Eq:GenGSTC2}, are now referred to as the \emph{extended GSTCs}, which can now be applied to the general fields scattering problem as illustrated in Fig.~\ref{fig:Problem}.

\subsection{Lorentz Oscillators}

Although the derivatives of the difference and average fields present in \eqref{Eq:GenGSTC1} and \eqref{Eq:GenGSTC2} can be of use and are amenable to the methods presented in this paper, we shall consider the analysis of an important subset which can be physically interpreted and understood, i.e., a Lorentzian oscillator. Let us consider that a collection of Lorentz resonators can describe the resonant response of a metasurface unit cell in the temporal- and spatial-frequency domain as, again using $\Ptf_z^z$ as an example,
\begin{align}\label{Eq:Lor_kx_w}
-\omega^2\Ptf_z^z + j\gamma\omega \Ptf_z^z  + \omega_0^2\Ptf_z^z = \epsilon_0\omega_p^2\Etf_{z,\text{av}}
\end{align}
where $\{\omega_0, \omega_p,\gamma\}$ are the resonant frequency, plasma frequency, and the damping coefficient, respectively, which all depend on the geometrical and electrical characteristics of the unit cell. Moreover, they also depend on the angle of incidence of the incoming plane-waves \cite{Part_1_Nizer_SD}. We can do a Taylor expansion, for instance, around the normal incidence (i.e. $k_0\sin\theta = k_y = 0$), so that
\begin{subequations}\label{Eq:gamma_omega_kx}
\begin{align}
\gamma &= \alpha_0 + \alpha_1 k_y + \alpha_2 k_y^2 + O(k_y^n) \\
\omega_p^2 &= \beta_0^2 +\beta_1 k_y  + \beta_2 k_y^2 + O(k_y^n) \\
\omega_0^2 &= \zeta_0^2 +\zeta_1 k_y  + \zeta_2 k_y^2 + O(k_y^n)
\end{align}
\end{subequations}
Using \eqref{Eq:gamma_omega_kx} in \eqref{Eq:Lor_kx_w}, as shown in \cite{Part_1_Nizer_SD}, we can relate the average fields with the polarization in the following form in the spatial frequency domain:
\begin{align}\label{Eq:Lor_kx_w_2}
\Ptf_z^z  = \epsilon_0 \frac{\chi^\zz_\ee}{(1 + j\xi_{\ee,1}^\zz k_y - \xi_{\ee,2}^\zz k_y^2)}\Etf_{z,\text{av}},
\end{align}
which upon an inverse spatial Fourier transform leads to~\cite{Part_1_Nizer_SD},
 \begin{align}\label{Eq: Lorentz Simplified}
      \xi_{\ee,2}^\zz \frac{\partial^2\Pt_z^z}{\partial y^2} +  \xi_{\ee,1}^\zz \frac{\partial \Pt_z^z}{\partial y} + \Pt_z^z = \epsilon_0 \chi_\ee^\zz \Et_{z,\text{av}}.
 \end{align}
Equation \eqref{Eq: Lorentz Simplified} is a simplification of \eqref{Eq:GenGSTC1} with only \nth{2} order terms on the left-hand side. A more complex form considers \nth{2} order derivatives of the average fields as well, and is considered in a general form in Sec. \ref{Sec: Generalized Surface Equations}. It is worth mentioning that for a symmetric spatial dispersion condition, the term $\xi_1$ in \eqref{Eq:Lor_kx_w_2} is zero. That is the case for the practical metasurface cells we will analyze later in this paper for simplicity and illustration. This form thus represents a very simple physically motivated case of a spatial dispersion, which is exhibited by a simple array of short conducting wires, as shown in \cite{Part_1_Nizer_SD} and to be used later in the example section of this work.

For general and more complex unit cell architectures (such as Huygens' structures, for example), we can assume that the surface polarization components are described using $N_L$ Lorentz spatial resonances, so that Eq. \eqref{Eq:Lor_kx_w_2} assumes the form,
\begin{align}\label{Eq:Lor}
\Ptf_z^z &=  \Ptf_{z_0}^z + \sum_{i=1}^{N_L} \Ptf_{z,i}^z \\
&= \epsilon_0\chi_{\ee_0}^\zz\Etf_{z,\text{av}} + \sum_{i=1}^{N_L}  \epsilon_0 \frac{\chi_{\ee,i}^\zz}{(1 + j\xi_{\ee,1,i}^\zz k_y - \xi_{\ee,2,i}^\zz k_y^2)}\Etf_{z,\text{av}}
\end{align}
Taking the inverse spatial Fourier transform,
\begin{align}\label{Eq:Ratio2}
\sum_{i=1}^{N_L} \left[ \xi_{\ee,2,i}^\zz \frac{\partial^2\Pt_{z,i}}{\partial y^2} +  \xi_{\ee,1,i}^\zz\frac{\partial\Pt_{z,i}}{\partial y}
+ \Pt_{z,i}\right] = \notag\\
\epsilon_0\chi_{\ee_0}^\zz\Et_{z,\text{av}} + \epsilon_0 \sum_{i=1}^{N_L} \chi_{\ee,i}^\zz \Et_{z,\text{av}}.
\end{align}
As before, we use $\Delta\Ht_{y,i}^z = j\omega \Pt_{z,i}^z$, then using superposition and the independence of the various polarization components, we get,
\begin{align} \label{Eq:DeltaHyL}
    \Delta\Ht_y^z &= \Delta \Ht_{y_0}^z + \sum_{i=1}^{N_L} \Delta \Ht_{y,i}^z = j\omega\Pt^z_{z_0} + \sum_{i=1}^{N_L} j\omega \Pt^z_{z,i}
\end{align}
\noindent defining,
\begin{equation}
    \Delta \Ht_{y,i}^z = j \omega \Pt_{z,i}^z
\end{equation}
\noindent where the \nth{0} order term is 
\begin{align} \label{Eq:0th}
\Delta \Ht_{y_0}^z = j\omega\epsilon_0\chi_{\ee_0}^\zz\Et_{z,\text{av}}
\end{align}
\noindent For each term $1 <i \leq N_L$ we can write an equation of the form,
\begin{align}\label{Eq:GenHy}
\xi_{\ee,2,i}^\zz \frac{d^2\Delta \Ht_{y,i}^z}{dy^2} +  \xi_{\ee,1,i}^\zz\frac{d\Delta \Ht_{y,i}^z}{dy} + \Delta \Ht_{y,i}^z = j \omega\epsilon_0 \chi_{\ee,i}^\zz \Et_\text{z,av}
\end{align}
Which describes the surface relationship between a component of $\Delta \Ht_y$ and $\Et_{z,\text{av}}$.\footnote{To make clear how the field differences are captured, the total field difference for a single component (such as $\Delta \Ht_y$) is comprised of three fundamental contributions each due to generation from an average field component as described in \eqref{Eq:DeltaHy}. Each of these contributions is in turn formed from contributions due to several resonances, see \eqref{Eq:DeltaHyL}. Each of the contributions, for example, $\Ht_{y,i}^z$, is independently generated and, when summed, forms a general solution to the field equation for $\Delta \Ht_y$ at the interface.}

To complete this formulation, we can  write each Lorentzian equation associated with the tensorial form of the susceptibilities and introduce a generalized differential operator $\L$,
\begin{subequations}\label{eq:lor}
\begin{align}
\L_{\ee,i}^{\alpha\beta} \cdot {\Delta\Ht_{\gamma,i}^\beta}   &= j\omega \epsilon_0 \chi^{\alpha\beta}_{\ee,i}{\Et}_{\beta,\text{av}}\label{eq:lora}\\
\L_{\mm,i}^{\alpha\beta} \cdot {\Delta\Et_{\gamma,i}^\beta}   &= j\omega \mu_0 \chi^{\alpha\beta}_{\mm,i}{\Ht}_{\beta,\text{av}}\label{eq:lorb}
\end{align}
\end{subequations}
where,
\begin{subequations}\label{Eq:Lop}
\begin{align}
    \L_{\ee,i}^{\alpha\beta} &= \left[\xi^{\alpha\beta}_{\ee,2,i}\frac{\partial^2 }{\partial y^2} + \xi^{\alpha\beta}_{\ee,1,i} \frac{\partial }{\partial y} +  1 \right]
    \\
    \L_{\mm,i}^{\alpha\beta} &= \left[\xi^{\alpha\beta}_{\mm,2,i}\frac{\partial^2 }{\partial y^2} + \xi^{\alpha\beta}_{\mm,1,i} \frac{\partial }{\partial y} +  1\right]
\end{align}
\end{subequations}
with $\alpha \in \{x,y,z\}$, $\beta \in \{x,y,z\}$ and $\gamma$ is defined by $\nh \times \{x,y,z\}$ with respect to value of $\alpha$.

\subsection{Generalized Surface Equations}
\label{Sec: Generalized Surface Equations}
The previous section shows the surface formulation for the case of \eqref{Eq:Lor} with a second-order denominator and constant numerator (no zeros). A more general (but still \nth{2} order equation) can be obtained by using a \nth{2} order polynomial in $k_y$ for the numerator of \eqref{Eq:Lor},
\begin{align}
\Ptf_z^z 
&= \epsilon_0\chi_{\ee_0}^\zz\Etf_{z,\text{av}} + \sum_{i=1}^{N_L}  \epsilon_0 \frac{\chi_{\ee,0,i}^\zz+ j\chi_{\ee,1,i}^\zz k_y - \chi_{\ee,2,i}^\zz k_y^2}{1 + j\xi_{\ee,1,i}^\zz k_y - \xi_{\ee,2,i}^\zz k_y^2}\Etf_{z,\text{av}}
\end{align}
This form incorporates spatial derivatives of the average fields, when transformed to the space domain, and can be related to the normal components of the surface polarizations in some special cases. This extension will later allow for a very simple test to validate the IE-GSTC-SD formulation using a well characterized unit cell with constant normal surface susceptibility.

By following the same procedure as used for the field differences on the left-hand side of \eqref{Eq:Ratio2} to handle the spatial derivatives of the average field, we define two more operators associated with each component,
\begin{subequations} \label{Eq:X}
\begin{align}
    \X_{\ee,i}^{\alpha\beta} &= \left[\chi^{\alpha\beta}_{ee,2,i}\frac{\partial^2 }{\partial y^2} + \chi^{\alpha\beta}_{ee,1,i} \frac{\partial }{\partial y} +  \chi_{\ee,0,i}^\zz \right]
    \\
    \X_{\mm,i}^{\alpha\beta} &= \left[\chi^{\alpha\beta}_{\mm,2,i}\frac{\partial^2 }{\partial y^2} + \chi^{\alpha\beta}_{\mm,1,i} \frac{\partial }{\partial y} +  \chi_{\mm,0,i}^\zz\right]
\end{align}
\end{subequations}
we can generalize \eqref{eq:lor} to 
\begin{subequations}\label{Eq:Lor_2_2_BC}
\begin{align}
\L_{\ee,i}^{\alpha\beta} \cdot {\Delta\Ht_{\gamma,i}^\beta}   &= j\omega \epsilon_0 \X^{\alpha\beta}_{\ee,i}{\Et}_{\beta,\text{av}}\label{eq:lora2}\\
\L_{\mm,i}^{\alpha\beta} \cdot {\Delta\Et_{\gamma,i}^\beta}   &= j\omega \mu_0 \X^{\alpha\beta}_{\mm,i}{\Ht}_{\beta,\text{av}}\label{eq:lorb2}
\end{align}
\end{subequations}

This generalized form can capture more complex angular dependence arising from the structure of the unit cell~\cite{Part_1_Nizer_SD}. We should note that we have assumed a \nth{2} order for both operators, but higher orders could be used. However, odd orders will produce reflection asymmetry and may be omitted for physical reasons. 

\section{IE-GSTC-SD formulation}

Our next task is to integrate the zero-thickness boundary condition of \eqref{Eq:Lor_2_2_BC} into the bulk Maxwell's equations, to develop a general-purpose field scattering solver accounting for spatial dispersion in an IE approach with one or more surfaces. Firstly, we formulate the field equations at the surface for a discretized surface by transforming equations such as \eqref{Eq:GenHy} into matrix equations suitable for incorporation into a system-level solution. Secondly, we form the propagation equations for the scattered fields generated by an implicit incident field, which requires propagation matrices for a self-consistent scattering solution. Finally, the system matrix and source vector must be formed to provide an equation to be solved. We will now take each of these tasks in turn.

\subsection{Surface Discretization}

The IE approach, in general, requires discretized surfaces and follows the Boundary Element Method (BEM) techniques\cite{chew2009integral,Method_Moments,stewart2019scattering}. These surfaces connect regions of homogeneous material properties, coupling the regions together through transmissive and reflective properties. Although a defined surface must surround a region, a portion of the surface may be at infinity (where the fields are assumed to be zero) or implicitly present and not actually modeled. The characteristics of the implicit surface will depend on the assumptions about the implicit excitation.

The problems addressed in this paper will be 2D scattering cases and the surfaces are curvilinear line elements, as illustrated in Fig.~\ref{fig:Problem}.  For surfaces which are to be physically modeled, we impose a discretization using a uniform segmentation of each surface. Each discrete element of the surface is characterized by a center position $\r_{\ptt,i} = [x_\ptt ,y_\ptt ,0]_i$, a length $\ell_i$ and a normal $\Nh_i$ which are collected into vectors such as $\rbb_\ptt = \bbmatrix \r_{\ptt,1} & \dots & \r_{\ptt,m} \ebmatrix$. The field quantities are stored in vectors of the same form (with $m$ surface elements):
\begin{align*}
    \Ev &= \bbmatrix \Efd_1 & \Efd_2 & \dots & \Efd_{m} \ebmatrix\\
    \Hv &= \bbmatrix \Hfd_1 & \Hfd_2 & \dots & \Hfd_{m} \ebmatrix\\
    \Sv_\Ft &= \bbmatrix \Ev^- & \Hv^- &\Ev^+ & \Hv^+  \ebmatrix
\end{align*}
The surface field vector $\Sv_\Ft$ holds the fields present on both sides of a surface ($+$ or $-$) and will be useful in the formulation of the IE problem \cite{smy2021iegstc}. The fields are stored sequentially along the surface and can be thought of as triplets of fields ($\Ef_i$ and $\Hf_i$). Therefore, facilitating the operators' use to implement the spatial derivatives in the general formulation given above.

\subsection{Surface Equations}

The GSTCs shown in \eqref{Eq:ConvGSTC} when formulated using matrix operators and omitting the bi-anisotropic terms for simplicity are
\begin{subequations}
\begin{align}
	\NT \Delta  \Efd &= j\omega \mu_0 \RT \Mfd\\
	\NT \Delta \Hfd  &= - j\omega \epsilon_0 \RT \Pfd
\end{align}
\end{subequations}
where both $\NT$ and $\RT$ extract the tangential components of a field, but $\RT$ incorporates the rotation induced by the cross-product. The addition of the GSTCs with constant angular independent susceptibilities (i.e., spatially non-dispersive structures) to the IE formulation has been reported extensively elsewhere, and we will refer to it as the standard IE-GSTC formulation~\cite{stewart2019scattering,smy2020IllOpen,GenBCEM}.

We will now develop a spatially dispersive implementation of the GSTCs within the IE framework (IE-GSTC-SD). If we assume a vertical surface at $x = 0$ running along the $y$ axis, then we have,
\begin{align}
    \NT =  \bbmatrix 0 & 1 & 0\\ 0 & 0 & 1 \ebmatrix
\end{align}
which extracts the tangential components of the field at the surface, for example $\E_T = \bbmatrix E_y &E_z \ebmatrix^T$
and 
\begin{align}
    \R_T 
    = \bbmatrix 0 & 0 & -1\\0 & 1 & 0 \ebmatrix
\end{align}

We now simplify the following derivations (without losing any generality) by only retaining the primary tangential terms and have,
\begin{align*}
    \chit_\ee  = \bbmatrix 0 & 0 & 0\\
                0&\chi^{yy}_\ee&0\\
                0&0&\chi^\zz_\ee\ebmatrix , 
    \quad \chit_\mm  = \bbmatrix 0 & 0 & 0\\
                   0&\chi^{yy}_\mm&0\\
                   0&0&\chi^\zz_\mm\ebmatrix
\end{align*}

To generalize the relationship between $\Delta \Hfd$ and $\Efd_\text{av}$ two generalized \emph{Lorentzian Operators} are introduced [$\X_\ee/\Lf_\ee$, $\X_\mm/\Lf_\mm$ see \eqref{Eq:Lor_2_2_BC}] and we use the relationship $\Pt = j \omega \Delta \Ht$ for each component,
\begin{align*}
    \L_\ee \Delta \Hfd  & = \X_\ee \Esfd_\text{av}, \quad    \L_\mm \Delta \Efd = \X_\mm  \Hfd_\text{av} 
\end{align*}
with
\begin{subequations}
\begin{align}
    \L_{\ee/\mm} &= \bbmatrix 0 & 0 & 0\\ 0 & \L^{yy}_{\ee/\mm} & 0\\0 & 0 & \L^\zz_{\ee/\mm}\ebmatrix \\
    \X_{\ee/\mm} &= \bbmatrix 0 & 0 & 0\\ 0 & \X^{yy}_{\ee/\mm} & 0\\0 & 0 & \X^\zz_{\ee/\mm}\ebmatrix 
\end{align}
\end{subequations}
and 
\begin{subequations}
\begin{align}
    \L_{\ee/\mm}^\ab &= 1 + \sum_{i=1}^{N_L} \L^\ab_{\ee/\mm,i} \\
    \X_{\ee/\mm}^\ab &= \epsilon_0\chi_{\ee/\mm,0}^\ab + \sum_{i=1}^{N_L}  \epsilon_0 \X_{\ee/\mm,i}^\ab
\end{align}
\end{subequations}
where $\L^\ab_{\ee/\mm,i}$ and $\X^\ab_{\ee/\mm,i}$ are defined by \eqref{Eq:Lop} and \eqref{Eq:X} respectively. 

For this simplified system, we have only one contribution to each component of $\Delta \Hfd$  and $\Delta \Efd$. For example choosing an excitation polarization of $\{\Et_z,\Ht_x,\Ht_y\}$ for the case of a vertical surface at $x = 0$ we can write,
\begin{subequations}\label{Eq:MatGSTC}
\begin{align}
    \L_{\ee}^\zz \Delta \Ht_y &= \epsilon_0 \X_{\ee}^\zz \Et_{z,\text{av}}\\
    \L_{\mm}^\yy \Delta \Et_y &= \mu_0 \X_{\mm}^\yy \Ht_{y,\text{av}}
\end{align}
\end{subequations}

The primary complication in the implementation of these equations is the presence of the spatial derivatives in the $\L$ and $\X$ operators defined by \eqref{Eq:Lop} and \eqref{Eq:X}. However, a similar issue was solved in \cite{GenBCEM} to allow for the incorporation of the terms involving the gradient of the normal component of the polarizations in \eqref{Eq:ConvGSTC}, and we will follow a similar approach here.

We define central difference operators for the $n^\text{th}$ segment triplet:
\begin{align}
    \frac{\partial \Psi_n}{\partial y}  &= \Df^{(1)} \bbmatrix \Psi_{n-1} \\ \Psi_{n}\\\Psi_{n+1}\ebmatrix, \quad
    \frac{\partial^2 \Psi_n}{\partial y^2}  &= \Df^{(2)} \bbmatrix \Psi_{n-1} \\ \Psi_{n}\\\Psi_{n+1}\ebmatrix
\end{align}
where $\Psi$ would be a field difference associated with a particular resonance, such as $\Delta \Et_{z,i}$, and obtain,
\begin{subequations}
\begin{align*}
    \Df^{(1)} &= \frac{1}{2} \left[\frac{\Psi_{n+1} - \Psi_{n}}{l_{n+1/2}} +  \frac{\Psi_{n} - \Psi_{n-1}}{l_{n-1/2}} \right]\\
    \Df^{(2)} &= \left[\frac{\Psi_{n+1} - \Psi_{n}}{l_{n+1/2}} -  \frac{\Psi_{n} - \Psi_{n-1}}{l_{n-1/2}} \right] {\left(\frac{l_{n-1/2} + l_{n+1/2}}{2}\right)^{-1}}
\end{align*}
\end{subequations}
which allows us to write,
\begin{subequations}
\begin{align}\label{eq:Ratior1}
\Lff_{\mm}^\tt &=  \U + \sum_{i=1}^{N_L} \left[\xi^\tt_{\mm,2,i}\Df^{(2)}_i + \xi^\tt_{\mm,1,i} \Df^{(1)}_i +  \U\right] \\
\Lff_{\ee}^\zz &=  \U + \sum_{i=1}^{N_L} \left[\xi^\zz_{\ee,2,i}\Df^{(2)}_i + \xi^{\zz}_{\ee,1,i} \Df^{(1)}_i +  \U\right] 
\end{align}
\end{subequations}
and
\begin{subequations}
\begin{align}\label{eq:Ratior}
\Xf_{\mm}^\yy &=  \mu_0\chi_{\mm_0}^{\zz} \notag\\
&\quad + \mu_0\sum_{i=1}^{N_L} \left[\chi^\tt_{\mm,2,i}\Df^{(2)}_i + \chi^\tt_{\mm,1,i} \Df^{(1)}_i +  \chi^\yy_{\mm,0,i}\U\right] \\
\Xf_{\ee}^\zz &=  \epsilon_0\chi_{\ee_0}^{\zz}\notag\\
&\quad + \epsilon_0\sum_{i=1}^{N_L} \left[\chi^\zz_{\ee,2,i}\Df^{(2)}_i + \chi^{\zz}_{\ee,1,i} \Df^{(1)}_i +  \chi^\zz_{\ee,0,i}\U\right] 
\end{align}
\end{subequations}
where $\U = [0,~1,~0]$. By applying this equation to each segment on the surface we can arrive at the surface operators,
\begin{align*}
    \Lv  = \bbmatrix \Lff_{\alpha\beta,1} & \dots & \0 \\
                         \0 &\ddots  &\0 \\
                         \0 & \0 & \Lff_{\alpha\beta,m} \\
    \ebmatrix,
    \Xv  = \bbmatrix \Xf_{\alpha\beta,1} & \dots & \0 \\
                         \0 &\ddots  &\0 \\
                         \0 & \0 & \Xf_{\alpha\beta,m} \\
    \ebmatrix,
\end{align*}
which are diagonal matrices created from the triplet operators and would operate respectively, on all of the field differences defined by,
\begin{align*}
    \Sv_{\Delta} = \bbmatrix \Delta E_{z,0}^0 & \cdots &\Delta H_{y,N_L}^{m\times N_L} \ebmatrix^T;
\end{align*}
and the surface scattered fields $\Sv_\Ft$ and applied incident fields $\Sv^i$ to obtain,
\begin{align}\label{Eq:SurfEq}
    \Lv\Sv_{\Delta}  &= \Xv (\Sv_\Ft + \Sv^i) 
\end{align}
which is the surface level equivalent of \eqref{Eq:MatGSTC}, a generalization of the GSTCs, and completely defines the tangential field relationships at the surface. 

\subsection{IE-GSTC-SD Field Equations}

As with the surface equations, the integral expressions derived from Maxwell's equations that capture the propagation need to be put into a discretized form. The EM fields radiated into free-space from electric and magnetic current sources, $\{\J,~\K\}$, can be generally expressed using an IE formulation as \cite{chew2009integral, Method_Moments,GenBCEM}:
\begin{subequations}\label{Eq:FieldProp}
\begin{align}
    \Efd^\st(\r) &= -j\omega\mu(\Lo \Jfd)(\r,\r') - (\Ro \Kfd)(\r,\r')\\
    \Hfd^\st(\r) &= -j\omega\epsilon(\Lo \Kfd)(\r,\r') + (\Ro \Jfd)(\r,\r'),
\end{align}
\end{subequations}
with $\r$ being the point of interest, $\r'$ the position of the source current; and  $\E^\st$ and $\H^\st$ the radiated (scattered) fields from the surface.\footnote{We will denote scattered or radiated fields due to the surface currents by the superscript $\st$ and total fields which include both scattered and incident fields by a lack of superscript. Hence, generally $\E = \E^\st + \E^\text{i}$ where $\E^\text{i}$ is the incident field, for example.}  The field operators are given by:
\begin{align*}
    (\Lo \Cfd)(\r, \r') &= \int_{\ell}[1+\frac{1}{k^2}\nabla\nabla\cdotp] [G(\r,\r')\Cfd(\r')] \,d\r'\\
    (\Ro \Cfd)(\r, \r') &= \int_{\ell}\nabla \times [G(\r,\r')\Cfd(\r')] \,d\r'
\end{align*}
with $\Cfd \in \{\Jfd,\Kfd\}$. $G(\r,\r')$ represents the Green's function which, for a 2D case, is given by the Hankel function of the \nth{2} kind and represents outwardly propagating radial waves. 

 Using this discretization, \eqref{Eq:FieldProp} is transformed into a set of algebraic equations relating surface currents $\Cv = \bbmatrix \Jv & \Kv\ebmatrix^T$ to the scattered fields at $\rbb_\St$,
\begin{align}\label{eq:BEMProp}
     \bbmatrix \Ev^\st(\rbb_\ptt) \\\Hv^\st(\rbb_\ptt) \ebmatrix & = 
    \bbmatrix - j\omega \mu \Lv(\rbb _\ptt, \rbb_\St)  & - \Rv(\rbb_\ptt, \rbb_\St)  \\ 
             \Rv(\rbb_\ptt, \rbb_\St)  & - j\omega \epsilon \Lv(\rbb_\ptt, \rbb_\St) \ebmatrix 
    \bbmatrix  \Jv_\St \\ \Kv_\St\ebmatrix  
\end{align}

If $\rbb_\ptt=\rbb_\St$ then this equation determines the self-propagation from every element to every other element and we can define a surface propagation operator,
\begin{align}
    \Pv = \bbmatrix - j\omega \mu \Lv(\rbb _\ptt, \rbb_\St)  & - \Rv(\rbb_\ptt, \rbb_\St)  \\ 
             \Rv(\rbb_\ptt, \rbb_\St)  & - j\omega \epsilon \Lv(\rbb_\ptt, \rbb_\St) \ebmatrix
\end{align}
and obtain,
\begin{align}\label{Eq:SurfProp}
    \Pv \Cv = \Sv_F.
\end{align}
\subsection{System Formulation}

The final task is to assemble the surface \eqref{Eq:SurfEq} and propagation equations \eqref{Eq:SurfProp} to solve for the unknowns in the scattering problem. The unknowns in the scattering problem can be identified as the element currents $\Cv$, the surface fields $\Sv_\Ft$, and the field differences associated with the Lorentz resonators $\Sv_\Delta$. To create a complete system matrix, we have to introduce equations to force the unknown currents to be tangential to the surface, to sum the field differences, and link them to the propagation equations. 

To force the surface currents to be tangential to the surface, we introduce an operator, 
\begin{align}
    \Nv_n \Cv = 0
\end{align}
this would impose the condition $\nh \cdot \C = 0$ for all elements. For our simple case of a vertical surface, this implies that $\Jt_x$ and $\Kt_y$ are zero for all elements. To link the propagation equations to surface equations, we define two new operators such that, 
\begin{align}
    \Lv_S \Sv_\Delta = \Dv (\Sv_\Ft + \Sv^i)
\end{align}
where $\Lv_S$ is a matrix that sums the field differences associated with each resonance to form the total field differences $\Delta \Et_z$ and $\Delta \Ht_y$ at every element -- $\Dv$ is the matrix that finds the difference of the fields on the two side of the surface. This equation thus links the propagating fields $\Sv_\Ft$ to the differences associated with the resonances present $\Sv_\Delta$. 

These equations are finally assembled into a system matrix equation, resulting in
\begin{align*}
    \bbmatrix 
        \Pv & -\I &\0\\
        \Nv_n & \0 &\0 \\
        \0 &-\Dv &  \Lv_S \\
        \0 &-\Xv &\Lv \\
    \ebmatrix
    \bbmatrix \Cv \\\Sv_\Ft\\\Sv_\Delta \ebmatrix
    = 
    \bbmatrix \0 \\ \Dv\Sv^i \\ \Xv \Sv^i \ebmatrix
\end{align*}
which is square and can be solved directly to determine the unknowns. Once $\Cv$ is known, the propagation matrix defined in \eqref{eq:BEMProp} can be used to compute the scattered fields in the entire or desired simulation region.

\section{Numerical Verification}

To numerically verify the IE-GSTC-SD formulation, we will use two methods using 2D Gaussian beam illumination of uniform metasurfaces. The first approach will compare a semi-analytical technique using Fourier Decomposition (FD) of the incident waves into plane wave components and then determine reflected and transmitted fields. The second method will take a specific form of SD (angular dependence) that can model an otherwise spatially non-dispersive metasurface with two constant susceptibilities -- one of which operates on a normal component of the magnetic field. This second approach allows for a direct comparison between the standard IE-GSTC \cite{smy2021iegstc} and the proposed IE-GSTC-SD formulations for the same surface.

\subsection{Fourier Decomposition Method}

Let us consider a case where an incident field is specified, and we wish to determine the scattered fields from a uniform spatially dispersive metasurface. Since it is a linear problem, we can use the principle of superposition by expressing the incident field as a sum of uniform plane waves \cite{Saleh_Teich_FP}. The two GSTCs for the case of a vertical surface with simple tangential susceptibilities and an assumed TE polarization (w.r.t the normal $x$) given by $\{E_z, H_x, H_y\}$ can be expressed in the spatial frequency domain as, 
\begin{subequations} \label{eq:SimpleGSTC}
\begin{align}
\Delta \Etf_z(k_y)  &=  j\omega\mu_0 \chi_\mm^\yy(k_y)\Htf_{y,\text{av}}(k_y)  \\
\Delta \Htf_{y}(k_y)  &=  j\omega \epsilon_0\chi^\zz_\ee(k_y)\Etf_{z,\text{av}}(k_y)  
\end{align}
\end{subequations}
Introducing $T = \Etf_t/\Etf_0$ and $R = \Etf_r/\Etf_0$, we can show that the reflection of single plane wave at an incident angle is given by,\cite{Part_1_Nizer_SD}
\begin{subequations}\label{Eq:RT_Tang}
\begin{align}
R(\theta) &= \frac{2jk_0\{ \cos\theta^2\chi_\mm^\yy-\chi_\ee^\zz\}}{\{jk_0\chi_\ee^\zz+2\cos\theta\}\{jk_0\cos\theta \chi_\mm^\yy  +2\}}\\
T(\theta) &= \frac{\cos\theta[4+ k_0^2  \chi_\mm^\yy \chi_\ee^\zz]}{\{jk_0\chi_\ee^\zz+2\cos\theta\}\{jk_0\cos\theta \chi_\mm^\yy  +2\}}%
\end{align}
\end{subequations}
where $\chi_\mm^\yy$ and $\chi_\ee^\zz$ are functions of the incidence angles.

The above equation gives the transmitted and reflected plane waves for a particular spatial component of a general incident field. These can then be integrated over all the $k_y$ components (propagating terms) of the incident field to construct the complete reflected and transmitted fields \cite{Saleh_Teich_FP}. Specifically, if we represent the incident field at the surface as $\Et_z^i(y)$, then we can transform this to the spatial Fourier domain using,
\begin{align}
    \Etf_z^i(k_y) = \mathcal{F}_y\{\Et_z^i(y)\}
\end{align}
where we have decomposed the incident field into plane wave components. We can then form the reflected and transmitted waves,
\begin{align}
    \Etf_z^r(k_y) = R(\theta) \Etf_z^i(k_y), \quad \Etf_z^t(k_y) = T(\theta) \Etf_z^i(k_y)
\end{align}
with $\sin \theta = k_y/k_0$. Using an inverse spatial Fourier transform, we can obtain the scattered fields at the surface,
\begin{align}
    \Et_z^r(y) = \mathcal{F}^{-1}\{\Etf_z^r(k_y)\},\quad
    \Et_z^t(y) = \mathcal{F}^{-1}\{\Etf_z^t(k_y)\}
\end{align}
This methodology can be easily implemented using discrete Fourier transforms (DFTs) and can now be used to validate the IE-GSTC-SD implementation.

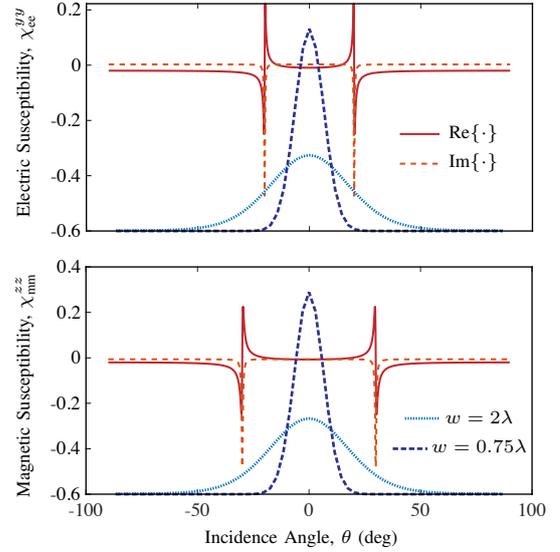
\begin{figure}[htbp]
    \centering
    \begin{overpic}[width = 0.8\linewidth]{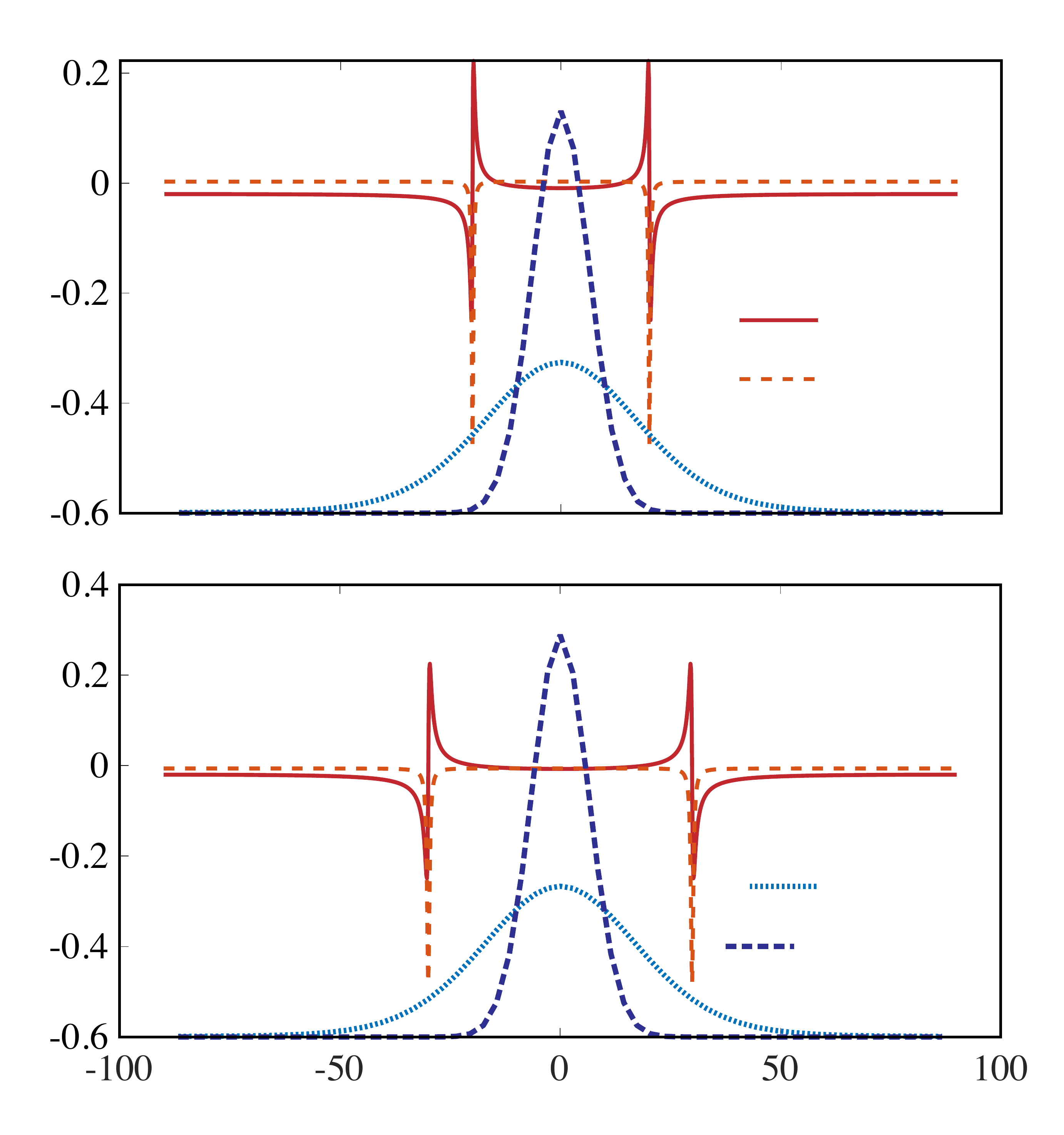}
    \put(48, 0){\scriptsize\htext{Incidence Angle, $\theta$ (deg)}}
    \put(0, 27){\scriptsize\vtext{Magnetic Susceptibility, $\chi_\mm^\zz$}}
    \put(0,75){\scriptsize\vtext{Electric Susceptibility, $\chi_\ee^\tt$}}
    \put(78, 71){\htext{\scriptsize Re$\{\cdot\}$}}
    \put(78, 66){\htext{\scriptsize Im$\{\cdot\}$}}
    \put(79, 21){\scriptsize \htext{$w = 2\lambda$}}
     \put(79, 16){\scriptsize \htext{$w = 0.75\lambda$}}
    \end{overpic} 
    \caption{Spatially dispersive metasurface described using electric and magnetic tangential susceptibilities, each with an arbitrarily chosen angular Lorentzian profile of \eqref{Eq:Lor_exm_Gb} for illustration. a) Electric susceptibility. b) Magnetic susceptibility. Also shown is the angular spectrum of two incident Gaussian beams (narrow and wide) at the surface for the simulations in Fig.~\ref{Fig:GaussianBeam}. Various surface susceptibility parameters are: $\chi_\ee^\zz(0) = -0.0092 + j0.0027$ and $\chi_\mm^\tt(0) = -0.0073 - j0.0062$, $\xi_{\ee,2}^\zz = -1.9317\times10^{-5} + j3.8635\times10^{-6}$ and $\xi_{\mm,2}^\tt = -9.1405\times 10^{-5} + j1.8281\times10^{-7}$.}
    \label{fig:chis}
\end{figure}

\subsection{2D Gaussian Beam Propagation}

Consider a uniform metasurface described using tangential susceptibilities only and with unit symmetry about both $x-$ and $y-$ axis. Assume that a single Lorentz resonator can describe it for both $\chi_\ee^\zz$ and $\chi_\mm^\nn$, as
\begin{subequations}\label{Eq:Lor_exm_Gb}
\begin{align}
     \chi_\ee^\zz(k_y) &= \chi_{\ee_0}^\zz +  \frac{\chi^\zz_{\ee,0}}{1  - \xi_{\ee,2}^\zz k_y^2}\\
    \chi_\mm^\tt(k_y) &= \chi_{\mm_0}^\tt +  \frac{\chi^\tt_{\mm,0}}{1  - \xi_{\mm,2}^\tt k_y^2}.
\end{align}
\end{subequations}
The Lorentzian parameters are next synthesized so that the surface reflection is symmetrical with respect to the incident angle ($\xi_{\ee/\mm,1} = \chi_{\ee/\mm,1}=0$) and that at normal incidence we have $T = 0.9 j$ and $R = j \sqrt{1-|T|^2}$ (arbitrarily chosen). To determine the Lorentzian parameters of \eqref{Eq:Lor_exm_Gb}, we can invert \eqref{Eq:RT_Tang} for $\theta = 0^\circ$ and find the two parameters, $\chi^\zz_{\ee}(0)$ and $\chi^\yy_{\mm}(0)$. We then chose to set $\chi_{\ee,0}^\zz = |\chi_\ee^\zz(0)|$ and  $\chi_{\ee_0}^\zz = \chi_\ee^\zz(0) -\chi_{\ee,0}^\zz$. The same procedure was used for $\chi_\mm^\yy$. Finally, we set $\xi_{\ee,2}^\zz$ and $\xi_{\mm,2}^\yy$, which produces resonances at specified incidence angles of $20^\circ$ and $30^\circ$, respectively (again arbitrarily chosen). These synthesized susceptibilities of the form \eqref{Eq:Lor_exm_Gb}, are shown as a function of incidence angle (or alternatively vs spatial frequency $k_y = k_0\sin\theta$) in Fig.~\ref{fig:chis}. As specified, one angular resonance is placed in each susceptibility component with symmetric response about $\theta=0^\circ$. We now wish to determine the scattered fields when it is illuminated with a 2D Gaussian beam, i.e. spatially broadband in $k_y$.

\begin{figure*}[t]
\centering
\begin{subfigure}{0.45\columnwidth}
		\centering
    \begin{overpic}[width=\columnwidth,grid=false,trim={0cm 0cm 0cm 0cm},clip]{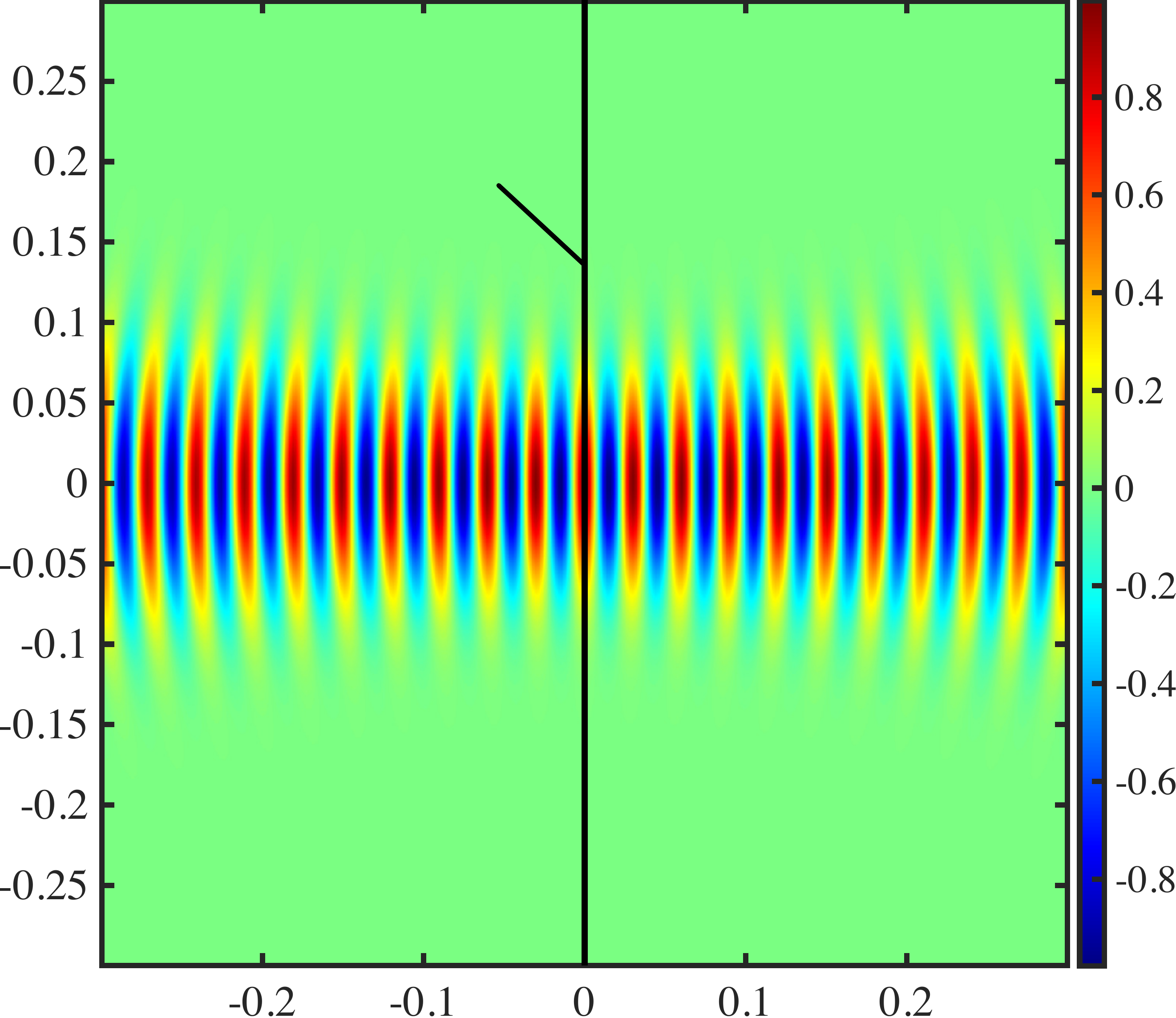}
    \put(50,94){\htext{\scriptsize \shortstack{Incident Field Re$ \{E_z^\text{inc.}\}$ \\  \textbf{2D Gaussian Beam}}}}
    \put(50,-6){\htext{\scriptsize $x$ (m)}}
    \put(-5,45){\vtext{\scriptsize $y$ (m)}}
    \put(-12,49){\vtext{\footnotesize\color{amber} \textsc{Wide Gaussian Beam}}}
             \put(33, 76){\htext{\tiny \shortstack{Waist \\ Location}}}
              \put(70, 20){\htext{\scriptsize $\boxed{w = 2\lambda}$}}
    \end{overpic}
\end{subfigure}
    \hspace{0.2cm}
\begin{subfigure}{0.45\columnwidth}
        \begin{overpic}[width=\columnwidth,grid=false,trim={0cm 0cm 0cm 0cm},clip]{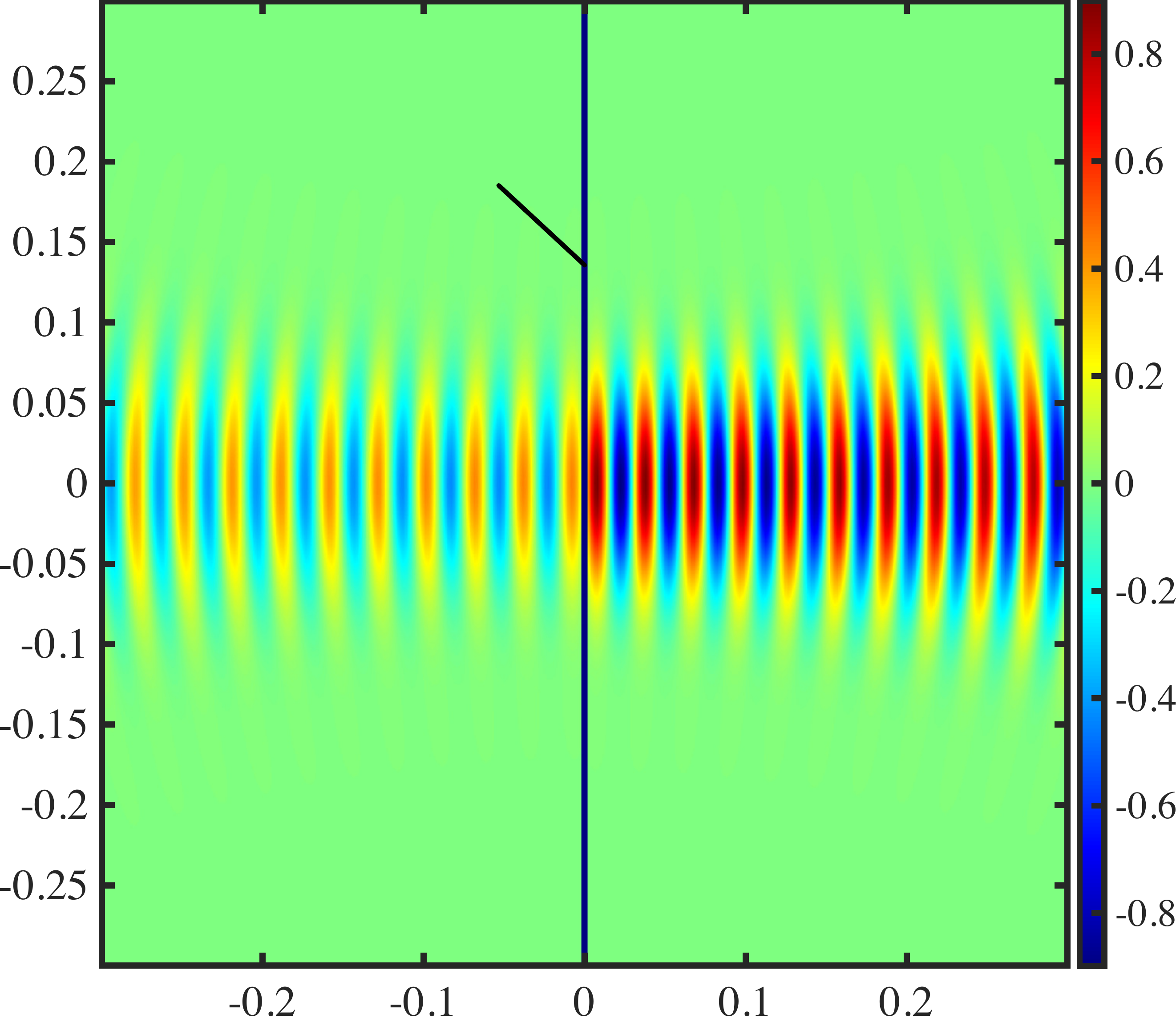}
    \put(48,96){\htext{\scriptsize \shortstack{Scattered Field Re$ \{E_z^\text{sct.}\}$, IE-GSTC\\ $\chi_\ee^\zz(\theta)= \chi_\ee^\zz$, \textbf{No Spatial Dispersion}}}}
    \put(50,-6){\htext{\scriptsize $x$ (m)}}
    \put(-5,45){\vtext{\scriptsize $y$ (m)}}
     \put(27, 70){\htext{\tiny \shortstack{Zero Thickness \\ Metasurface}}}
    \end{overpic} 
\end{subfigure}
    \hspace{0.2cm}
\begin{subfigure}{0.45\columnwidth}    
     \begin{overpic}[width=\columnwidth,grid=false,trim={0cm 0cm 0cm 0cm},clip]{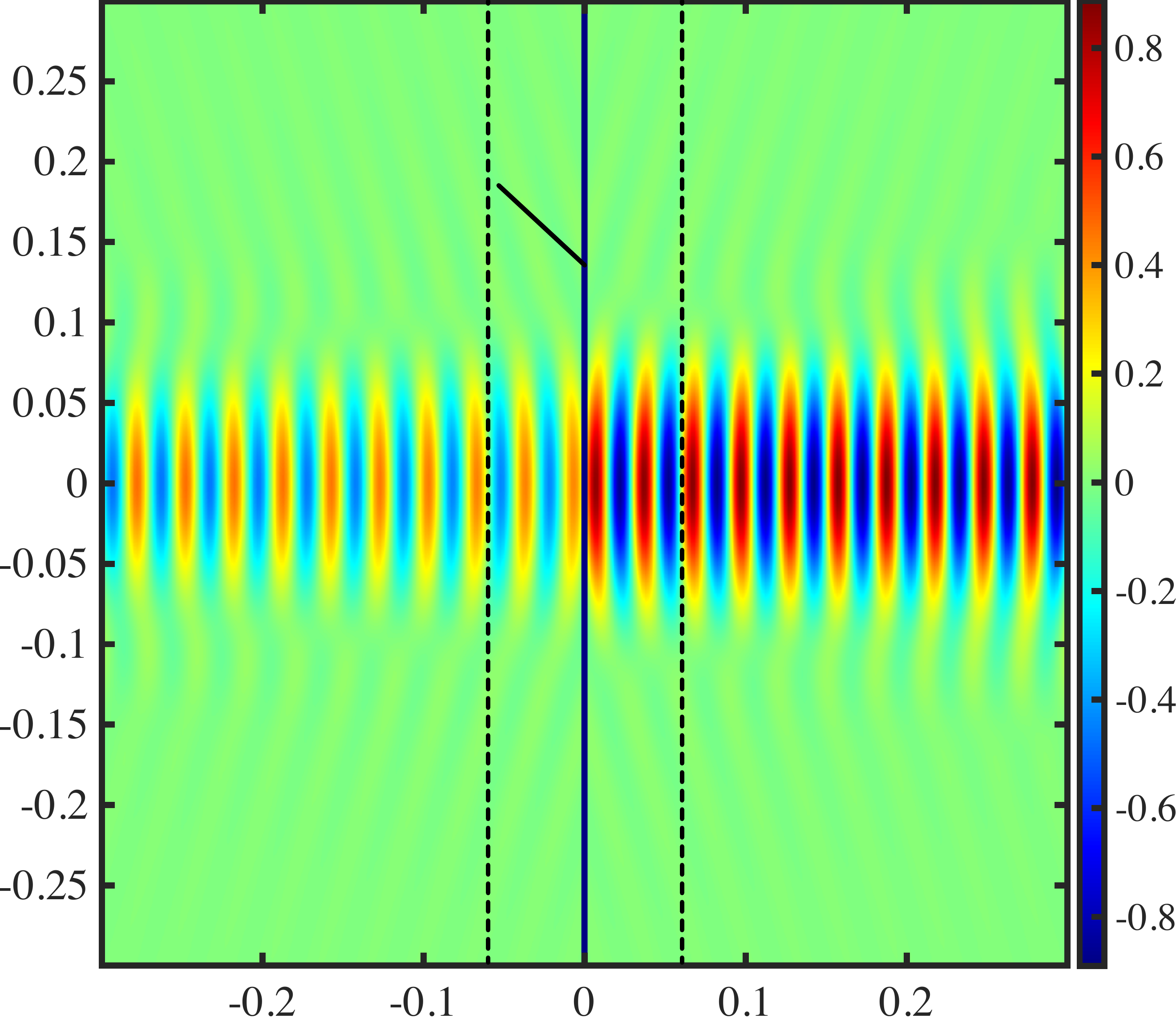}    
    \put(49,96){\htext{\scriptsize \shortstack{Scattered Field Re$ \{E_z^\text{sct.}\}$, IE-GSTC-SD\\ $\{\chi_\ee^\zz,\chi_\mm^\yy\}(\theta)$, \textbf{Spatial Dispersion}}}}
    \put(50,-6){\htext{\scriptsize $x$ (m)}}
    \put(-5,45){\vtext{\scriptsize $y$ (m)}}
         \put(27, 70){\htext{\tiny \shortstack{Zero Thickness \\ Metasurface}}}
    \end{overpic} 
\end{subfigure}
    \hspace{0.2cm}
\begin{subfigure}{0.45\columnwidth}    
    \begin{overpic}[width=\columnwidth,grid=false,trim={0cm 0cm 0cm 0cm},clip]{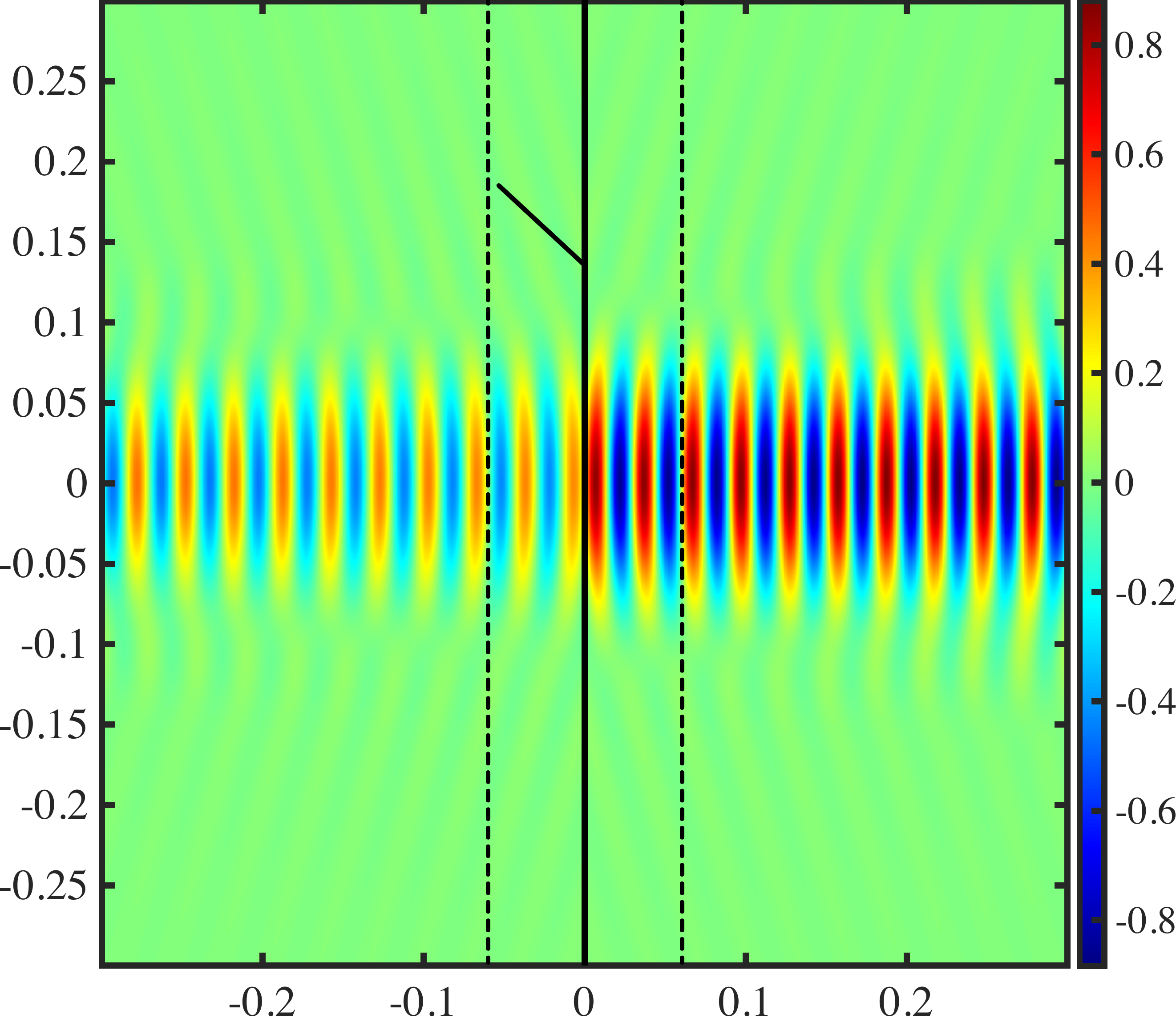}    
    \put(50,96){\htext{\scriptsize \shortstack{Scattered Field Re$ \{E_z^\text{sct.}\}$\\ $\{\chi_\ee^\zz,\chi_\mm^\yy\}(\theta)$, \textbf{Fourier Decomposition}}}}
    \put(50,-6){\htext{\scriptsize $x$ (m)}}
    \put(-5,45){\vtext{\scriptsize $y$ (m)}}
         \put(27, 70){\htext{\tiny \shortstack{Zero Thickness \\ Metasurface}}}
    \put(106,49){\vtextf{\footnotesize\color{amber} \textsc{\shortstack{Weak Spatial \\ Dispersion}}}}
    \end{overpic}
\end{subfigure}
\\
\vspace{0.45cm}
\begin{subfigure}{0.45\columnwidth}
    \begin{overpic}[width=\columnwidth,grid=false,trim={0cm 0cm 0cm 0cm},clip]{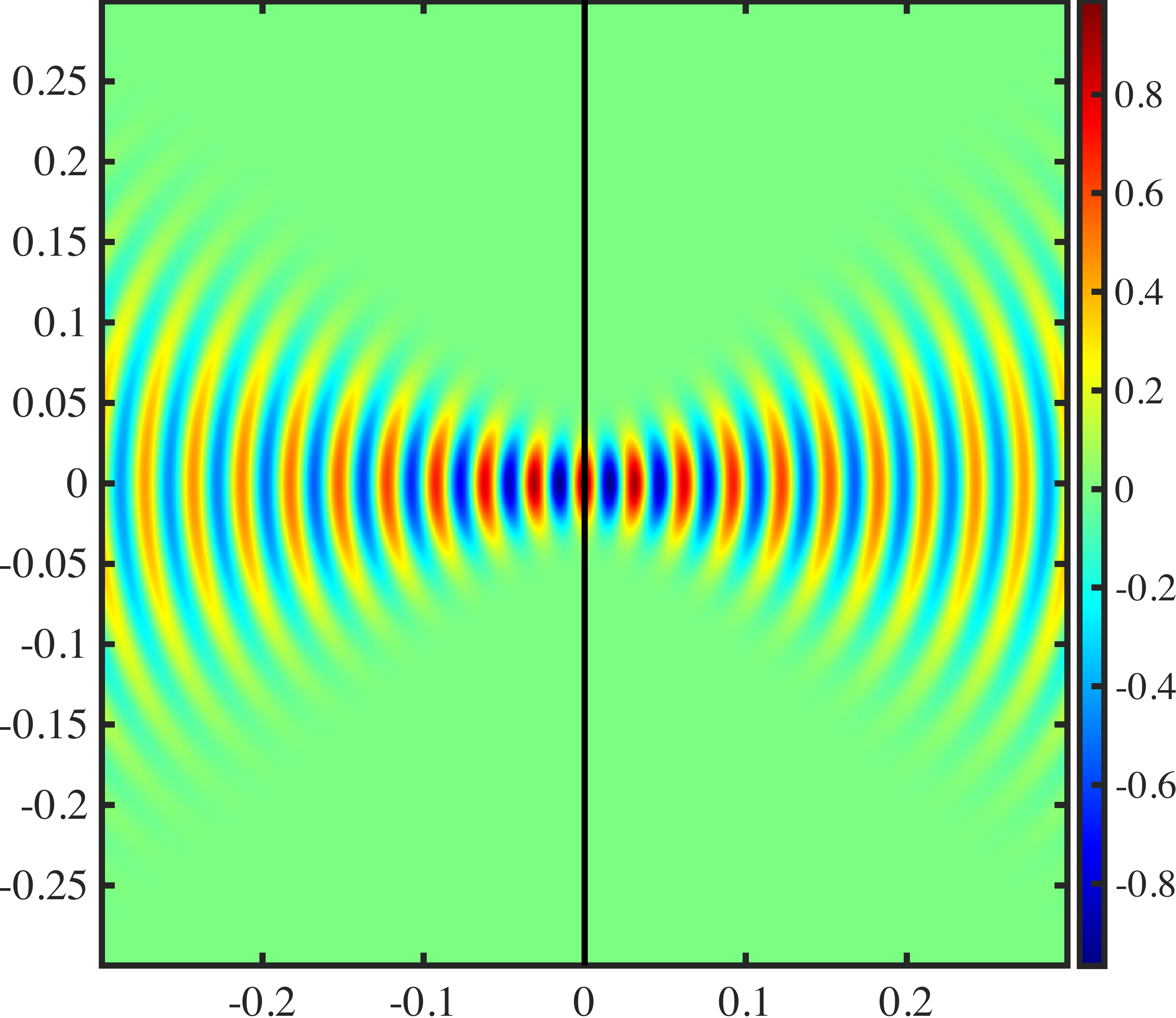}
    \put(50,-6){\htext{\scriptsize $x$ (m)}}
    \put(-5,45){\vtext{\scriptsize $y$ (m)}}
      \put(-12,49){\vtext{\footnotesize\color{cobalt} \textsc{Narrow Gaussian Beam}}}
      \put(70, 20){\htext{\scriptsize $\boxed{w = 0.75\lambda}$}}
    \end{overpic}\vspace{0.2cm}\caption{}
\end{subfigure}
    \hspace{0.2cm}
\begin{subfigure}{0.45\columnwidth}
    \begin{overpic}[width=\columnwidth,grid=false,trim={0cm 0cm 0cm 0cm},clip]{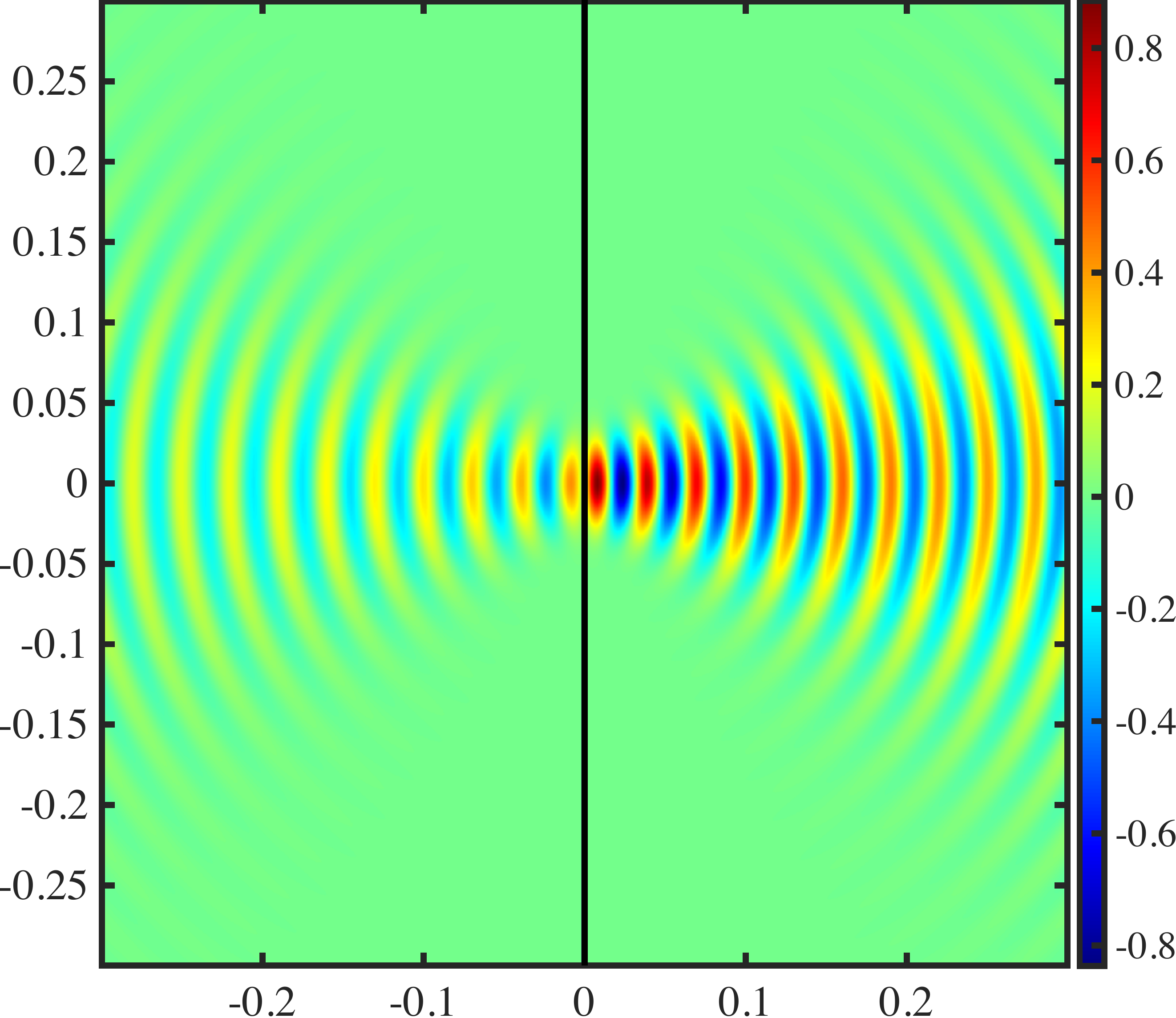}
    \put(50,-6){\htext{\scriptsize $x$ (m)}}
    \put(-5,45){\vtext{\scriptsize $y$ (m)}}
    \end{overpic}\vspace{0.2cm}\caption{}
\end{subfigure}
    \hspace{0.2cm}
\begin{subfigure}{0.45\columnwidth}    
     \begin{overpic}[width=\columnwidth,grid=false,trim={0cm 0cm 0cm 0cm},clip]{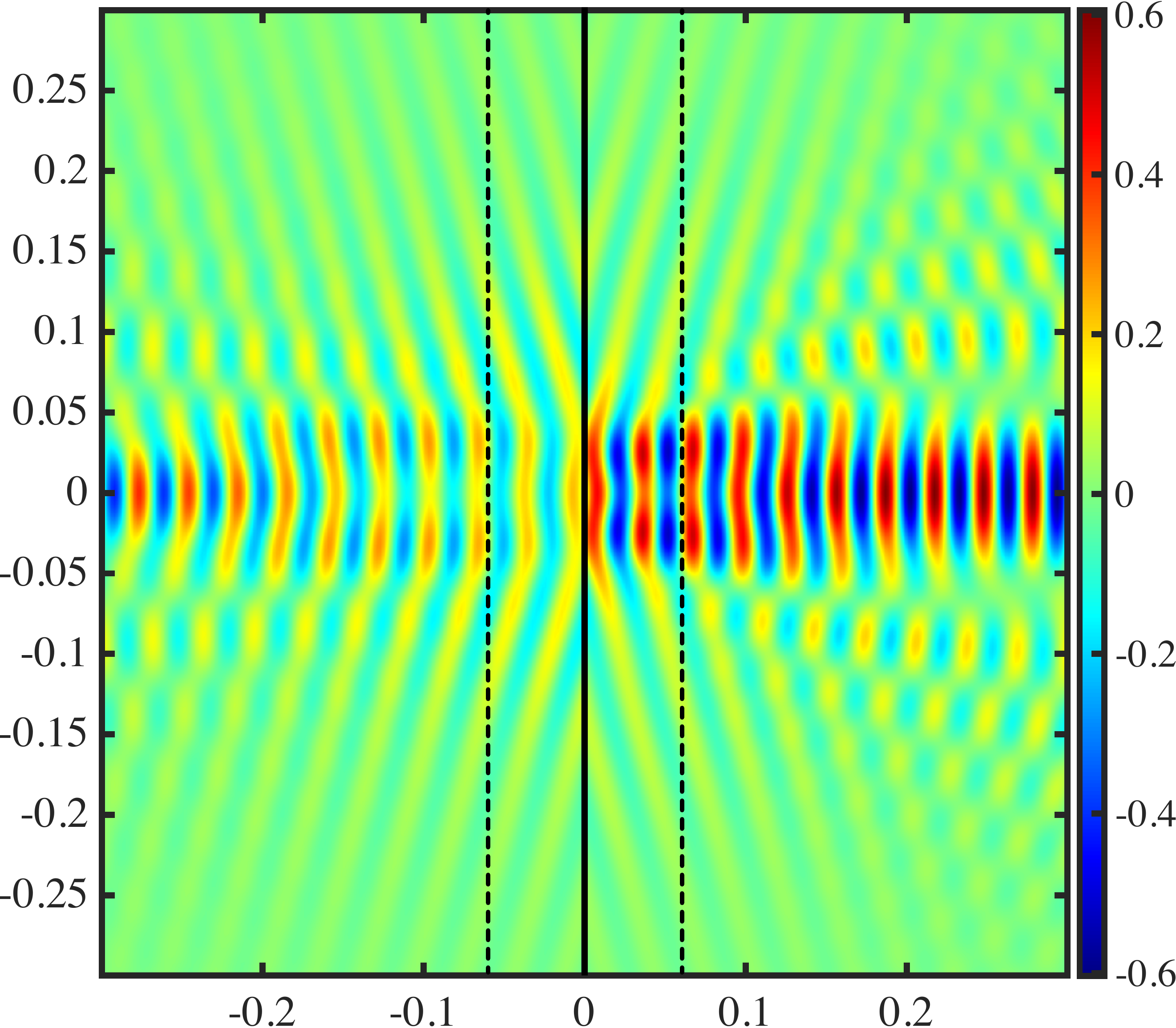}    
    \put(50,-6){\htext{\scriptsize $x$ (m)}}
    \put(-5,45){\vtext{\scriptsize $y$ (m)}}
    \end{overpic}\vspace{0.2cm}\caption{}
\end{subfigure}
    \hspace{0.2cm}
\begin{subfigure}{0.45\columnwidth}    
    \begin{overpic}[width=\columnwidth,grid=false,trim={0cm 0cm 0cm 0cm},clip]{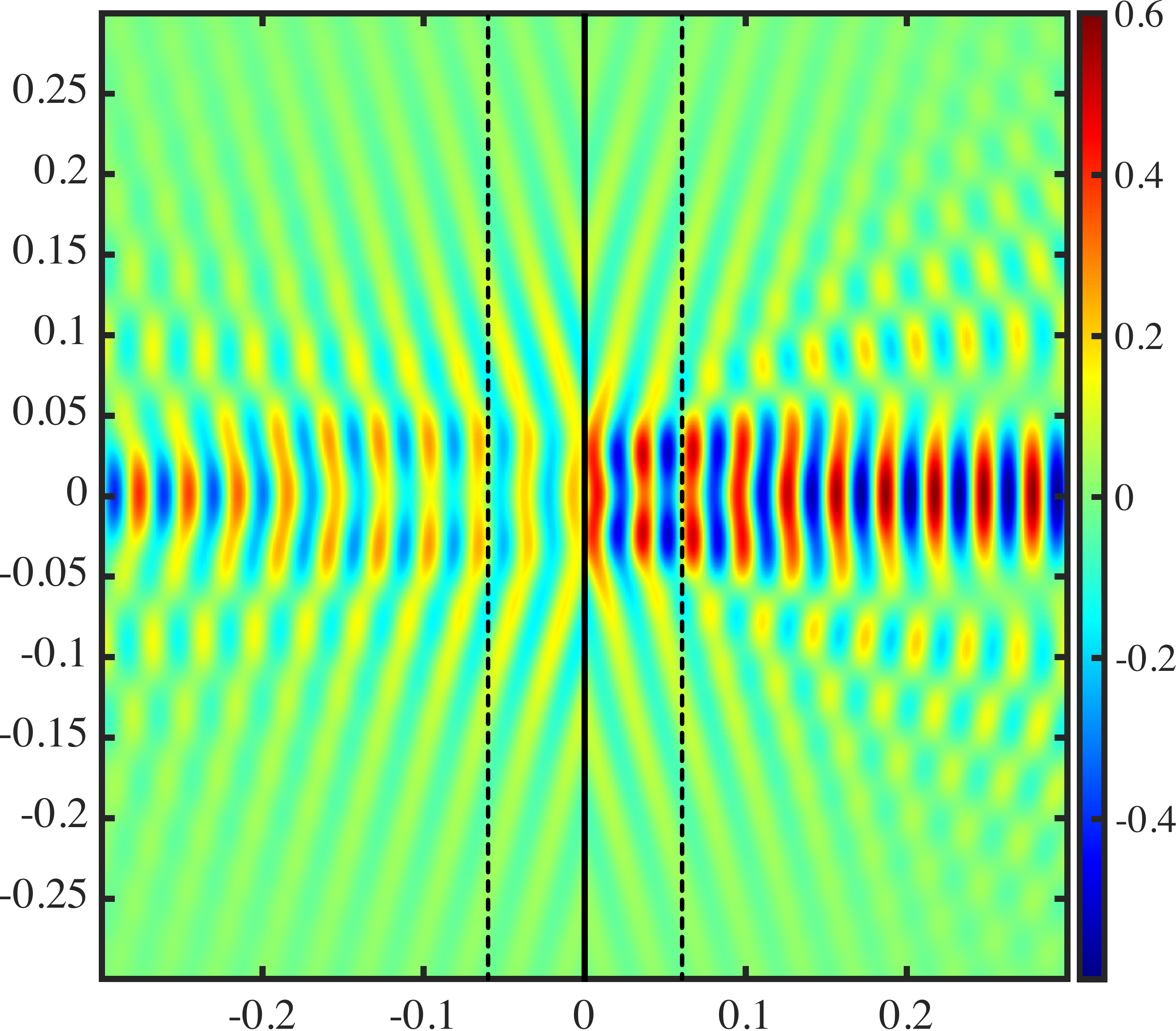}    
    \put(50,-6){\htext{\scriptsize $x$ (m)}}
    \put(-5,45){\vtext{\scriptsize $y$ (m)}}
    \put(106,49){\vtextf{\footnotesize\color{cobalt} \textsc{\shortstack{Strong Spatial \\ Dispersion}}}}
    \end{overpic}\vspace{0.2cm}\caption{}
\end{subfigure}
\caption{Surface scattering from a uniform metasurface with susceptibilities shown in Fig. \ref{fig:chis}, when excited with a normally incident 2D Gaussian beam. The Gaussian beam waist is positioned at the surface at $x=0$, and the two cases are presented for waists of $2\lambda$ (wide beam) and $0.75\lambda$ (narrow beam). a) Incident fields. b) Scattered fields from a non-dispersive surface with angle-independent surface susceptibilities. (c) Proposed IE-GSTC-SD for a spatially dispersive metasurface characterized by angle-dependent surface susceptibilities shown in Fig.~\ref{fig:chis}. (d) Scattered fields computed from the semi-analytical Fourier decomposition method. The simulation surface is $40 \lambda$ long and operating frequency $f = 60~$GHz.} \label{Fig:GaussianBeam}
\end{figure*}
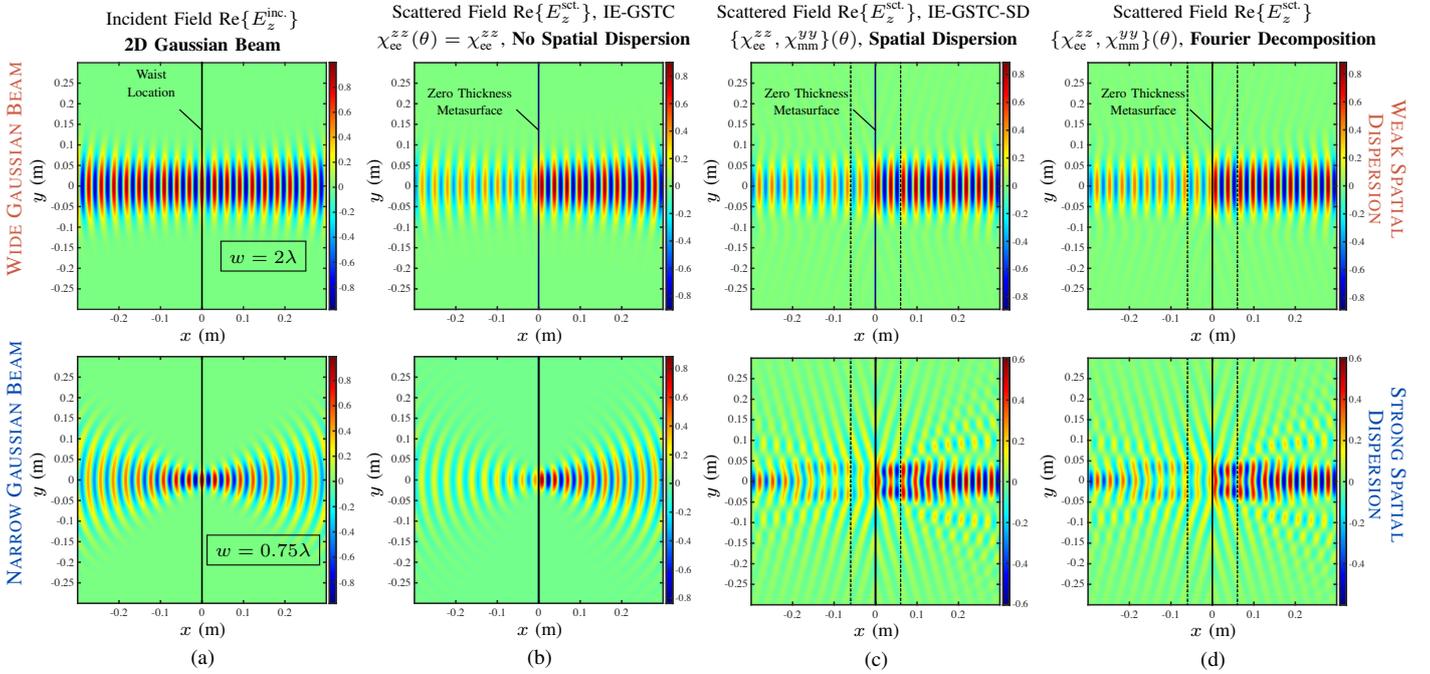

A 2D Gaussian beam (GB) is a solution of the Paraxial Helmholtz equation, with $\partial/\partial z = 0$, so that there is no spread along the $z-$direction and the propagation is confined within the $x-y$ plane only. The various field components of a 2D Gaussian beam propagating normally to the surface along $x-$axis can be easily shown to be:
\begin{subequations}
\begin{align}
     E_z(x,y) &= E_0 \sqrt{\frac{j w^2}{\frac{2x}{k_0} + j w^2}}\exp\left\{-j\frac{y^2}{\frac{2x}{k_0} + jw^2}\right\} e^{-j k_0 x}\\
    H_y(x,y) &= E_0 \frac{-j}{\mu_0\omega_0}\sqrt{\frac{jw^2}{\frac{2x}{k_0} +jw^2}}\times \label{eq:2dGEz}\notag\\
     &\quad \left(\frac{-1}{2 x + j k_0 w^2} + \frac{2 j k_0 y^2}{(2z + j k w^2)^2} - j k \right)\times\notag\\
     &\quad \exp\left\{-j \frac{y^2}{\frac{2x}{k_0} +  j w^2}\right\} e^{-j k_0 x}\\
    H_x(x,y) &= E_0 \frac{2 x}{\mu_0\omega_0}\frac{\sqrt{j w^2}}{(\frac{2x}{k_0} + j w^2)^{1.5}}\times\notag\\
    & \quad \exp\left\{-j\frac{y^2}{\frac{2x}{k_0} + jw^2}\right\}e^{-j k_0 x}
\end{align}
\end{subequations}
where $w$ is the width of the beam at the waist $x = 0$, $k_0 = \omega/c$ is the free-space wave-number. The magnitude of the Gaussian beam with waists of $w = 2\lambda$ and $w = 0.75\lambda$, respectively are also shown in Fig.~\ref{Fig:GaussianBeam}(a). Moreover, their angular spectra at the metasurface location are also shown as a function of incidence angle $\theta$ in Fig.~\ref{fig:chis}.\footnote{The incident field given by \eqref{eq:2dGEz} was transformed to the spatial domain and then $\sin \theta = k_y/k_0$ was used to translate this into the angular domain.} The angular content of the two beams can be compared to the position of the resonances in the susceptibilities, and it can be seen that the wider spatial beam content lies within the two resonances, and little spatial dispersion is expected. On the other hand, the narrow spatial beam has a substantial amount of energy at angles at or above the resonances, and we expect to see a substantial distortion of this beam as it interacts with the metasurface.

\begin{figure}[htbp]
    \centering
    \vspace*{0.5cm}
    \begin{subfigure}{0.5\columnwidth}    
    \begin{overpic}[width = \linewidth]{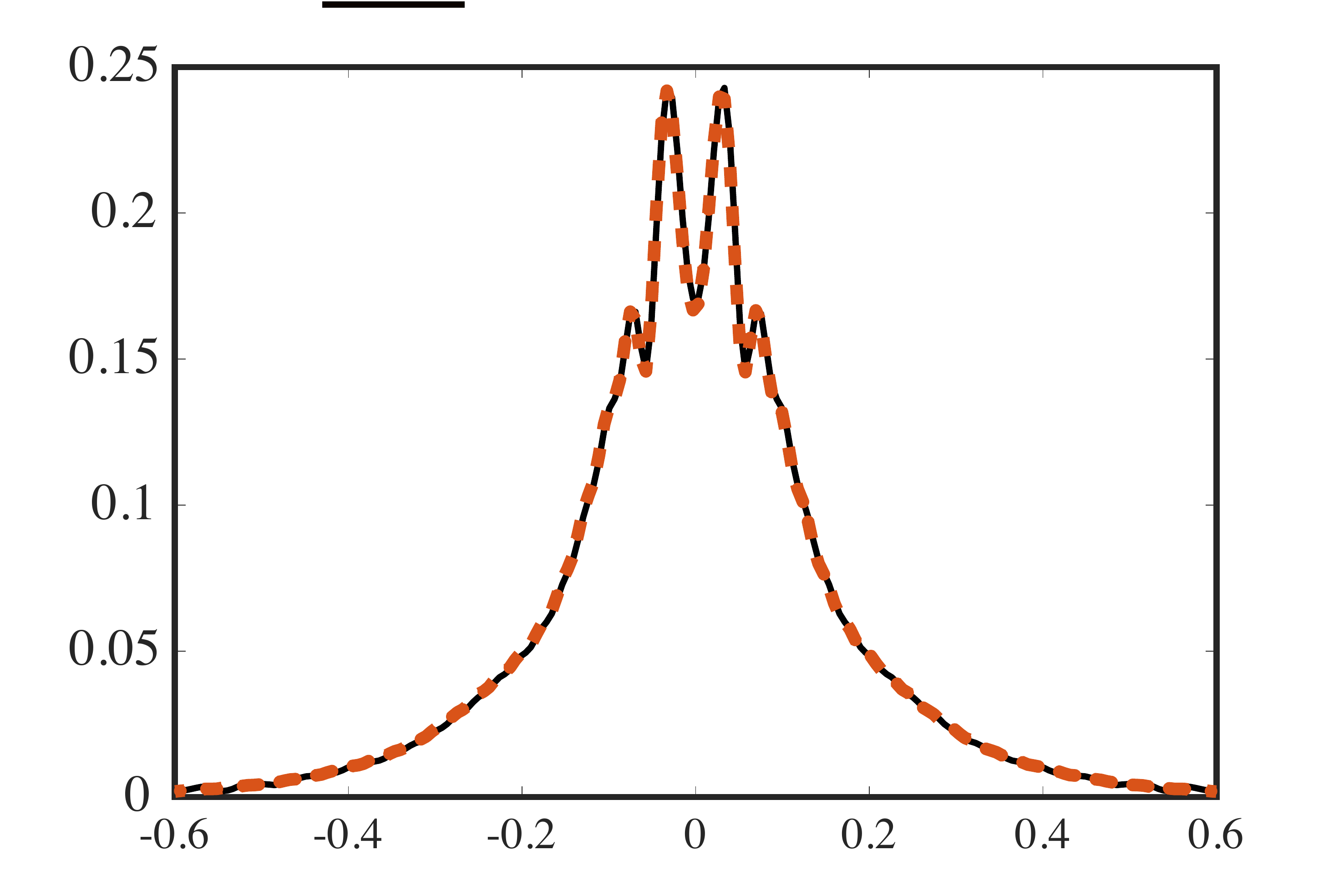}
    \put(53,-2){\scriptsize \htext{$y$~(m)}}
        \put(1, 35){\scriptsize \vtext{Reflection, $|E_z^\text{tot.}(-x_0)|$}}
    \put(55, 65){\scriptsize\htext{\scriptsize IE-GSTC-SD }}
    \end{overpic} \caption{}
    \end{subfigure}\hfill
    \begin{subfigure}{0.5\columnwidth}    
   \begin{overpic}[width = \linewidth]{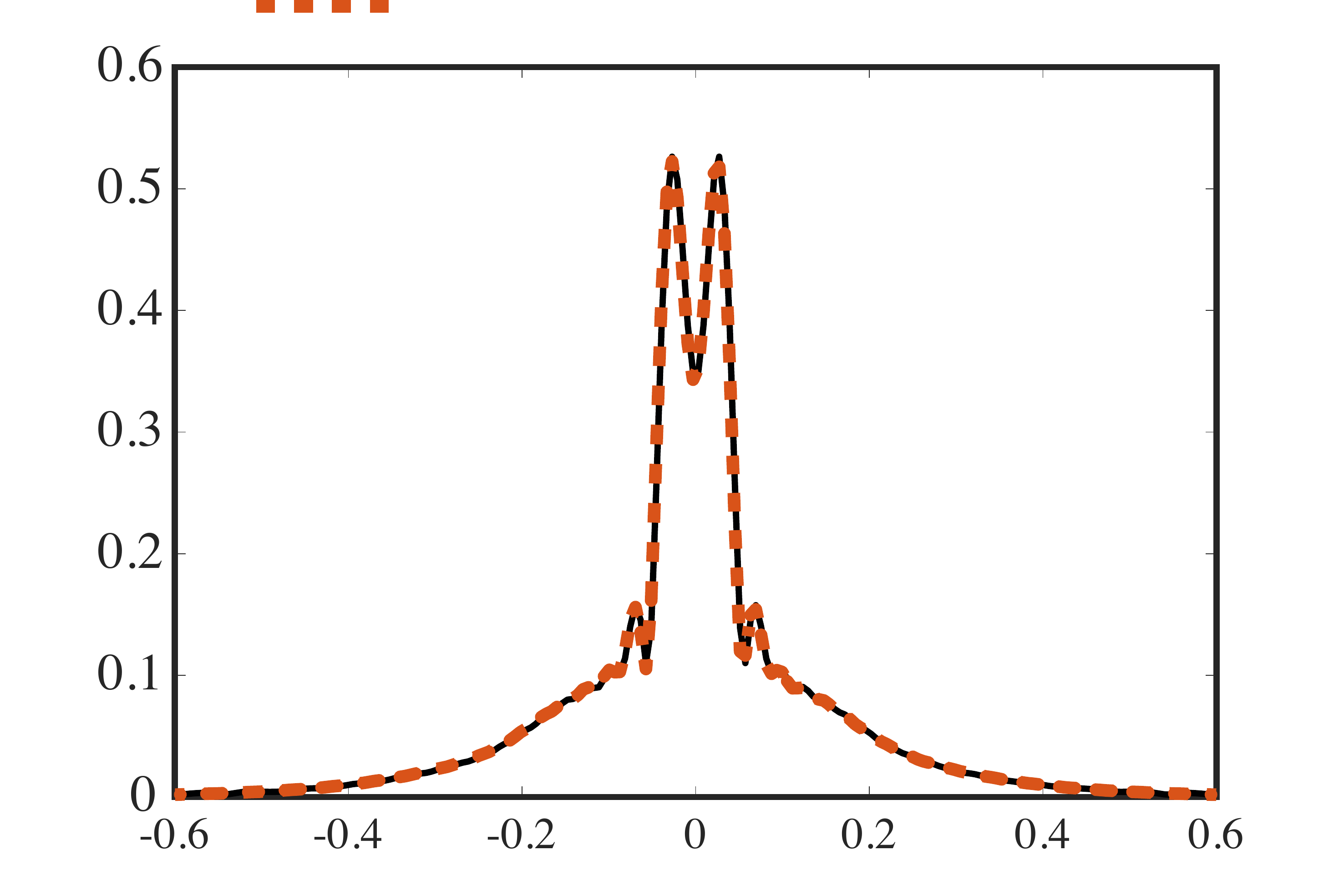}
   \put(53,-2){\scriptsize \htext{$y$~(m)}}
    \put(1, 33){\scriptsize \vtext{Transmission, $|E_z^\text{tot.}(+x_0)|$}}
        \put(58, 65){\scriptsize\htext{\scriptsize Fourier Decomposition}}
    \end{overpic}
    \caption{}
    \end{subfigure}
    \caption{Comparison of the total fields measured along the observation lines (at $x=\pm 2\lambda$) shown in Fig.~\ref{Fig:GaussianBeam}, for the case of narrow Gaussian beam between IE-GSTC-SD and the semi-analytical Fourier decomposition method, in a) the reflection and b) the transmission region.}
    \label{fig:GBRT}
\end{figure}
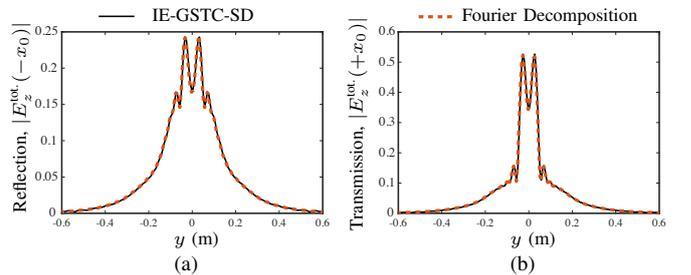

The first case presented in Fig. \ref{Fig:GaussianBeam} (top four plots) is for the moderately wide Gaussian beam ($w = 2 \lambda$) with, as we have seen, little angular content. As would be expected, the reflected fields for all three cases are quite similar, as only in the presence of significant values of $k_y$ (high angular content) are the methods expected to differ. For the FD and the IE-GSTC-SD, we do see some marginal angular dispersion of the reflected and transmitted wave and an excellent match between the two. In Fig. \ref{Fig:GaussianBeam}(b) (top) with no SD with $\xi_{\ee,2}^\zz = \xi_{\mm,2}^\tt$ = 0, we essentially see a simple reflection/transmission of the Gaussian beam recreating the shape of original incident fields. Some distortion in the FD and IE-GSTC-SD fields is present, indicating a small amount of spatial dispersion. Nevertheless, the match between the two methods is excellent.

 \begin{figure}[t]
		\centering
\begin{subfigure}{0.7\columnwidth}
		\centering
		\vspace*{0.5cm}
		\begin{overpic}[width=\linewidth,grid=false,trim={0cm 0cm 0cm 0cm},clip]{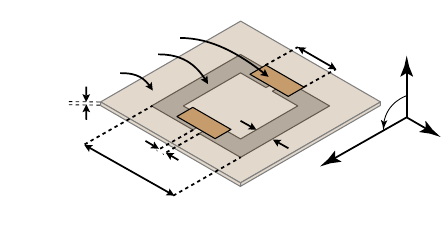}
				\put(10, 31){\htext{\scriptsize 0.025}}
				\put(23, 14){\htext{\scriptsize 2.5}}
				\put(22, 42){\htext{\scriptsize bottom copper}}
				\put(30, 47){\htext{\scriptsize top copper}}
				\put(20, 38){\htext{\scriptsize PI film}}
				\put(32, 18){\htext{\scriptsize 0.2}}
				\put(55, 30){\htext{\scriptsize 0.4}}
				\put(75, 45){\htext{\scriptsize $\ell$}}
				\put(5,5){\scriptsize \textit{(Units are mm)}}
				\put(73.5,27){\htext{\scriptsize $\Lambda$}}
				\put(68, 16){\htext{\scriptsize $y$}}
				\put(101, 22){\htext{\scriptsize $z$}}
				\put(91, 44){\htext{\scriptsize $x$}}
				\put(87, 22){\htext{\scriptsize $\theta$}}
				\put(20, 55){\htext{\footnotesize \color{amber}\shortstack{\textbf{\textsc{Loop Resonator Unit Cell}}\\ $\chib{ee}{zz}$,~$\chib{mm}{xx}$}}}
			\end{overpic}\caption{}
\end{subfigure}
\begin{subfigure}{\columnwidth}
		\centering
		\vspace*{0.5cm}
		\begin{overpic}[width=\linewidth,grid=false,trim={0cm 0cm 0cm 0cm},clip]{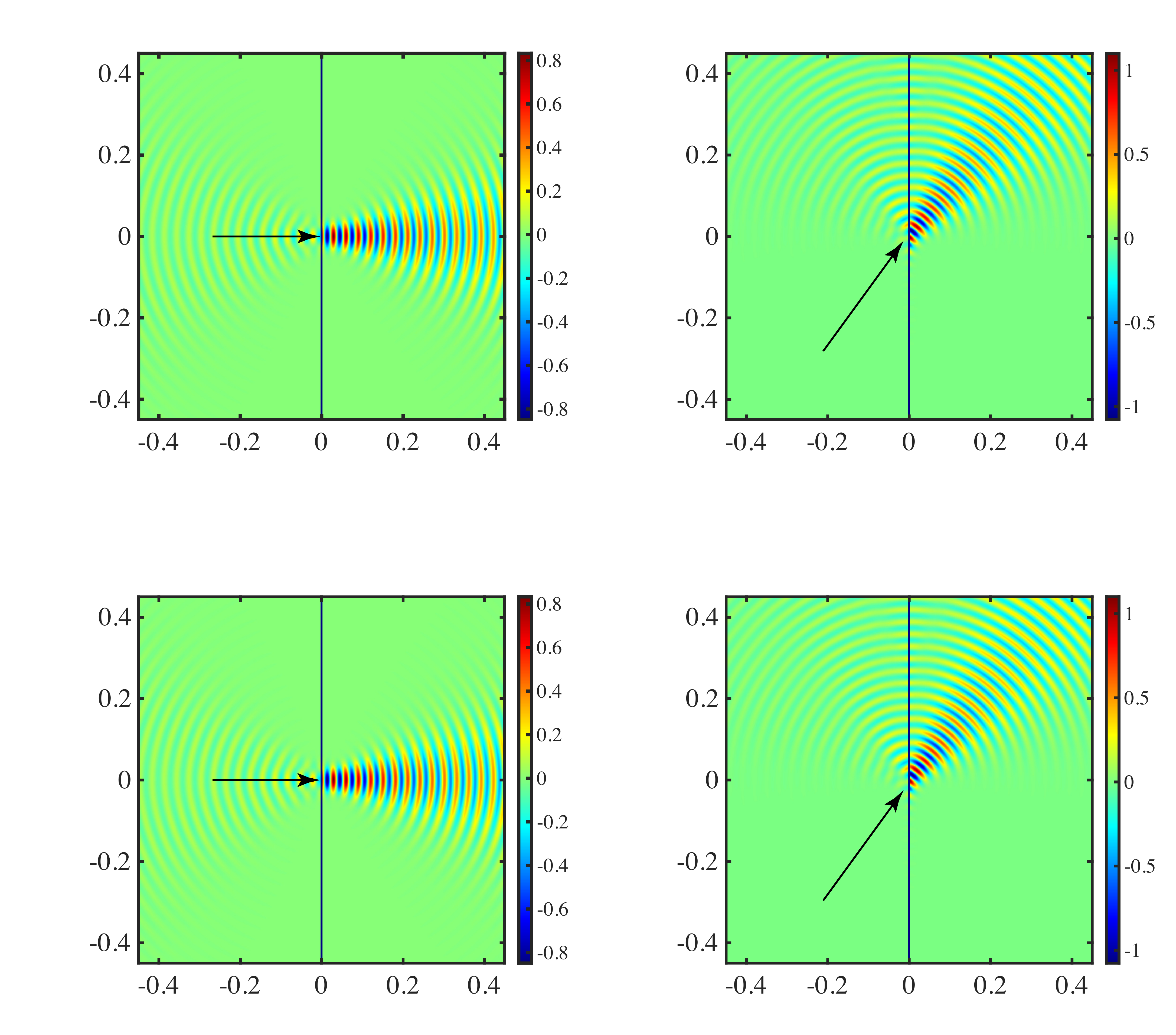}
				\put(27,0){\scriptsize \htext{$x$~(m)}}
				\put(77,0){\scriptsize \htext{$x$~(m)}}
				\put(6, 21){\scriptsize \vtext{$y$~(m)}}
				\put(55, 21){\scriptsize \vtext{$y$~(m)}}
				\put(55, 67){\scriptsize \vtext{$y$~(m)}}
				\put(6, 67){\scriptsize \vtext{$y$~(m)}}
				\put(2, 67){\color{amber} \footnotesize \vtext{Non-dispersive $\chib{ee}{zz}$, $\chib{mm}{xx}$}}
				\put(2, 21){\color{cobalt} \footnotesize \vtext{Dispersive $\chia{ee}{zz}$, IE-GSTC-SD}}
				\put(27, 43){\color{ltblu}\scriptsize \htext{$w = 0.75~\lambda,~\theta_\text{inc.} = 0^\circ$}}
				\put(77, 43){\color{ltblu}\scriptsize \htext{$w = 0.75~\lambda,~\theta_\text{inc.} = 45^\circ$}}
				\put(29, 87){\scriptsize \htext{Re$\{E_z^\text{sct.}(x,y)\}$}}
				\put(77, 87){\scriptsize \htext{Re$\{E_z^\text{sct.}(x,y)\}$}}
			\end{overpic}\caption{}
\end{subfigure}
\caption{Comparison of the scattered field generated from a loop resonator unit cell-based metasurface when excited with a 2D Gaussian beam, using two independent methods of spatially non-dispersive tangential and normal surface susceptibilities (Case 1)~\cite{smy2021iegstc, VilleFloq}, and an equivalent spatially dispersive model with angle-dependent tangential surface susceptibilities only (Case 2). a) The unit cell configuration. b) Scattered fields obtained using standard IE-GSTC for Case 1 and proposed IE-GSTC-SD for Case 2, for normal incidence and oblique incidence. The operating frequency is 60~GHz, and various susceptibility values are tabulated in Tab.~\ref{Tab:Parameters}.} \label{Fig:LoopCellResults}
\end{figure}
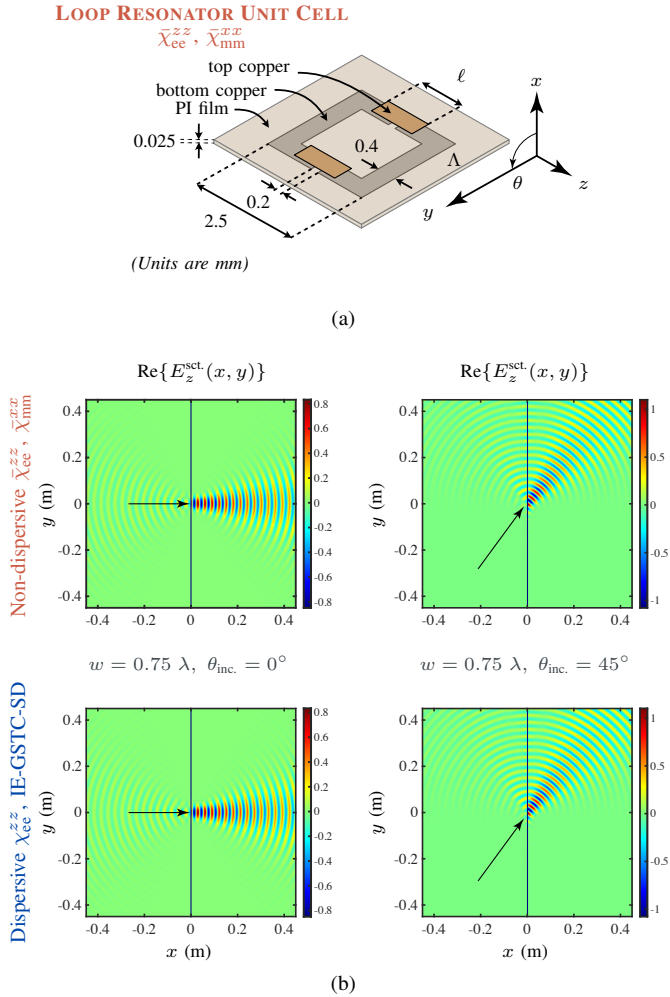

The second case (shown in the bottom four plots) presents a much clearer spatial dispersion effect. The Gaussian beam now has a small waist of $0.75\lambda$, and angular content is very obvious in the incident field shown in Fig.~\ref{Fig:GaussianBeam} bottom. A very strong distortion of the Gaussian beam is observed in both reflection and transmission, indicating a major re-arrangement of various spatial frequencies resulting from spatial dispersion. Again, an excellent match is seen with the FD method, indicating the correct implementation of various spatial derivatives in the extended GSTCs. In contrast, if we were to model this surface by ignoring the spatial dispersion, with constant $\chi^\zz_\ee$ and $\chi^\mm_\tt$ and equal to their nominal values at normal incidence, we observe a simple reflection and transmission of the incident Gaussian beam, which naturally is not correct. To further confirm the match between FD and IE-GSTC-SD, the transmitted and reflected fields for the case of a broadband $0.75\lambda$ beam are compared along the observations lines (dotted lines on the plot) $2\lambda$ away from the surface and are shown in Fig.~\ref{fig:GBRT}. They are practically identical. This result thus indicates the agreement of two independent methods, IE-GSTC-SD and FD, for capturing the spatial dispersion due to the angular resonance and is an initial validation of the proposed IE-GSTC-SD methodology.

\begin{table}[b]
\fontsize{7pt}{10pt}\selectfont
\centering
\caption{Surface Susceptibility Parameters for the Loop Cell of Fig.~\ref{Fig:LoopCellResults}}
\label{Tab:Parameters}
\setlength{\tabcolsep}{3pt}
\begin{tabular}{ccccc}
\toprule
\multirow{2}{*}{Case \#} & \multirow{2}{*}{$\chi_{\ee,0}^\zz$} & \multirow{2}{*}{$\chi_{\mm,0}^{\tt}$} & \multirow{2}{*}{$\bar{\chi}_\mm^\nn$} & $\chi_{\ee,2}^\zz$ \\
& & & & ($\times 10^{-7}$) \\
\midrule
1: $\chi_{\ee,0}^\zz$ and $\chi_\mm^\nn \ne 0$ & $0.0013$ &  $0$ & $0.0241 - j0.0131$ & -\\
2: $\chi_{\ee,0}^\zz$, $\chi_{\mm,0}^\tt$ and $\chi_{\ee,2}^\zz$& $0.0013$ & $0$  & - & $5.49 - j2.98$\\
\bottomrule
\end{tabular}
\end{table}

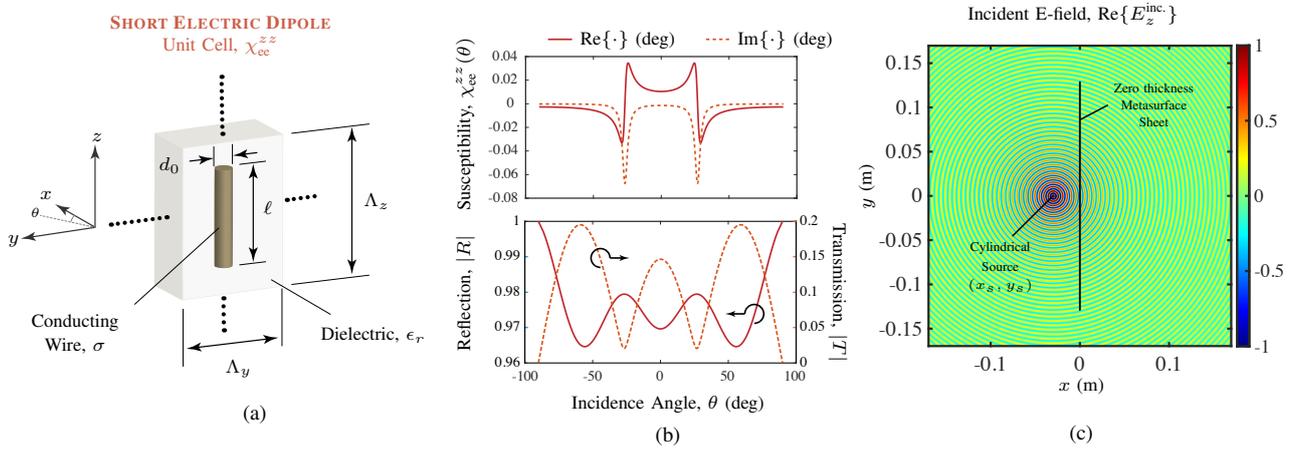
\begin{figure*}[htbp]
		\centering
		\begin{subfigure}{0.6\columnwidth}
		\centering
		\hspace*{-1.5cm}
		\begin{overpic}[width=0.9\columnwidth,grid=false,trim={0cm 0cm 0cm 0cm},clip]{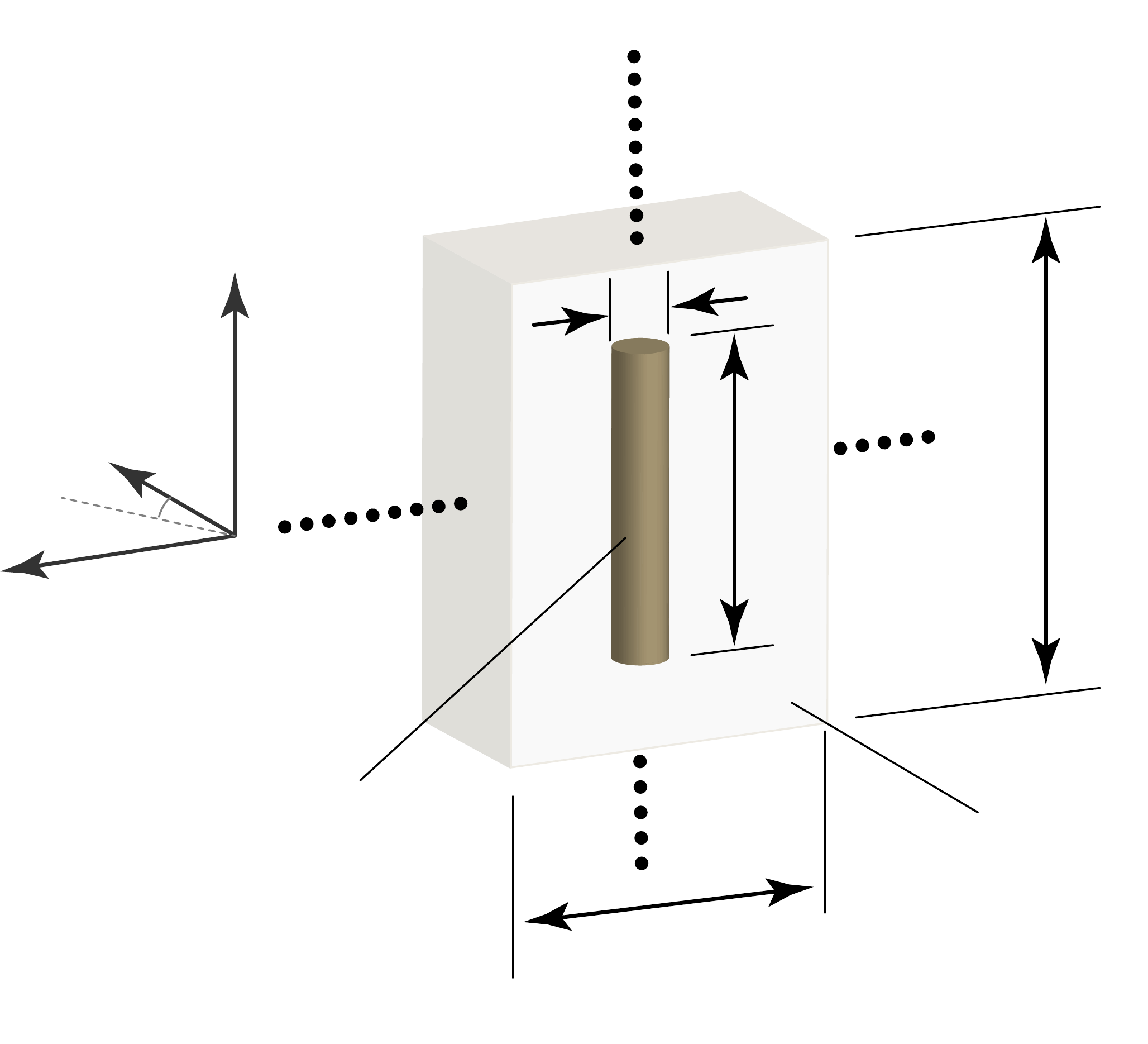}
				\put(68, 48){\htext{\scriptsize $\ell$}}
				\put(97, 13){\htext{\scriptsize Dielectric, $\epsilon_r$}}
				\put(15, 12){\htext{\scriptsize \shortstack{Conducting \\ Wire, $\sigma$}}}
				\put(-2, 40){\htext{\scriptsize $y$}}
				\put(7, 52){\htext{\scriptsize $x$}}
				\put(21, 68){\htext{\scriptsize $z$}}
				\put(4, 47){\htext{\tiny $\theta$}}
				\put(41, 60){\htext{\scriptsize $d_0$}}
				\put(98, 50){\htext{\scriptsize $\Lambda_z$}}
				\put(60, 4){\htext{\scriptsize $\Lambda_y$}}
				\put(55, 95){\htext{\scriptsize \color{amber}\shortstack{\textbf{\textsc{Short Electric Dipole}}\\ Unit Cell, $\chia{ee}{zz}$}}}
		\end{overpic} \caption{}
\end{subfigure}
\begin{subfigure}{0.6\columnwidth}
				\begin{overpic}[width=\columnwidth,grid=false,trim={0cm 0cm 0cm 0cm},clip]{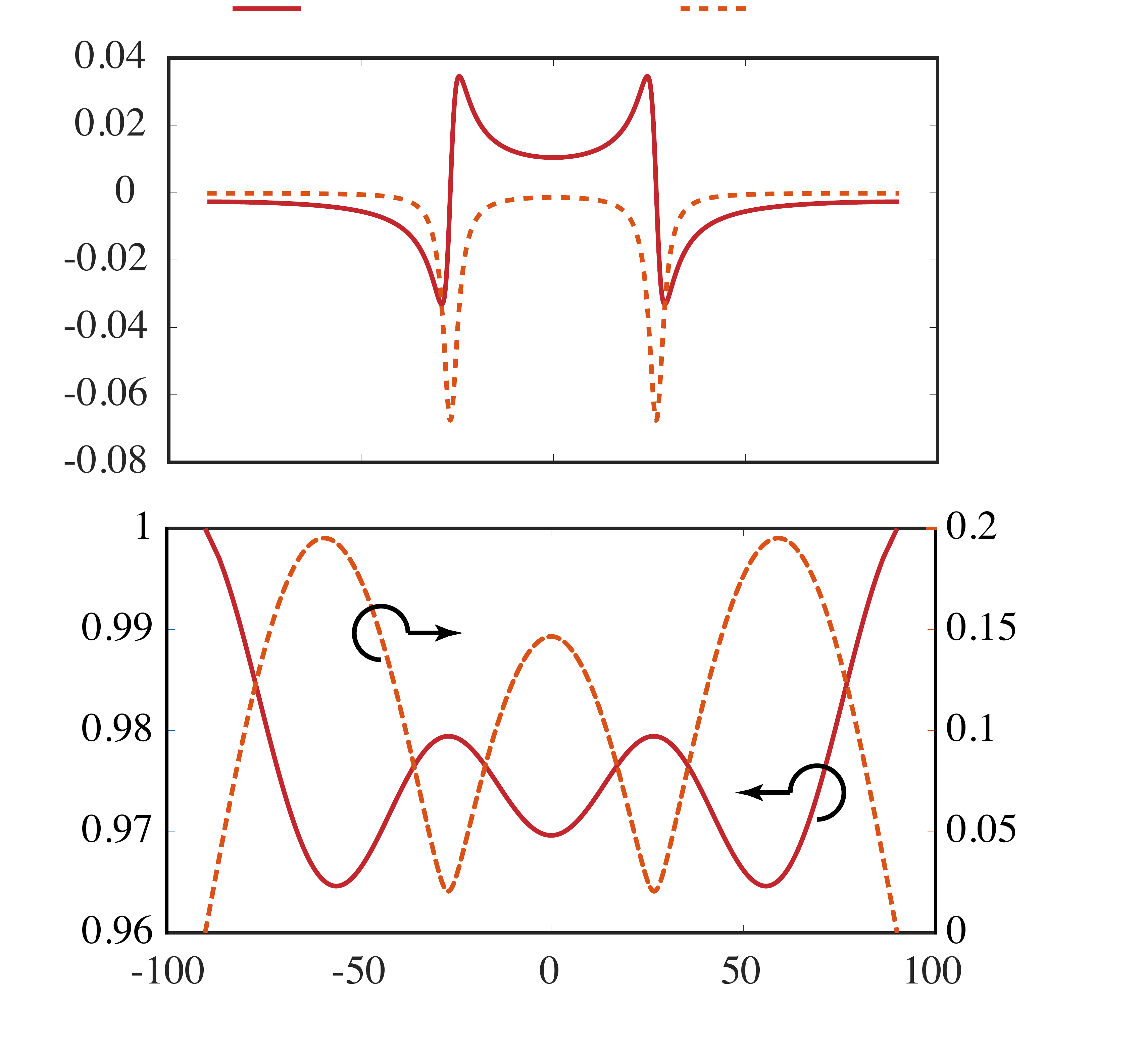}
    \put(0, 70){\scriptsize \vtext{Susceptibility, $\chi^\zz_\ee(\theta)$}}
    \put(50, 0){\scriptsize \htext{Incidence Angle, $\theta$ (deg)}}
    \put(92, 28){\scriptsize \vtextf{Transmission, $|T|$}}
    \put(0, 28){\scriptsize \vtext{Reflection, $|R|$}}
    \put(40, 91){\scriptsize \htext{Re$\{\cdot\}$ (deg)}}
    \put(79, 91){\scriptsize \htext{Im$\{\cdot\}$ (deg)}}
	\end{overpic} 
	\caption{}
\end{subfigure}	
\begin{subfigure}{0.6\columnwidth}
    \begin{overpic}[width=\linewidth,grid=false,trim={0cm 0cm 0cm 0cm},clip]{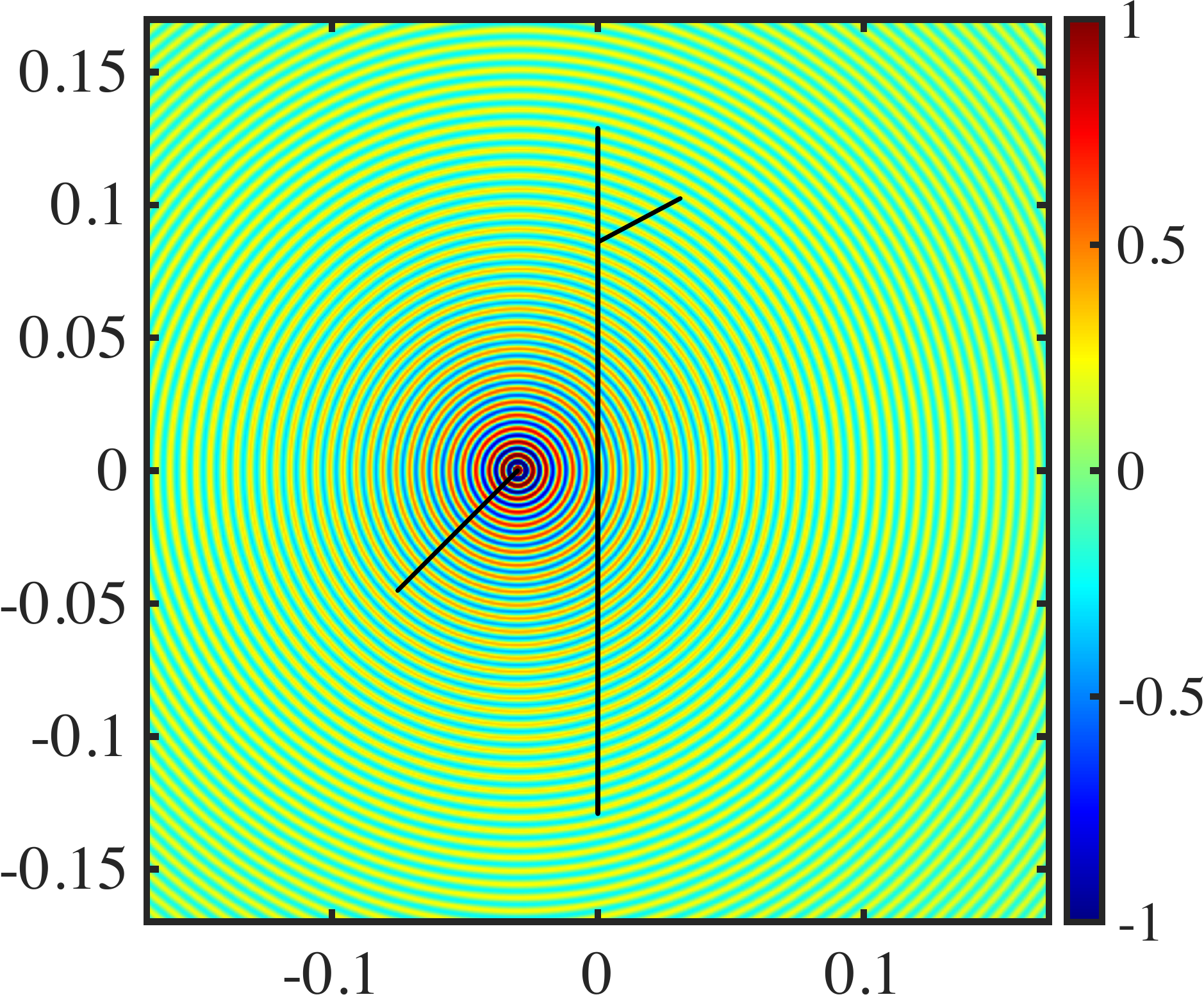}
    \put(48, 88){\scriptsize\htext{Incident E-field, Re$\{E_z^\text{inc.}\}$}}
    \put(50, -5){\scriptsize\htext{$x$~(m)}}
    \put(-3, 45){\scriptsize\vtext{$y$~(m)}}
    \put(68, 65){\tiny\htext{\shortstack{Zero thickness \\ Metasurface \\ Sheet}}}
    \put(30, 25){\tiny\htext{\shortstack{Cylindrical \\ Source \\ $(x_s, y_s)$}}}
    \end{overpic}
    \vspace{0.05cm}\caption{}
\end{subfigure}	
\caption{An example case of a practical spatially dispersive metasurface composed of a periodically arranged short conducting wire in the $x-z$ plane. a) The unit cell configuration. b) The angle dependent surface susceptibility $\chi^\zz_\ee(\theta)$ obtained using \eqref{Eq:Lor} with fitted parameters from FEM-HFSS~\cite{Part_1_Nizer_SD}, and the corresponding transmission and reflection obtained using \eqref{Eq:RT_Tang}. c) The simulation setup of a finite-sized flat metasurface excited with a cylindrical wave at a fixed frequency of 60~GHz. The fitted surface parameters are $\chi_{\ee,0}^\zz =  -5.866 \times 10^{-4}$, $\chi_{\ee,2}^\zz = 0.0104 - j0.0014$, $\xi_{\ee,1}^\zz = 0$, $\xi_{\ee,2}^\zz = (-30.0 + j4.79)\times 10^{-7}$ and $\chi_{\mm}^\tt = 0$. The wire parameters are: $d_0 = 0.2$~mm, $\ell = 2.5$~mm, $\Lambda_y = 2.15$~mm, $\Lambda_z = 4.3$~mm and conductivity $\sigma = 5.8\times 10^4$~S/m.}
\label{fig:WireExample}
\end{figure*}

\subsection{Spatially Non-dispersive Unit Cell with Normal Susceptibilities}

For a second verification of the proposed IE-GSTC-SD, method, let us consider a practical unit cell as shown in Fig.~\ref{Fig:LoopCellResults}. It is formed from a resonant loop structure consisting of a Metal-Insulator-Metal (MIM) capacitor printed on a thin dielectric slab, as shown in Fig.~\ref{Fig:LoopCellResults}(a). It has been shown that such a unit cell structure can be modeled using one constant tangential surface susceptibilities $\chib{ee}{zz}$ and a normal susceptibility component $\chib{mm}{xx}$, with no spatial dispersion~\cite{smy2021iegstc, VilleFloq}. Including the gradient of the normal fields in \eqref{Eq:ConvGSTC}, leads to a single field equation governing the field scattering and is given by
\begin{align}\label{eq:HyN}
\Delta \Ht_{y}  &=  j\omega \epsilon_0\chib{ee}{zz} \Et_{z,\text{av}}  + 
\mu_0\chib{mm}{yy}\frac{\partial \Ht_{x,\text{av}}}{\partial y}
\end{align}
which incorporates the effect of $\chi^\nn_\mm$. As shown in Part-1 \cite{Part_1_Nizer_SD}, we can model this uniform structure using purely \emph{tangential} surface susceptibility $\chia{ee}{zz}$, dependent on the angle of incidence $\theta$, and thus spatially dispersive, i.e.
\begin{align}\label{Eq:SpecialCase}
\chi_\ee^\zz(k_y) = \bar{\chi}_\text{ee}^{zz} + \left(\frac{\bar{\chi}^\nn_\mm}{k_0^2}\right) k_y^2 = \chi_{\ee,0}^\zz + \chi_{\ee,2}^\zz k_y^2
\end{align}
which is of the form \eqref{Eq:Ratio}, with a second order polynomial in the numerator and zeroth order denominator. Therefore, a standard constant dipolar normal surface susceptibility component accounts for a particular form of angular scattering from a uniform metasurface given by \eqref{Eq:SpecialCase}. However, it also presents an opportunity to verify the spatial dispersion modeling used in the IE-GSTC-SD, as it can be compared to the standard IE-GSTC formulation using constant surface susceptibilities. 

Such a demonstration is presented in Fig. \ref{Fig:LoopCellResults} for the two cases of Gaussian beam illumination (normal and oblique incidence) on a uniform surface described by the two differently formulated surfaces. The surface parameters for the first case which uses constant susceptibilities, one tangential $\bar{\chi}_{\ee,0}^\zz$ and one normal $\bar{\chi}_\mm^\nn \ne 0$, were reported in \cite{VilleFloq,GenBCEM} and are given in Tab.~\ref{Tab:Parameters}. The second case uses only one spatially dispersive tangential component, ${\chi}_\ee^\zz$, described by two parameters $\chi_{\ee,0}^\zz$ and $\chi_{\ee,2}^\zz$, also tabulated in Tab.~\ref{Tab:Parameters}.

These two conditions are shown in Fig. \ref{Fig:LoopCellResults}(b) for the two cases of normal incidence and $45^\circ$, respectively. In both cases, the predicted fields are essentially identical between the two methods. This result verifies that the SD methodology introduced into the GSTC framework successfully models the angular dependence of a physical cell using SD tangential components of the susceptibility in which the angular dependence is due to a strongly dominant normal component of the susceptibility. Naturally, this demonstration also validates the IE-GSTC-SD implementation further.

\section{Field Scattering from Finite-Sized Metasurfaces}

\subsection{Wire Dipole Based Unit Cell}

To demonstrate the importance of spatial dispersion and capability of the proposed IE-GSTC-SD framework to model angular scattering from physical unit cells, we will consider an example of a simple unit cell based on a short conducting dipole, which exhibits spatial dispersion. The basic unit cell used to form a 2D surface is shown in Fig.~\ref{fig:WireExample}(a) and consists of a short segment of wire of length $\ell$ and with a finite conductivity $\sigma$, placed inside free-space. This is then periodically arranged along $y-$ and $z-$axis with periods of $\Lambda_z$ and $\Lambda_y$ to form a surface lying in the $y-z$ plane. Due to symmetry considerations and assumed TE mode excitation, this cell can be shown to be modeled using two tangential susceptibilities, $\chia{ee}{zz}$ and $\chia{mm}{yy}$ only, which are angle-dependent. This unit cell is very simple yet very insightful, as it only exhibits a single Lorentzian resonance which dominantly depends on the length $\ell$ of the wire, thus acting as a perfect testbed for the IE-GSTC-SD framework.

In Part-1 of this work, this cell was characterized using HFSS to obtain its transmission and reflection characteristics as a function of frequency $\omega$ and angle of incidence, $\theta$ or spatial frequency $k_y$~\cite{Part_1_Nizer_SD}. Fig.~\ref{fig:WireExample}(b) presents the extracted susceptibilities for a fixed frequency of $f = 60$~GHz and the fitted model with a single Lorentzian using \eqref{Eq:Lor}.  These susceptibilities produce an angular dependent reflectivity $R(\theta)$ and transmission $T(\theta)$ which can be obtained from the susceptibilities using \eqref{Eq:RT_Tang}, and are also shown in Fig. \ref{fig:WireExample}(b). The reflectivity is close to unity and exhibits a small amount of angular dependence varying from about 0.96 to 1. The effect of the two resonances present in the susceptibilities is more pronounced in the reflectivity, $R$ dropping from a maximum of 0.2 at $55^\circ$ to a clear minimum at $30^\circ$ to nearly zero and also a drop off at high angles. 

\subsection{Finite-Sized Vertical Metasurface}

To demonstrate the basic field transformation phenomena of this surface and to allow for a detailed comparison with a commercial full-wave simulator like Ansys FEM-HFSS, a vertical surface along the $y-$axis, consisting of 60 units cells was formed and illuminated with a line source ($f = 60$ GHz) located at $\mathbf{r}_\text{s}= \{x_\text{s} = -30,y_\text{s} = 0\}$~mm.  The incident field produced by the line source is given by,
\begin{subequations}
\begin{align}
    E_{\text{i},y}(\mathbf{r}) &=  E_0\frac{H_{0}^{(2)}\left(k|\mathbf{r}-\mathbf{r}_\text{s}|\right)}{H_{0}^{(2)}\left(k|\mathbf{r}_\text{s}|\right)} \\
    H_{\text{i},x}(\mathbf{r}) &=E_0\frac{j(z-z_\text{s}) H_{1}^{(2)}\left(k|\mathbf{r}-\mathbf{r}_\text{s}|\right)}{\eta|\mathbf{r}|H_{0}^{(2)}\left(k|\mathbf{r}_\text{s}|\right)}\\
    H_{\text{i},z}(\mathbf{r}) &=-E_0\frac{j(x-x_\text{s}) H_{1}^{(2)}\left(k|\mathbf{r}-\mathbf{r}_\text{s}|\right)}{\eta|\mathbf{r}|H_{0}^{(2)}\left(k|\mathbf{r}_\text{s}|\right)}
\end{align}\label{Eq:ReflectorEi}%
\end{subequations}
where $H_{\{0,1\}}^{(2)}$ are Hankel functions of the second kind, of orders 1 and 2, and $E_0$ is the peak field amplitude. The surface is simply placed in free space and the simulation set up, and the source position is shown in Fig.~\ref{fig:WireExample}(c). 

\begin{figure*}[htbp]
\centering
		\begin{subfigure}{0.6\columnwidth}
		\centering
        \begin{overpic}[width=\linewidth,grid=false,trim={0cm 0cm 0cm 0cm},clip]{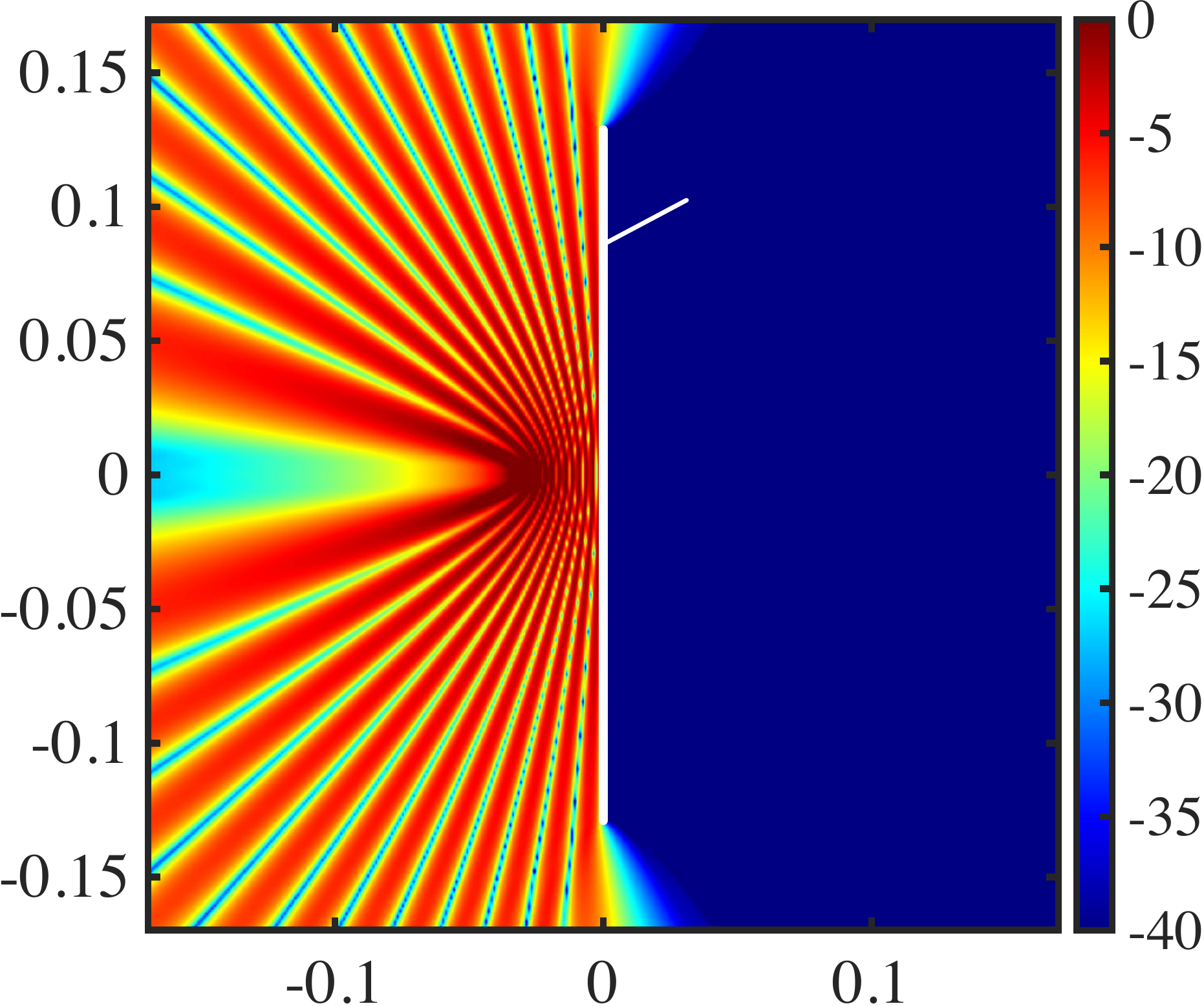}
    \put(50,87){\scriptsize \htext{\textbf{IE-PEC}  -- $|E_z^\text{tot.}|$ dB}}
    \put(50,-5){\scriptsize\htext{ $x$ (m)}}
    \put(-3,43){\scriptsize\vtext{ $y$ (m)}}
    \put(68, 73){\color{white}\scriptsize\htext{ \shortstack{PEC Surface}}}
    \end{overpic}\vspace{0.2cm}\caption{}
    \end{subfigure}
    \hspace{0.35cm}
		\begin{subfigure}{0.6\columnwidth}
		\centering
        \begin{overpic}[width=\linewidth,grid=false,trim={0cm 0cm 0cm 0cm},clip]{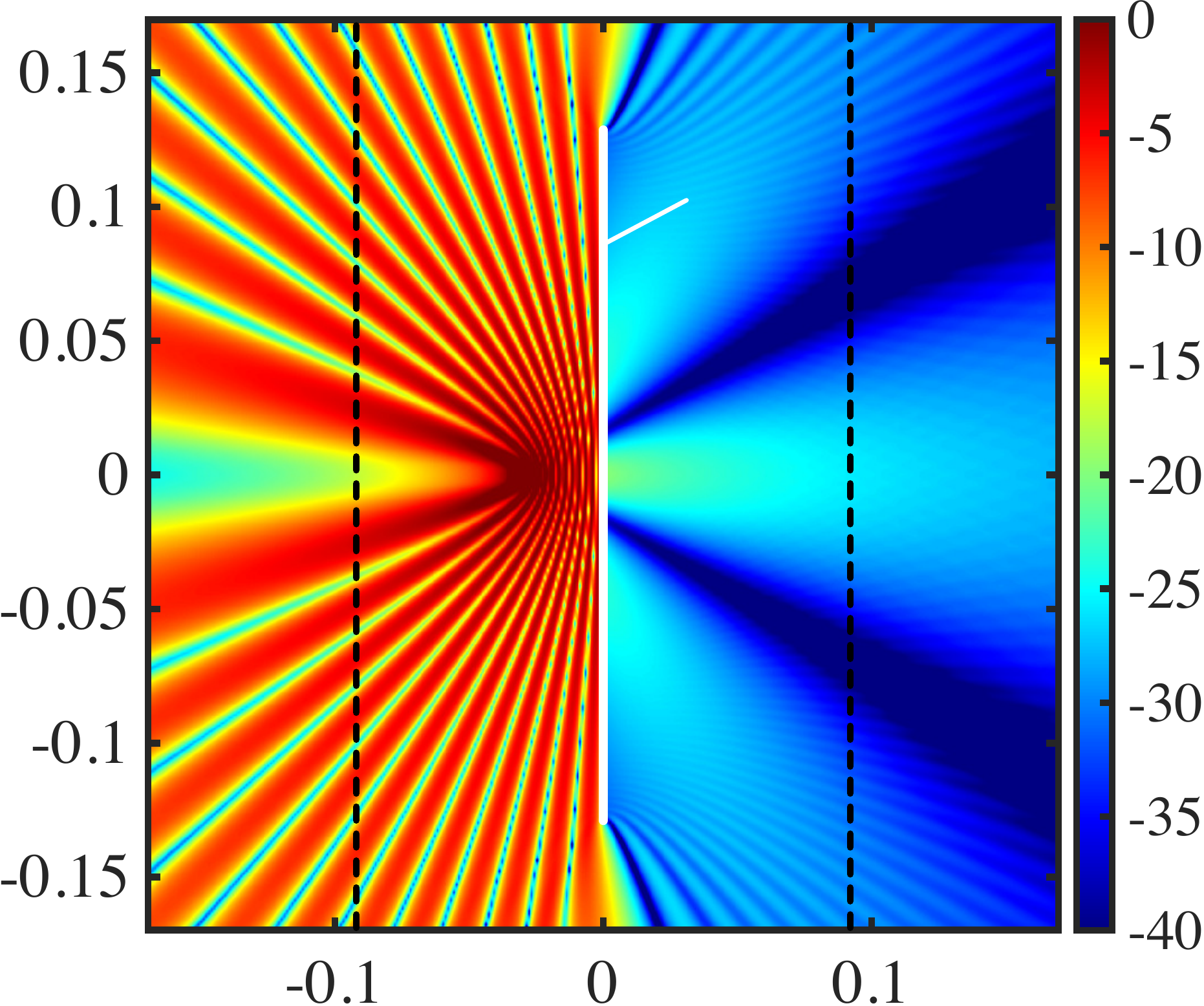}
    \put(50,87){\scriptsize \htext{\textbf{IE-GSTC-SD}  -- $|E_z^\text{tot.}|$ dB}}
    \put(50,-5){\scriptsize\htext{ $x$ (m)}}
    \put(-3,43){\scriptsize\vtext{ $y$ (m)}}
    \put(66, 70){\color{white}\scriptsize\htext{ \shortstack{Zero Thickness \\ Sheet}}}
    \end{overpic}\vspace{0.2cm}\caption{}
    \end{subfigure}
    \hspace{0.35cm}
    		\begin{subfigure}{0.6\columnwidth}
		\centering
    \begin{overpic}[width=\linewidth,grid=false,trim={0cm 0cm 0cm 0cm},clip]{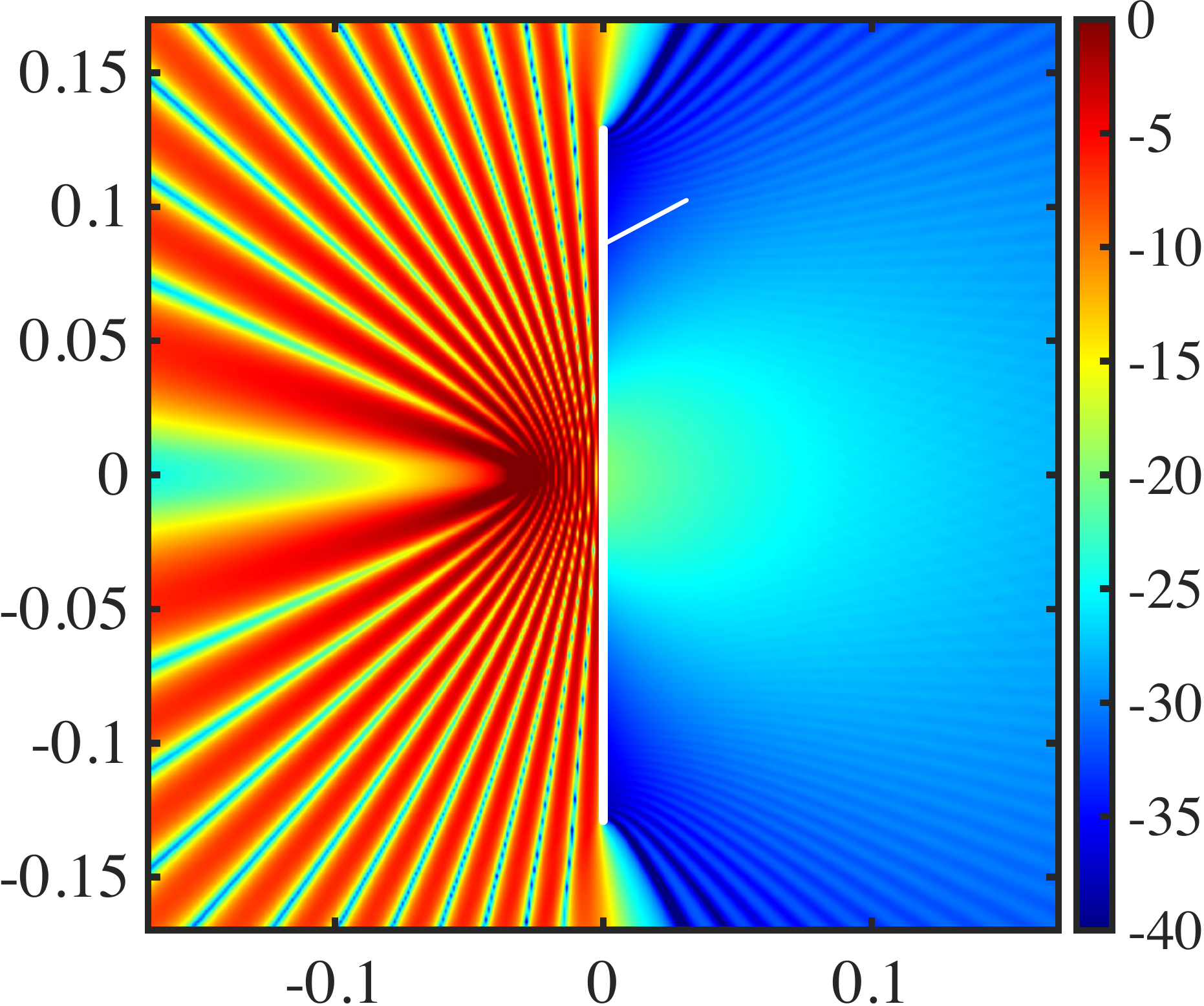}
    \put(50,87){\scriptsize \htext{\textbf{IE-GSTC}  -- Non-dispersive -- $|E_z^\text{tot.}|$ dB}}
    \put(50,-5){\scriptsize\htext{ $x$ (m)}}
    \put(-3,43){\scriptsize\vtext{ $y$ (m)}}
     \put(68, 73){\color{white}\scriptsize\htext{ \shortstack{Non-dispersive \\ Sheet Model}}}
    \end{overpic}\vspace{0.2cm}\caption{}
    \end{subfigure}
\caption{Total field scattered from a finite-sized flat surface excited by a cylindrical source located at $\mathbf{r}_\text{s}= \{x_\text{s} = -30,y_\text{s} = 0\}$~mm. a) Reference PEC case. b) Spatially dispersive zero thickness metasurface (10 div/$\lambda$) of length equivalent to 120 wires. c) A fictitious spatially non-dispersive zero thickness sheet. Simulation parameters: $N = 120$ wires, wire spacing $\Lambda = 1.25$~mm, operating frequency $60$~GHz.} \label{Fig:WireResults1}
\end{figure*}
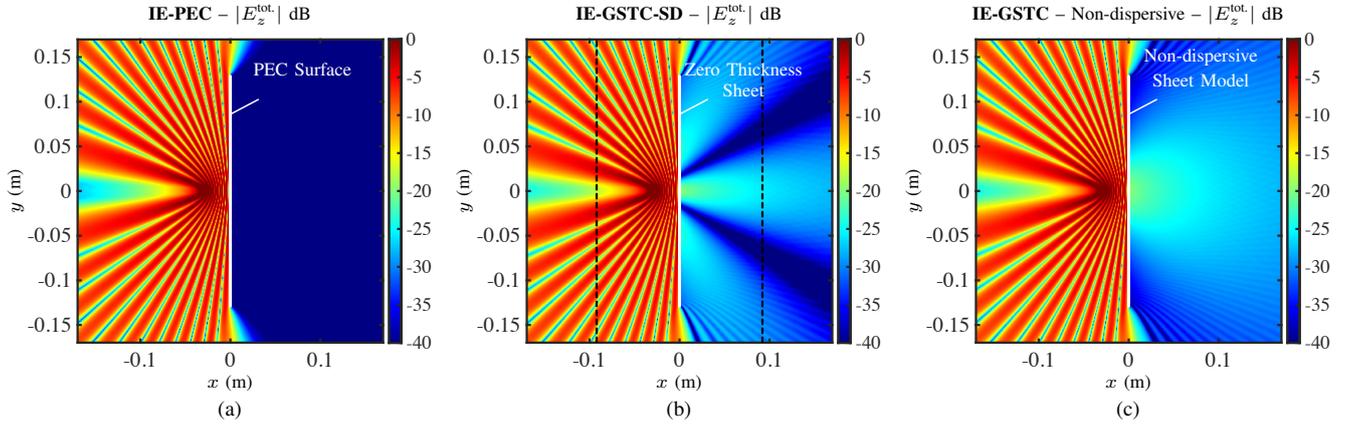

The basic behavior of the surface is presented in Fig.~\ref{Fig:WireResults1} where the total field $E_z$ is shown in the log scale to emphasize its low amplitude features. As a reference case, Fig.~\ref{Fig:WireResults1}(a) first shows the total fields generated by a Perfect Electric Conductor (PEC) of the same size, which naturally generates zero transmission through the surface and exhibits finite diffraction at the two edges. Fig.~\ref{Fig:WireResults1}(b) next shows the fields generates by the short conducting wire surface. Although the reflected fields are similar to those of a PEC, the transmitted fields are significantly transformed by spatial dispersion. The angular filtering is seen due to the drop in transmission at $30^\circ$ and the increased transmission at higher angles due to the spatial dispersion effect. A high level of transmission for incident flux at $45$--$65^\circ$, interferes with the edge diffraction to produce quite a characteristic field pattern. In contrast, Fig.~\ref{Fig:WireResults1}(c) shows the behavior of a uniform constant surface where the spatial dispersion is set to zero. We observe that the surface is highly reflective as expected, but there is no spatial filtering of the transmitted field in any significant way. Edge diffraction is present but minimal due to the incident flux at the edges arriving obliquely.

\begin{figure*}[htbp]
\centering
		\begin{subfigure}{0.6\columnwidth}
		\centering
    \begin{overpic}[width=\linewidth,grid=false,trim={0cm 0cm 0cm 0cm},clip]{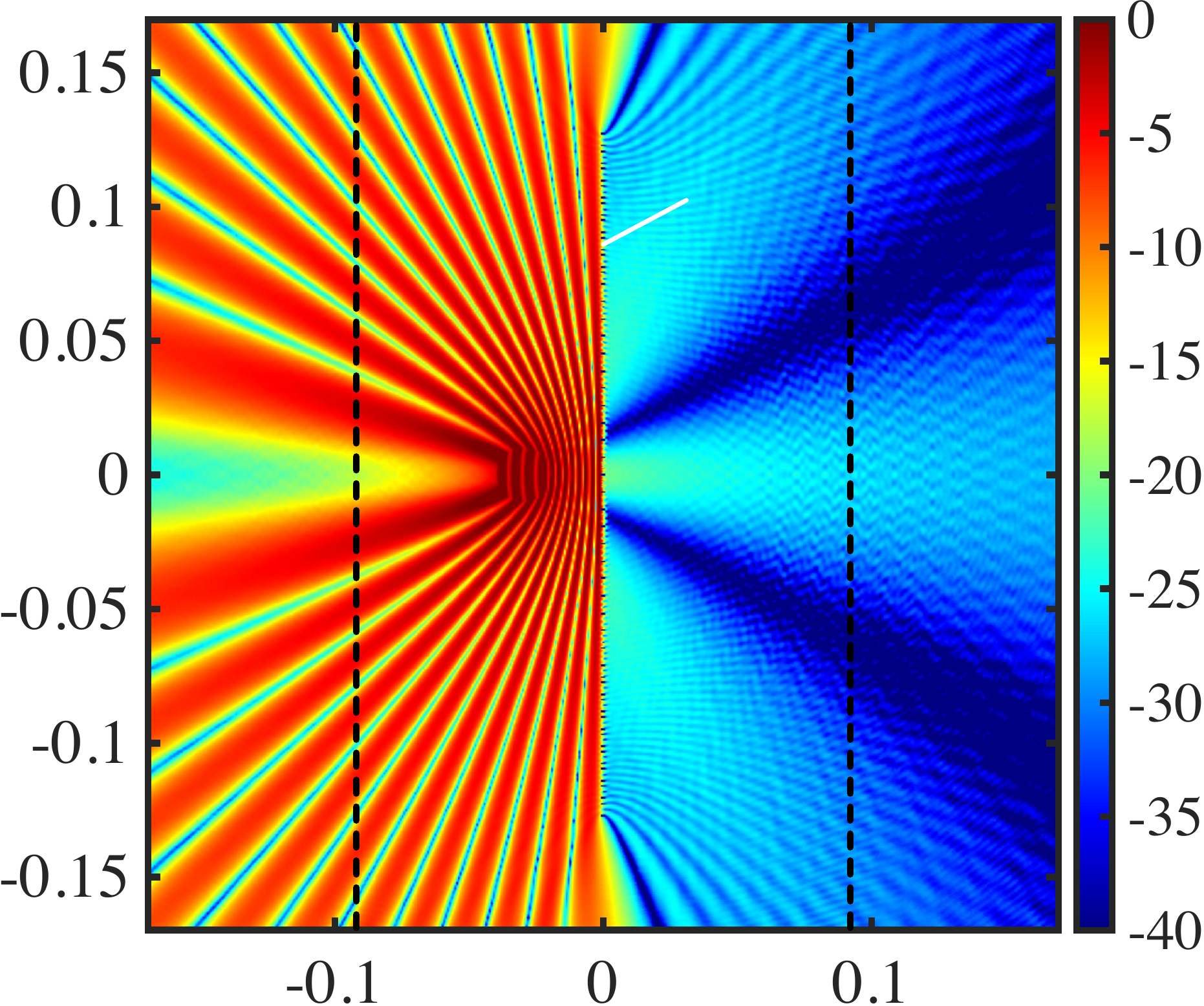}
    \put(50,87){\scriptsize \htext{\textbf{FEM-HFSS}  -- $|E_z^\text{tot.}|$ dB}}
    \put(50,-5){\scriptsize\htext{ $x$ (m)}}
    \put(-3,43){\scriptsize\vtext{ $y$ (m)}}
     \put(65, 65){\color{white}\scriptsize\htext{ \shortstack{Wire Array \\ Surface}}}
    \end{overpic}\vspace{0.2cm}\caption{}
    \end{subfigure}
    \hspace{0.35cm}
		\begin{subfigure}{0.6\columnwidth}
		\centering
 \begin{overpic}[width=\linewidth,grid=false,trim={0cm 0cm 0cm 0cm},clip]{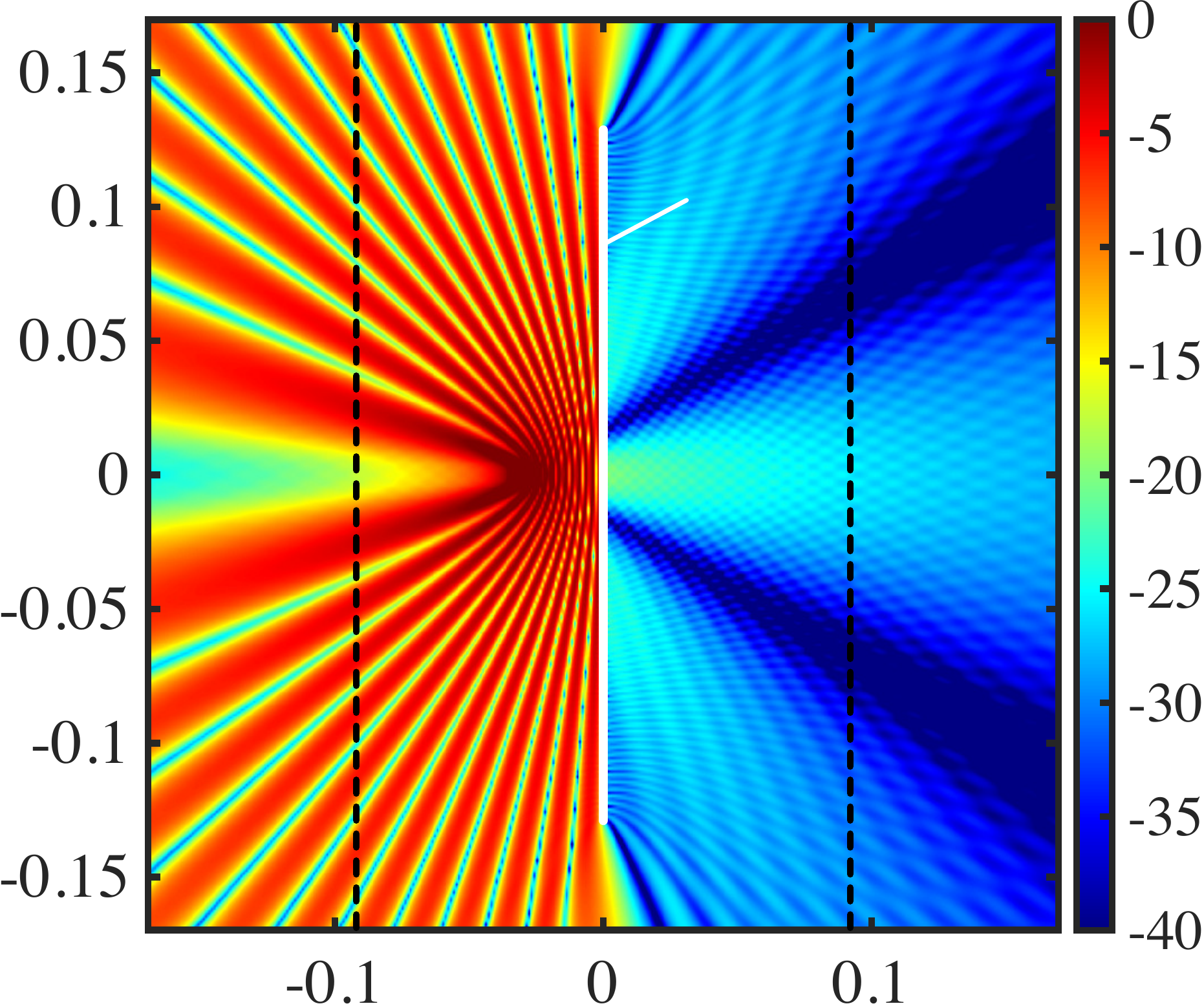}
    \put(50,87){\scriptsize \htext{\textbf{IE-GSTC-SD  -- Quantized} -- $|E_z^\text{tot.}|$ dB}}
    \put(50,-5){\scriptsize\htext{ $x$ (m)}}
    \put(-3,43){\scriptsize\vtext{ $y$ (m)}}
     \put(70, 65){\color{white}\scriptsize\htext{ \shortstack{Zero Thickness \\ Sheet}}}
     \put(68, 20){\color{white}\scriptsize\htext{ \shortstack{Observation \\ Lines \\ $x=\pm x_0$}}}
    \end{overpic}\vspace{0.2cm}\caption{}
    \end{subfigure}
    \hspace{0.35cm}
    		\begin{subfigure}{0.6\columnwidth}
		\centering
    \begin{overpic}[width=\linewidth,grid=false,trim={0cm 0cm 0cm 0cm},clip]{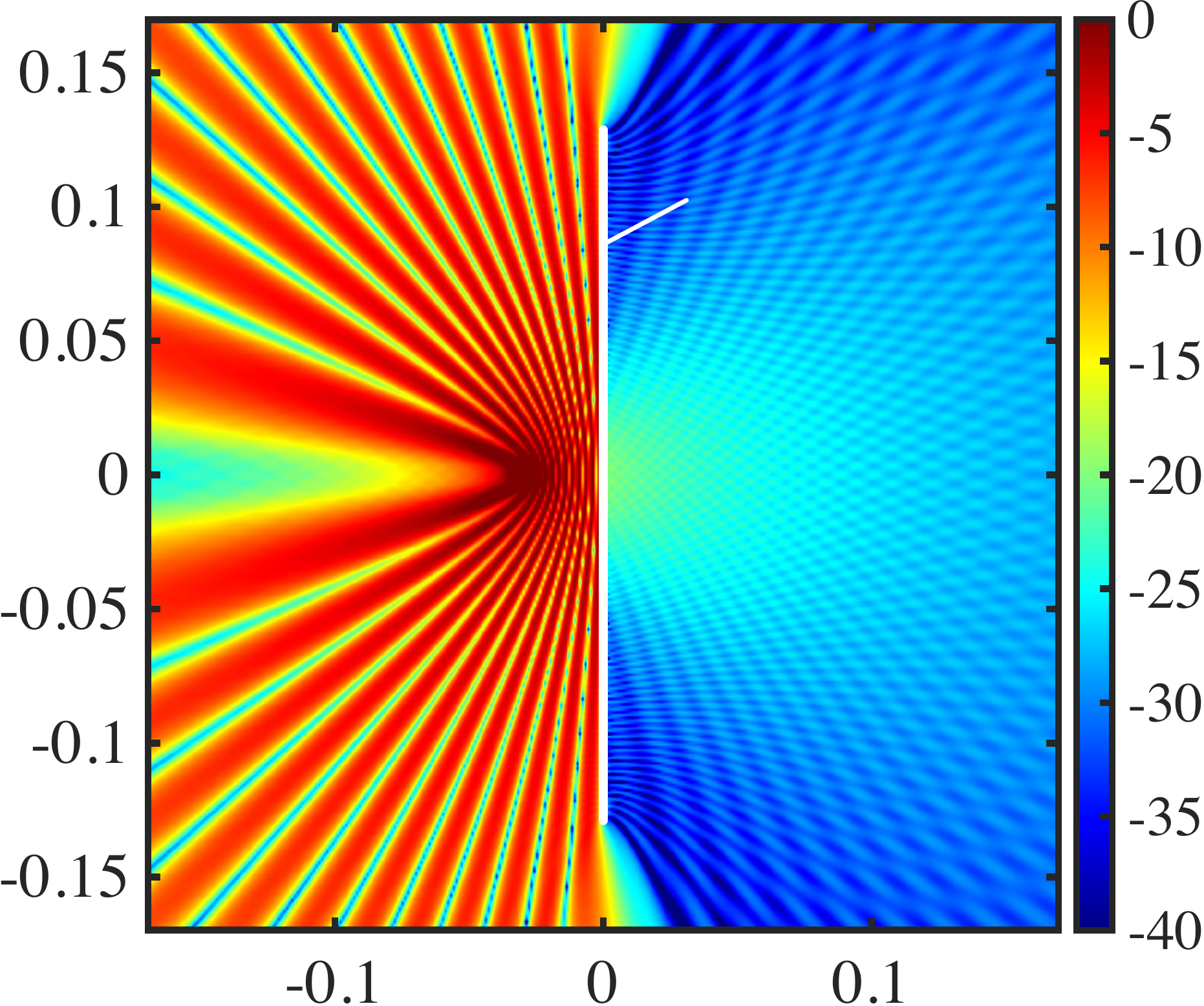}
    \put(54,87){\scriptsize\htext{\textbf{IE-GSTC} -- Non-dispersive/Quantized -- $|E_z^\text{tot.}|$ dB}}
    \put(50,-5){\scriptsize\htext{ $x$ (m)}}
    \put(-3,43){\scriptsize\vtext{ $y$ (m)}}	
        \put(70, 65){\color{white}\scriptsize\htext{ \shortstack{Non-Dispersive \\ Surface}}}
    \end{overpic}\vspace{0.2cm}\caption{}
\end{subfigure}
\caption{Total field scattered from a finite-sized flat surface composed of $N$ short wire dipoles, excited by a cylindrical source located at $\mathbf{r}_\text{s}= \{x_\text{s} = -30,y_\text{s} = 0\}$~mm. a) FEM-HFSS computed fields. b) Spatially dispersive equivalent zero thickness metasurface with quantized surface currents (1 div/cell) c) A fictitious spatially non-dispersive zero thickness sheet with quantized surface currents. Simulation parameters are same as those of Fig.~\ref{Fig:WireResults1}.} \label{Fig:WireResults2}
\end{figure*}
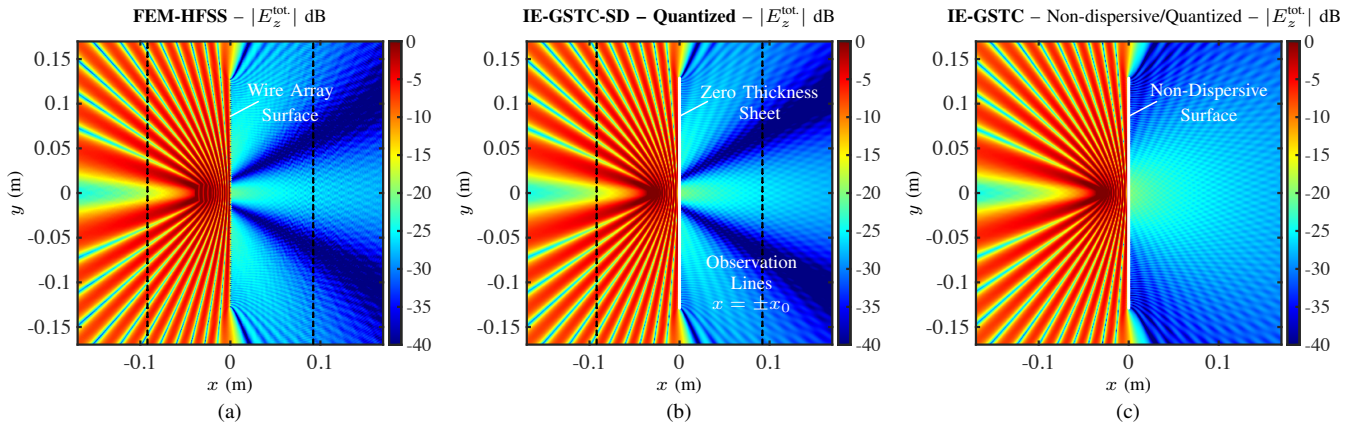

 These results clearly show that spatial dispersion strongly shapes the transmitted fields and, to some degree, the reflected ones. However, it remains to show that the fields depicted in Fig.~\ref{Fig:WireResults1}(b) are indeed correct solutions of the surface phenomena. To assess this, we built the entire volumetric surface in Ansys HFSS and full-wave simulated it. This result is shown in Fig.~\ref{Fig:WireResults2}(a), which has a very similar structure to the results obtained using IE-GSTC-SD in Fig.~\ref{Fig:WireResults1}(b). The IE-GSTC-SD has captured the structure of the fields and interference of the edge diffraction very well. Hence, confirming that this equivalent zero thickness model is correctly capturing the angular scattering of the metasurface structure\footnote{It can be noted that HFSS simulation took several hours to complete on a high-end server, where all possible symmetry boundary conditions were exploited, and yet it suffered from limited convergence. In contrast, the IE-GSTC-SD simulation took less than a minute on a desktop workstation once the angular-dependent surface susceptibilities are retrieved from unit cell simulations in HFSS, which were themselves computationally inexpensive.}.

However, there is an interesting feature in the fields obtained from the HFSS simulations related to the subtle interference pattern imposed on the fields. Although this pattern is at the limit of the HFSS simulation and meshing ability to resolve, it is present and thus investigated next. The simulation in Fig.~\ref{Fig:WireResults1}(b) shows none of this fine internal field structure, and spatial dispersion alone does not seem a credible explanation. However, the wire dipole cell is not deeply sub-wavelength and is relatively large with $\Lambda_y/\lambda \approx 0.6$, as compared to the unit cell of Fig.~\ref{Fig:LoopCellResults}(a), which is about $\lambda/10$ and thus safely with deep sub-wavelength periodicity. Therefore, the presence of a small dipole resonator in a large cell provides impetus to propose that the metasurface will act as a periodic surface with a finite number of current sources, as opposed to a sheet of continuous currents. By default, the simulation in Fig.~\ref{Fig:WireResults1}(b) used a discretization of 10 divisions per wavelength and models the surface current sources as quasi-continuous over the entire unit cell and the surface -- as would be standard in a BEM approach -- thus producing the smooth fields presented. 

To investigate this phenomenon, we modified the BEM methodology and placed the total current present in the unit cell on a single segment at the position of the wire only. The unit cell had five surface segments across its width; therefore, on 4 of the segments, the surface currents were set to zero and the total cell current placed on the middle segment. The result of this quantization of the surface currents is shown Fig.~\ref{Fig:WireResults2}(b). The effect is quite dramatic, unveiling the sought-out subtle interference pattern. Unlike the FEM method of HFSS, the interference pattern produced by the IE-GSTC-SD is very well defined due to the intrinsic nature of the method, which has a pure and simple description of the geometry. Figure \ref{Fig:WireResults2}(c) shows the generated fields when the spatial dispersion is switched off in BEM. The angular filtering features, as expected disappeared, while retaining the fine interference patterns only. Therefore, confirming that this interference pattern is solely due to current quantization and not spatial dispersion.

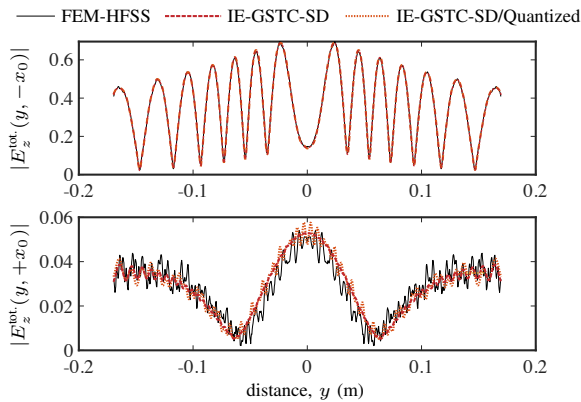
\begin{figure}[htbp]
\centering
    \begin{overpic}[width=0.8\linewidth,grid=false,trim={0cm 0cm 0cm 0cm},clip]{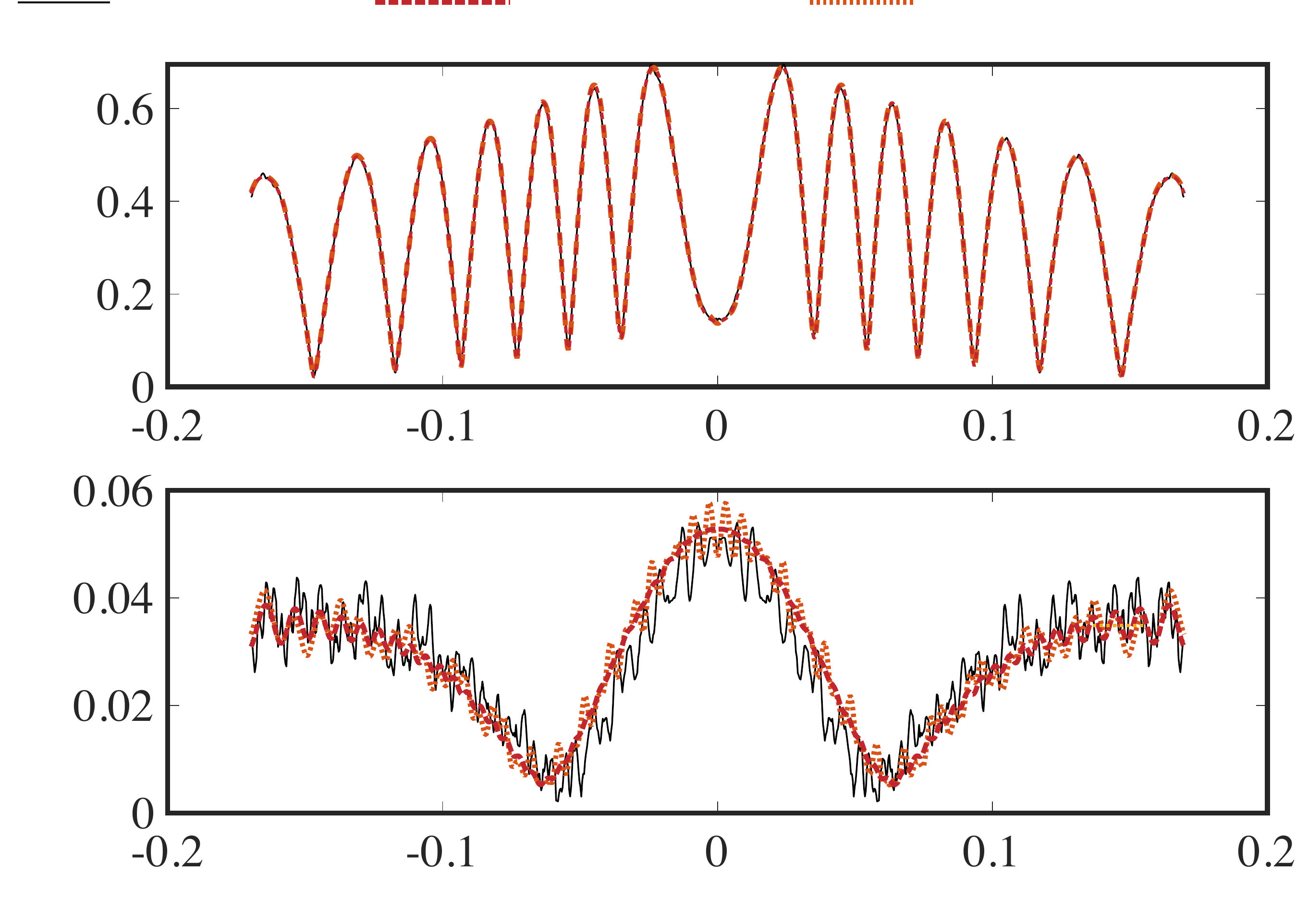}
        \put(55, 0){\scriptsize\htext{distance, $y$ (m)}}
        \put(2, 53){\scriptsize\vtext{$|E_z^\text{tot.}(y, -x_0)|$}}
        \put(2, 20){\scriptsize\vtext{$|E_z^\text{tot.}(y,+x_0)|$}}
        \put(18, 70){\scriptsize\htext{FEM-HFSS}}
        \put(50, 70){\scriptsize\htext{IE-GSTC-SD}}
         \put(89, 70){\scriptsize\htext{IE-GSTC-SD/Quantized}}
    \end{overpic}
\caption{Comparison of total electric fields measured on the two observation lines in the reflection and transmission region of the finite-sized metasurface of Fig.~\ref{Fig:WireResults2}. The observation lines are located at $x_0 = \pm 0.085$ m.}\label{fig:WireLine}
\end{figure}

To further investigate the match between the FEM HFSS simulation and the IE-GSTC-SD results, the fields were plotted for two vertical lines $\pm 0.085$~m removed from the surface (arbitrarily chosen but in the macroscopic region). These fields can be seen in Fig.~\ref{fig:WireLine}. Fields are presented for the three cases: FEM-HFSS, IE-GSTC-SD, and the quantized IE-GSTC-SD. The IE-GSTC-SD fields match the HFSS fields well, but of course, have none of the high-frequency variations. The use of the quantization method introduces the expected high-frequency modulation, and it is of a similar magnitude as those in the HFSS results. The details in the interference pattern do not match, but this is not really to be expected due to the small magnitude of the variation and the limits of HFSS simulation. Although this method of introducing the effect of discrete sources is somewhat ad hoc, it appears to confirm the source of the interference pattern and is quite successful.

\subsection{Finite-Sized Curvilinear Metasurfaces}

The final two examples will involve comparing Ansys FEM-HFSS and the proposed IE-GSTC-SD simulations for more geometrically complex structures created using the simple wire unit cell. We will consider two configurations: 1) an open hexagonal-based structure formed of three sides, and 2) an open semi-circular structure. For both simulations, the incident illumination will be a plane-wave at $60$~GHz. 

\begin{figure*}[t]
\centering
		\begin{subfigure}{0.5\columnwidth}
		\centering
    \begin{overpic}[width=\linewidth,grid=false,trim={0cm 0cm 0cm 0cm},clip]{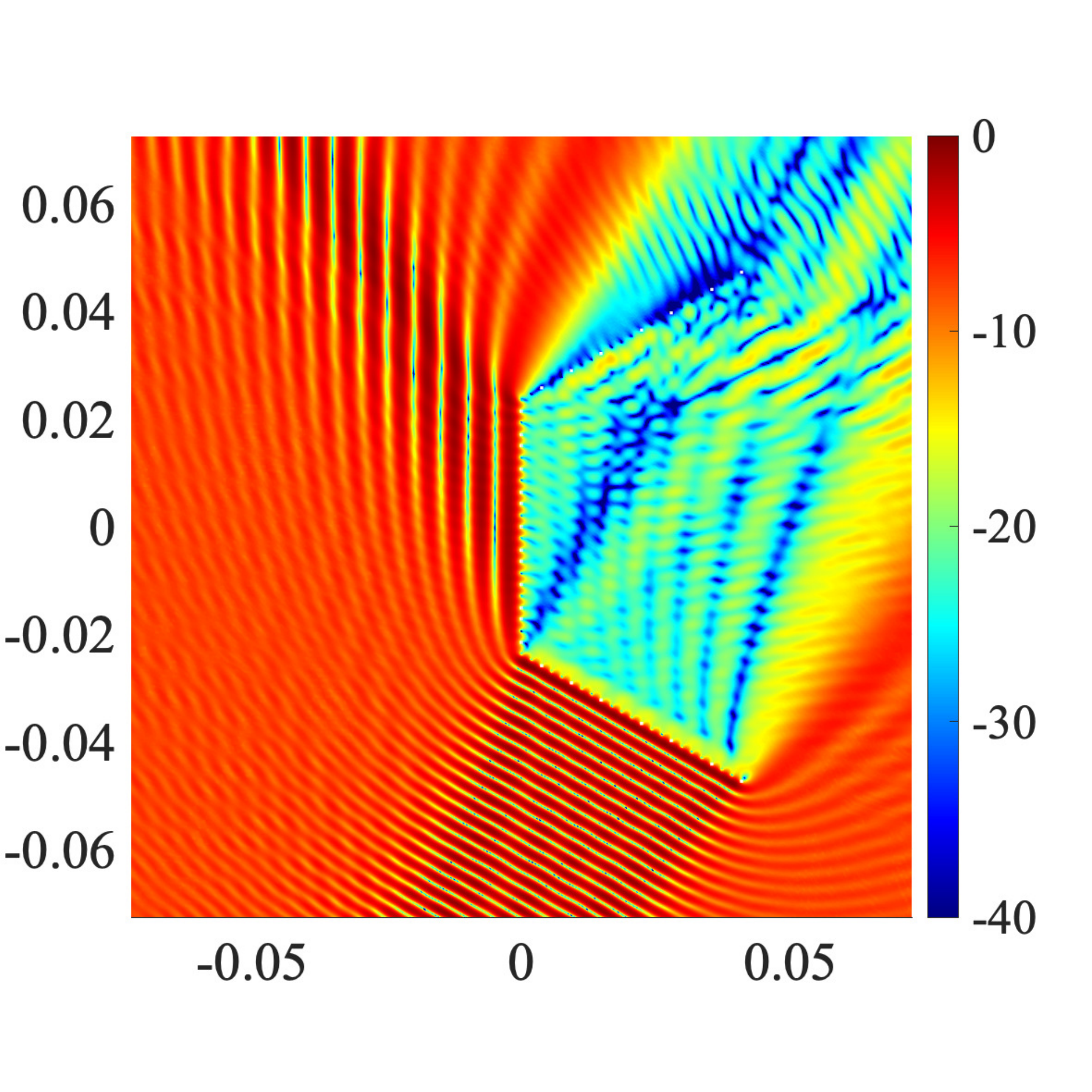}
    \put(50,92){\scriptsize \htext{\textbf{FEM-HFSS}  -- $|E_z^\text{tot.}|$ dB}}
    \put(50,5){\scriptsize\htext{ $x$ (m)}}
    \put(-3,50){\scriptsize\vtext{ $y$ (m)}}
     \put(28, 75){\color{white}\scriptsize\htext{ \shortstack{Wire Array \\ Surface}}}
    \put(32, 25){\color{white}\scriptsize\htext{ \shortstack{Incident \\ Plane-wave}}}
    \end{overpic} \caption{}
		\end{subfigure}
		\begin{subfigure}{0.5\columnwidth}
		\centering    
 \begin{overpic}[width=\linewidth,grid=false,trim={0cm 0cm 0cm 0cm},clip]{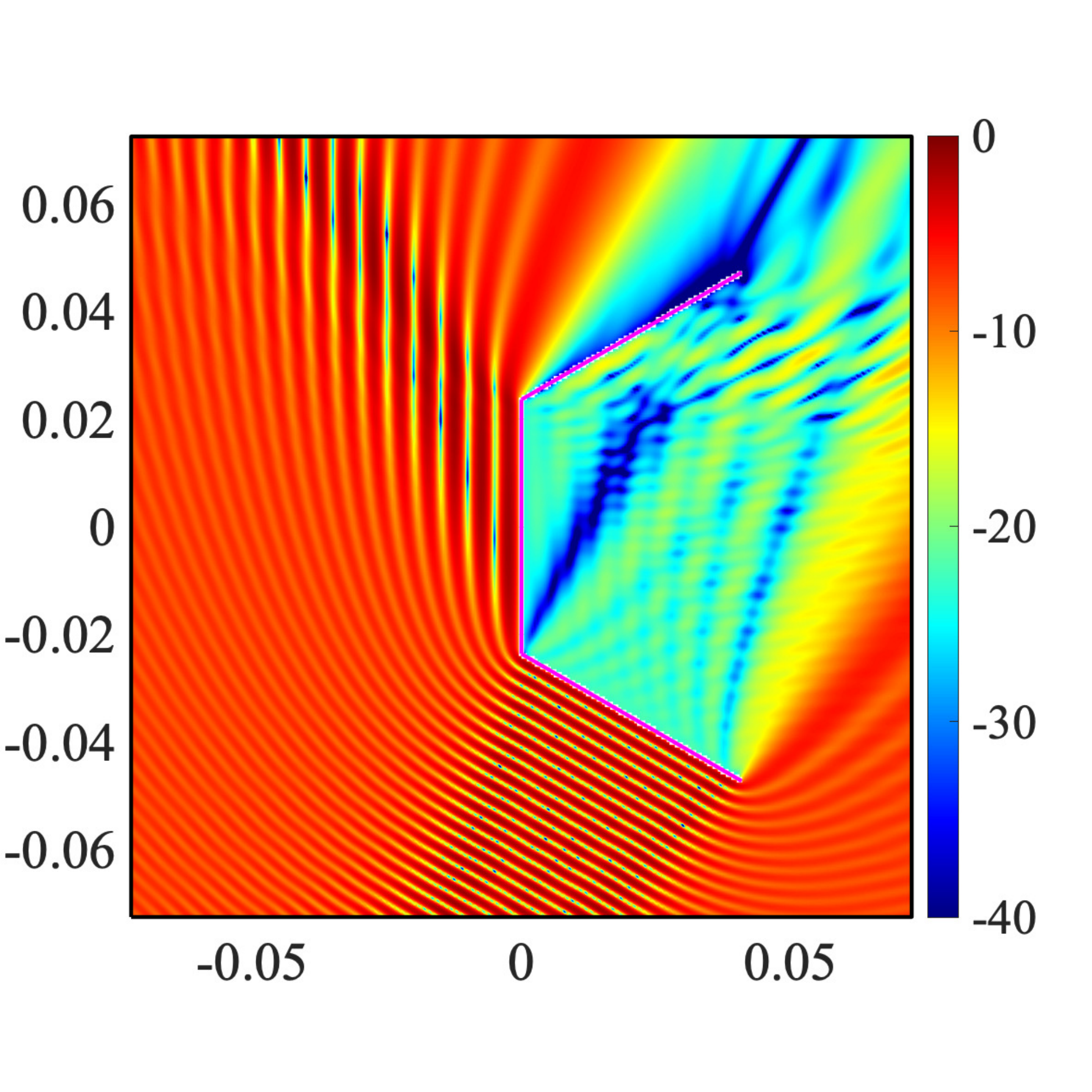}
    \put(50,92){\scriptsize\htext{\textbf{IE-GSTC-SD}  -- $|E_z^\text{tot.}|$ dB}}
    \put(50,5){\scriptsize\htext{ $x$ (m)}}
    \put(-3,50){\scriptsize\vtext{ $y$ (m)}}
         \put(32, 75){\color{white}\scriptsize\htext{ \shortstack{Zero Thickness \\ Sheet}}}
    \end{overpic} \caption{}
		\end{subfigure}
    		\begin{subfigure}{0.5\columnwidth}
		\centering
    \begin{overpic}[width=\linewidth,grid=false,trim={0cm 0cm 0cm 0cm},clip]{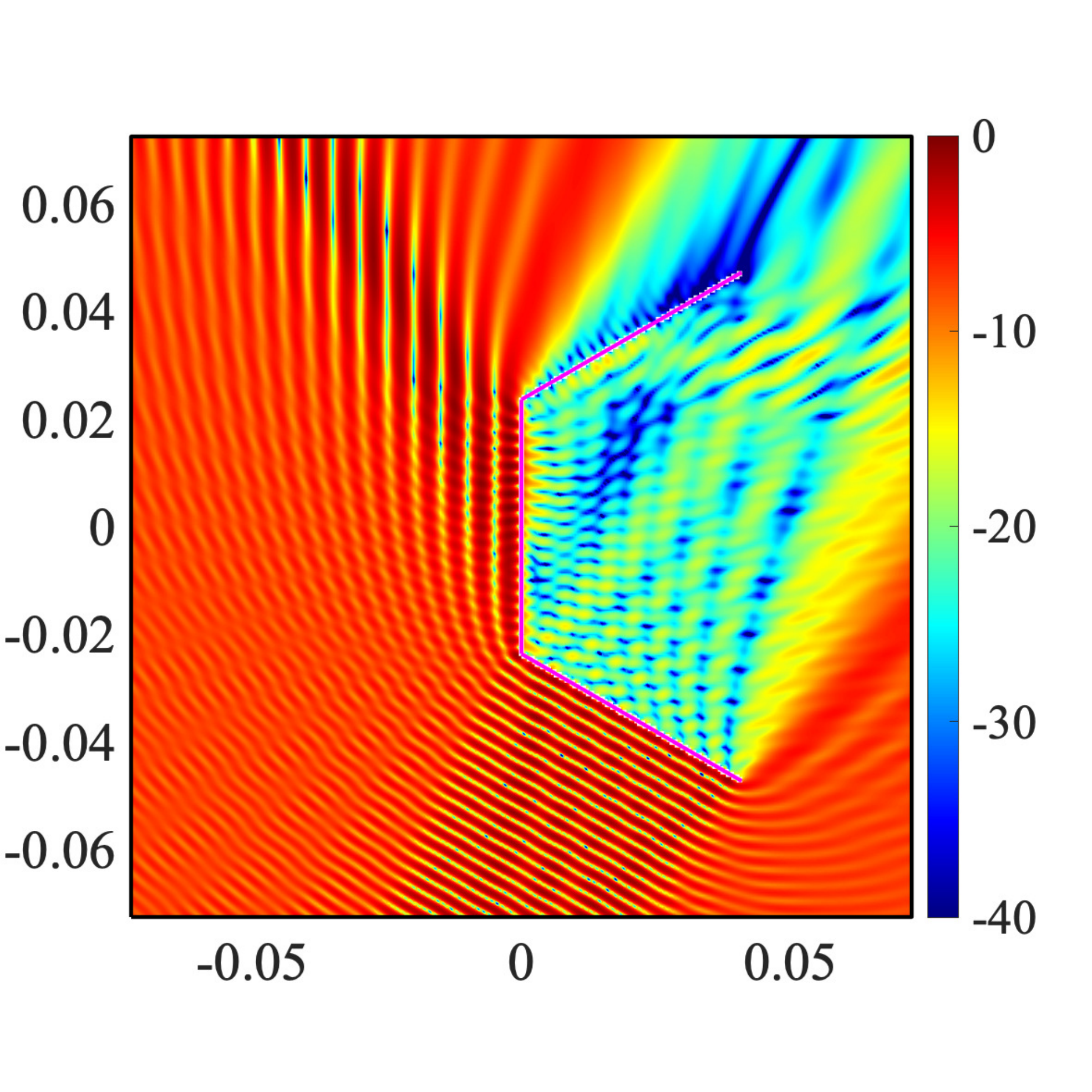}
    \put(50,92){\scriptsize\htext{\textbf{IE-GSTC-SD -- Quantized}  -- $|E_z^\text{tot.}|$ dB}}
    \put(50,5){\scriptsize\htext{ $x$ (m)}}
    \put(-3,50){\scriptsize\vtext{ $y$ (m)}}
        \put(34, 77){\color{white}\scriptsize\htext{ \shortstack{Discrete Array \\ Sources}}}
    \end{overpic}\caption{}
		\end{subfigure}
		\begin{subfigure}{0.5\columnwidth}
		\centering
\begin{overpic}[width=\linewidth,grid=false,trim={0cm 0cm 0cm 0cm},clip]{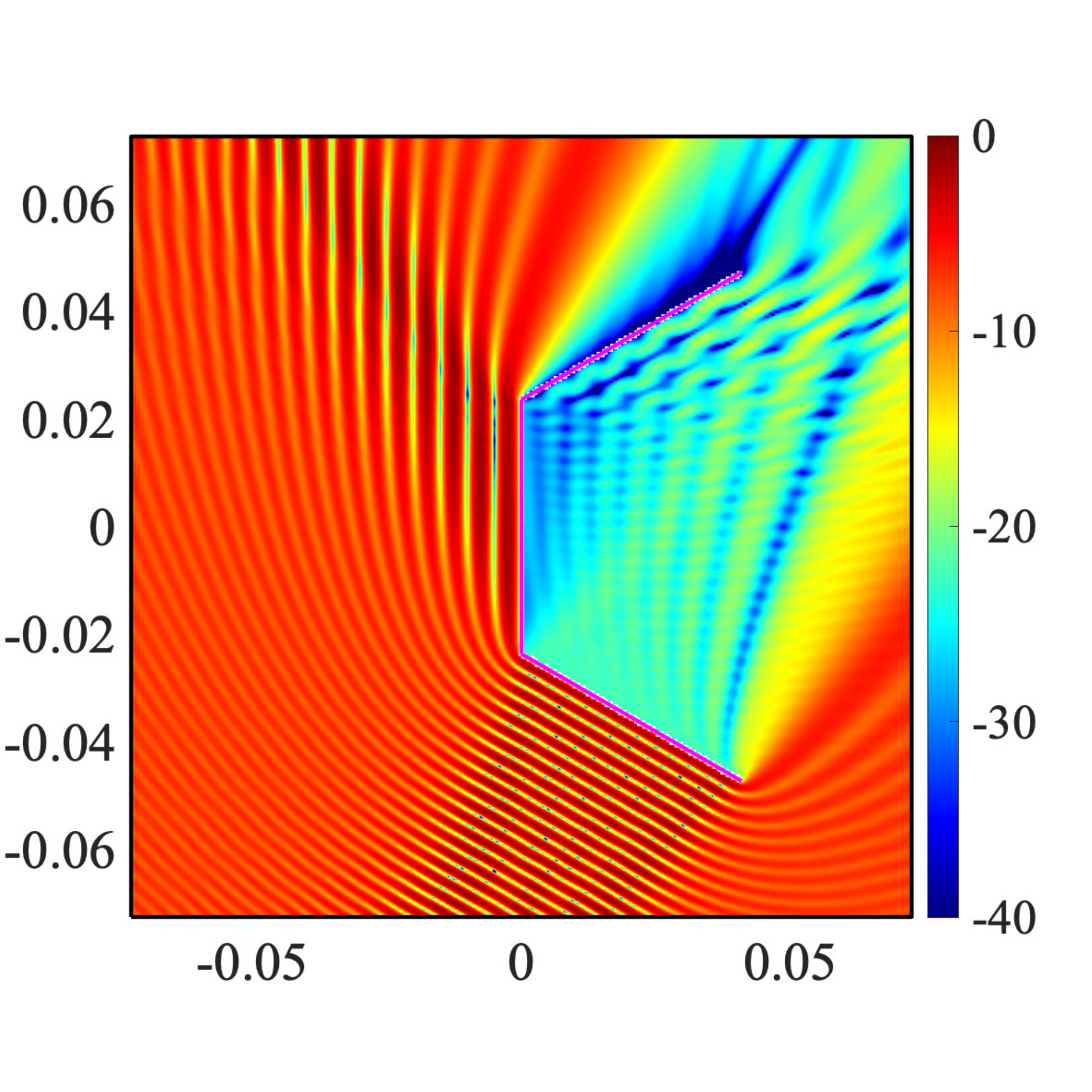}
    \put(50,92){\scriptsize\htext{\textbf{IE-GSTC} -- Non-dispersive -- $|E_z^\text{tot.}|$ dB}}
    \put(50,5){\scriptsize\htext{ $x$ (m)}}
    \put(-3,50){\scriptsize\vtext{ $y$ (m)}}
        \put(34, 77){\color{white}\scriptsize\htext{ \shortstack{Non-Dispersive \\ Surface}}}
    \end{overpic} 
\caption{}
		\end{subfigure}
\caption{Total fields produced by a 3-sided hexagonal structure, using an array of short wire dipole and excited with a uniform plane-wave: a) FEM-HFSS. b) Dispersive zero thickness sheet model using IE-GSTC-SD with continuous sheet currents (10 div/$\lambda$).  c) Dispersive zero thickness sheet model using IE-GSTC-SD but with discrete sources matching the number of wires (i.e., 1 div/cell). d) Fictitious non-dispersive zero thickness sheet with constant tangential surface susceptibilities. Simulation parameters: each side with $N= 63$~wires of radius $d_0 = 0.2$~mm, and length $\ell = 2.5$~mm, separated by $\Lambda = 2.15$~mm, operating frequency $60$~GHz and angle of incidence $60^\circ$ measured from $x-$axis. }\label{Fig:WireHex}
\end{figure*}

Simulations of the hexagonal-based structure are shown in Fig.~\ref{Fig:WireHex}. The structure consist of three sides of 21 cells and facet length of $s = 0.0473$~m. In this simulation, the incident plane wave travels left to right at $60^\circ$ measured from horizontal, striking the bottom facet of the open surface at normal incidence. Fig.~\ref{Fig:WireHex}(a) shows the fields obtained from HFSS simulations, which exhibit a complex field pattern due to the reflection from two exposed facets, transmission through the facets, and the intrinsic angular filtering of the wire structure. The facet in the shadow of the incident field also has a strong effect as it reflects back the field transmitted through the bottom and side facets. Some strong interference patterns are present due to the multiple reflections present in the interior of the hexagonal. We also see evidence of the individual wires acting as discrete sources in creating a more subtle interference pattern of the field in the interior of the open hexagon.

The second Fig.~\ref{Fig:WireHex}(b) presents a basic IE-GSTC-SD simulation of the structure with continuous current distribution accounting for spatial dispersion. It can be seen that this simulation produces a very close match with the HFSS results. It captures all of the basic features of transmission and reflection, including the interference patterns from the complicated reflection in the interior. Once again, to recreate the interior fine interference features, the currents were quantized, and the generated field patterns are shown in Fig.~\ref{Fig:WireHex}(c). As before, this quantization is speculative and ad hoc; however, as with the simulation of the vertical surface, we see the successful creation of a subtle interference pattern (particularly in the transmitted fields) as observed in the HFSS fields. Finally, to emphasize the importance of spatial dispersion, Fig.~\ref{Fig:WireHex}(d) presents the same simulation but with no spatial dispersion. The internal field structure due to complex interference is simply lost.

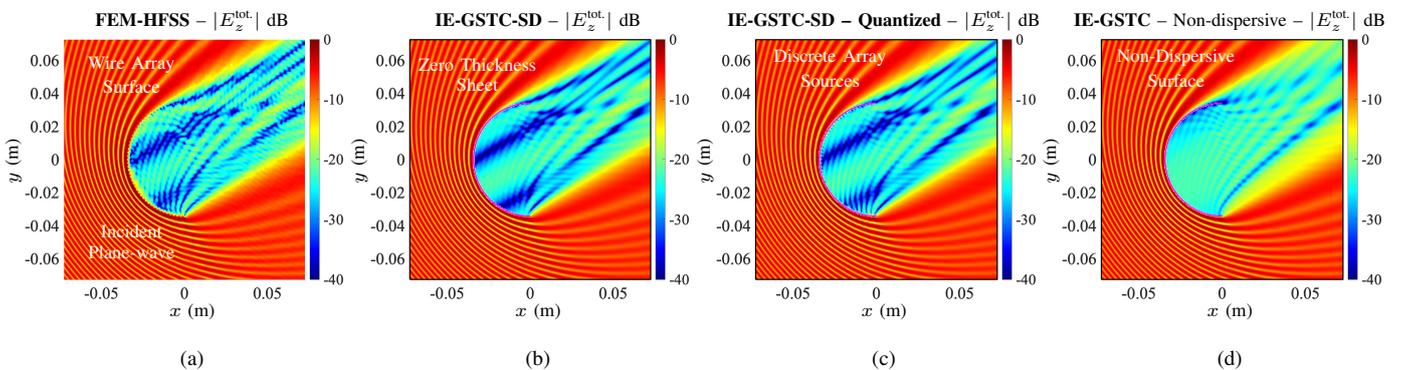
\begin{figure*}[b]
		\centering
		\begin{subfigure}{0.5\columnwidth}
		\centering
    \begin{overpic}[width=\linewidth,grid=false,trim={0cm 0cm 0cm 0cm},clip]{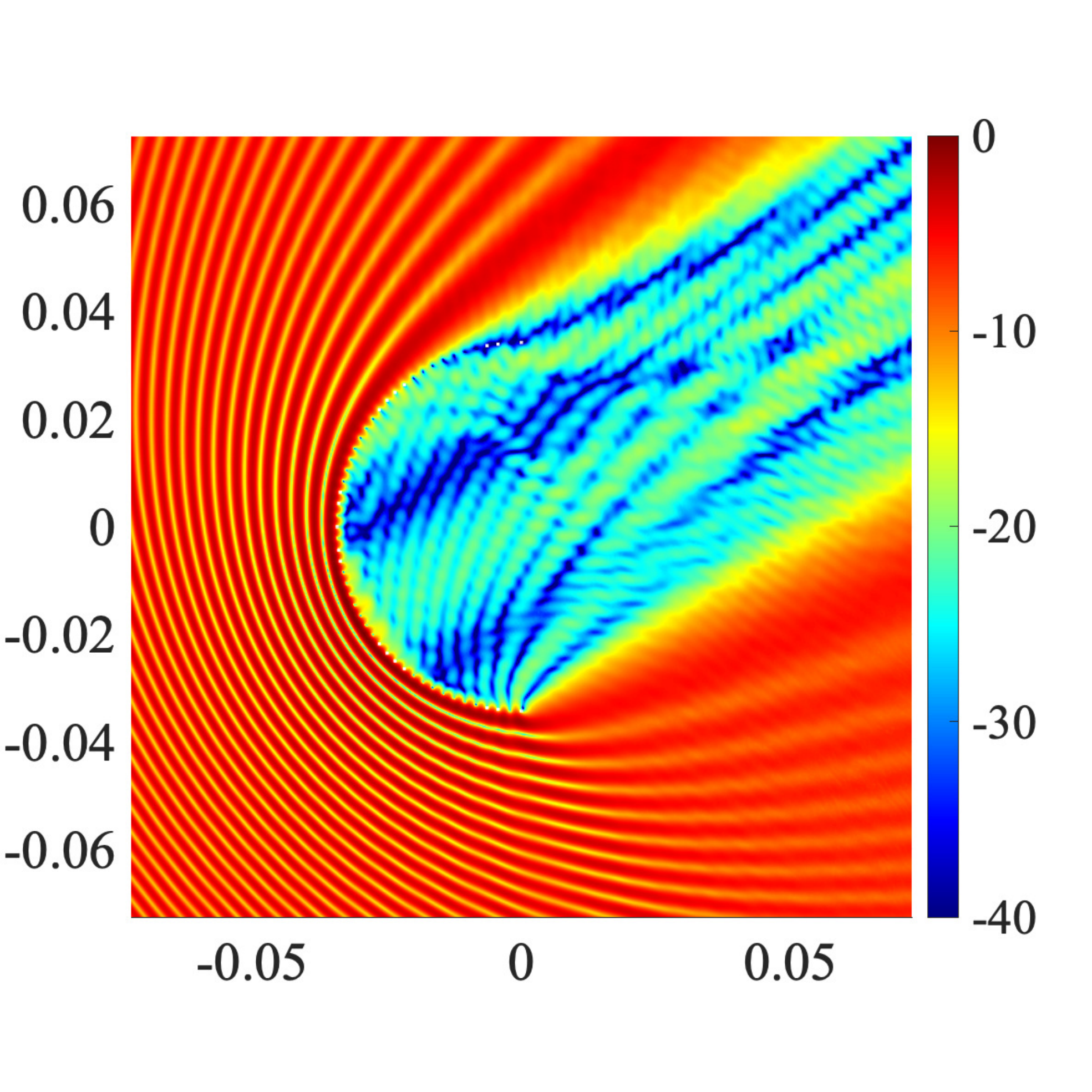}
    \put(50,92){\scriptsize \htext{\textbf{FEM-HFSS}  -- $|E_z^\text{tot.}|$ dB}}
    \put(50,5){\scriptsize\htext{ $x$ (m)}}
    \put(-3,50){\scriptsize\vtext{ $y$ (m)}}
     \put(32, 75){\color{white}\scriptsize\htext{ \shortstack{Wire Array \\ Surface}}}
    \put(32, 25){\color{white}\scriptsize\htext{ \shortstack{Incident \\ Plane-wave}}}
    		\end{overpic} \caption{}
		\end{subfigure}
		\begin{subfigure}{0.5\columnwidth}
		\centering
		 \begin{overpic}[width=\linewidth,grid=false,trim={0cm 0cm 0cm 0cm},clip]{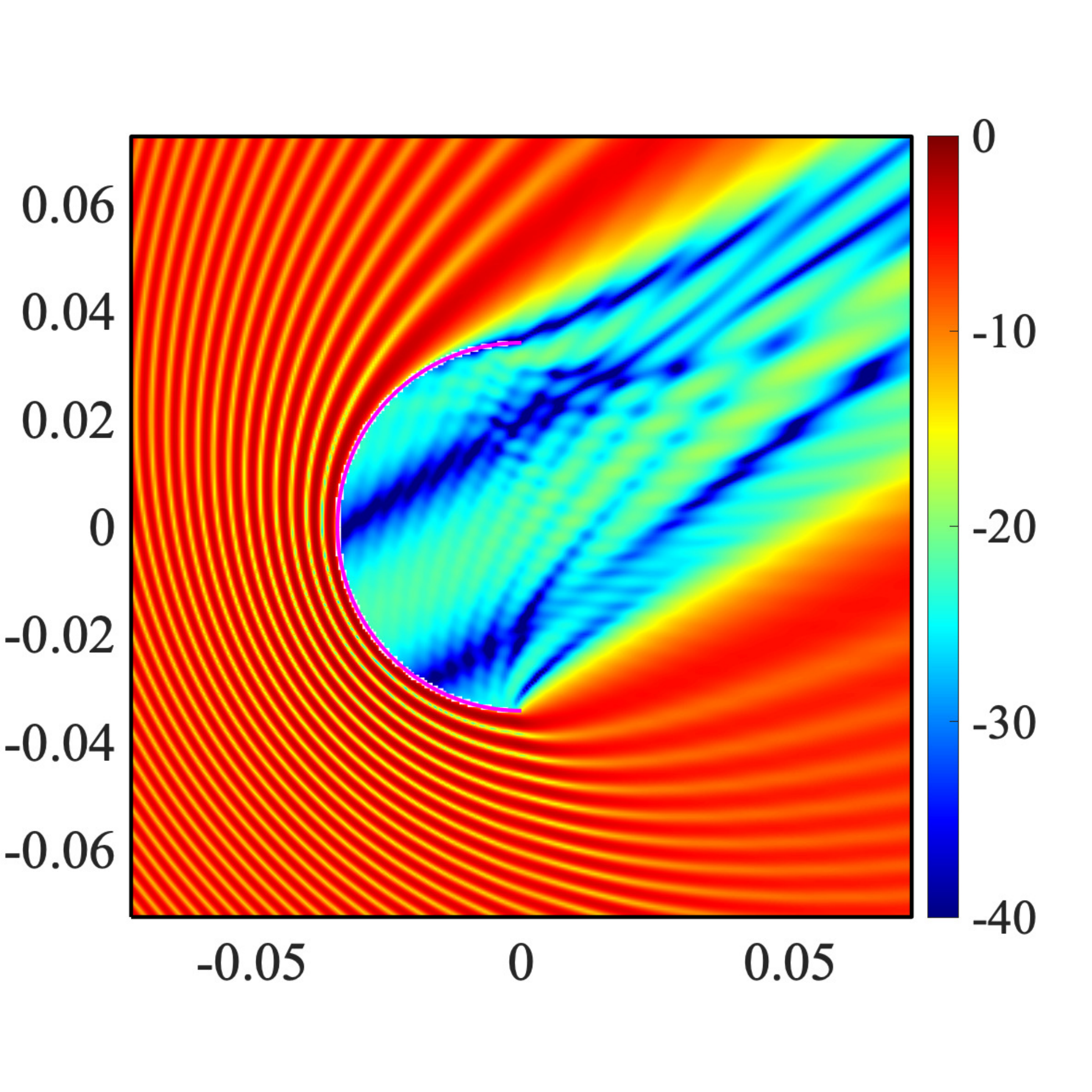}
    \put(50,92){\scriptsize\htext{\textbf{IE-GSTC-SD}  -- $|E_z^\text{tot.}|$ dB}}
    \put(50,5){\scriptsize\htext{ $x$ (m)}}
    \put(-3,50){\scriptsize\vtext{ $y$ (m)}}
         \put(32, 75){\color{white}\scriptsize\htext{ \shortstack{Zero Thickness \\ Sheet}}}
    		\end{overpic} \caption{}
		\end{subfigure}
		\begin{subfigure}{0.5\columnwidth}
		\centering
		    \begin{overpic}[width=\linewidth,grid=false,trim={0cm 0cm 0cm 0cm},clip]{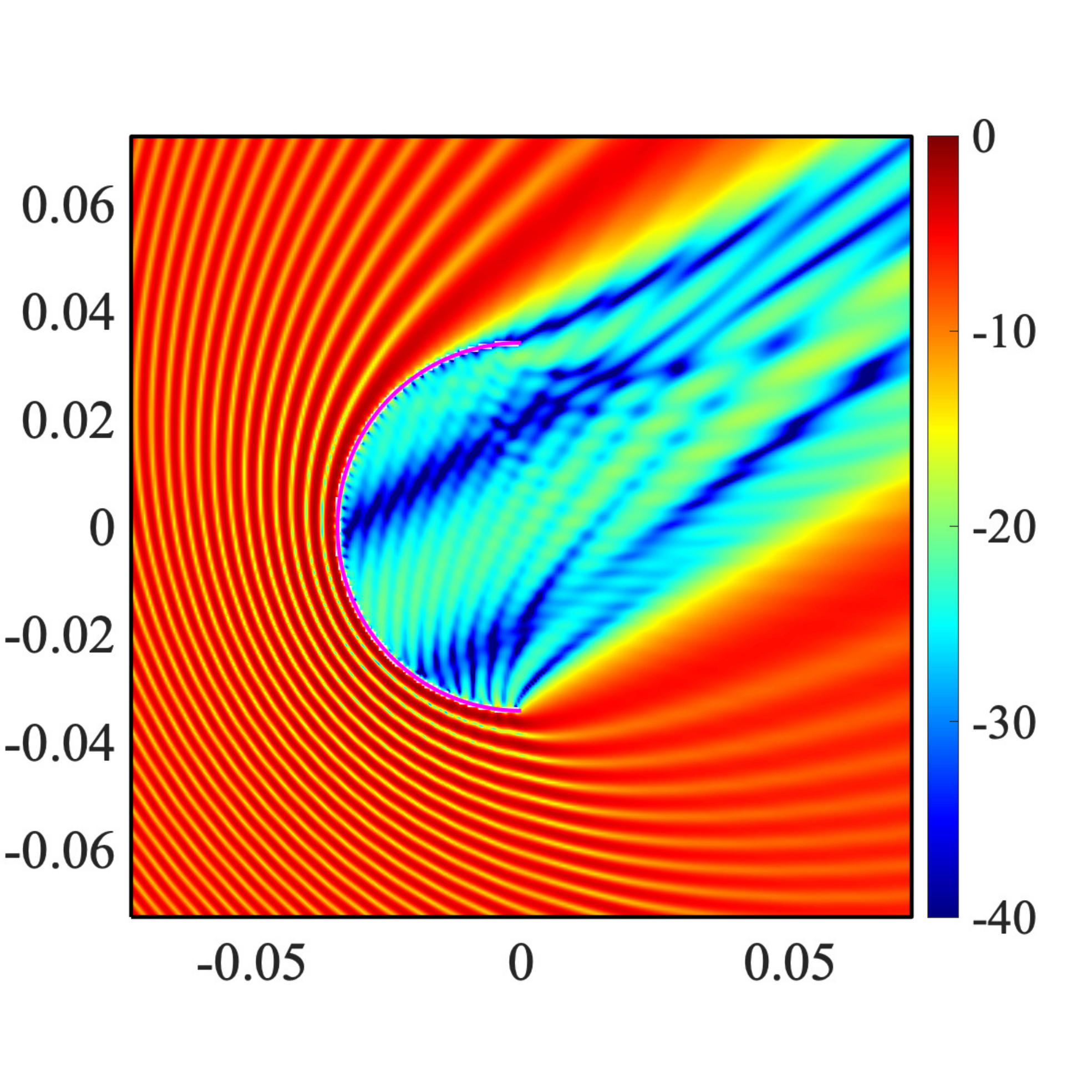}
    \put(50,92){\scriptsize\htext{\textbf{IE-GSTC-SD -- Quantized}  -- $|E_z^\text{tot.}|$ dB}}
    \put(50,5){\scriptsize\htext{ $x$ (m)}}
    \put(-3,50){\scriptsize\vtext{ $y$ (m)}}
        \put(34, 77){\color{white}\scriptsize\htext{ \shortstack{Discrete Array \\ Sources}}}
    		\end{overpic} \caption{}
		\end{subfigure}
		\begin{subfigure}{0.5\columnwidth}
		\centering
		 \begin{overpic}[width=\linewidth,grid=false,trim={0cm 0cm 0cm 0cm},clip]{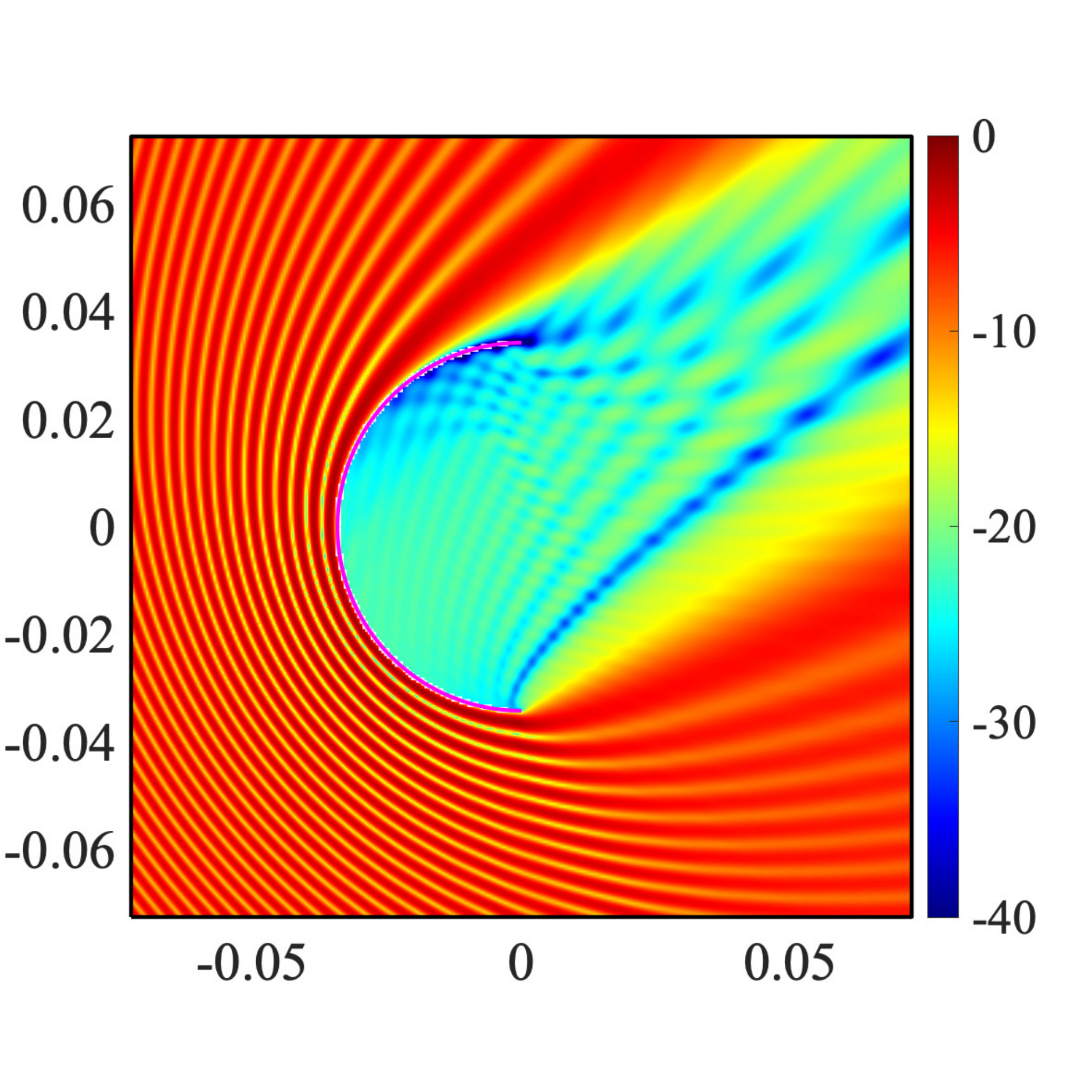}
    \put(50,92){\scriptsize\htext{\textbf{IE-GSTC} -- Non-dispersive -- $|E_z^\text{tot.}|$ dB}}
    \put(50,5){\scriptsize\htext{ $x$ (m)}}
    \put(-3,50){\scriptsize\vtext{ $y$ (m)}}
        \put(34, 77){\color{white}\scriptsize\htext{ \shortstack{Non-Dispersive \\ Surface}}}
    		\end{overpic} \caption{}
		\end{subfigure}						

		\caption{Total fields produced by a curved semi-circular metasurface formed using an array of short wire dipoles and excited with a uniform plane-wave: a) FEM-HFSS. b) Dispersive zero thickness sheet model using IE-GSTC-SD with continuous sheet currents (10 div/$\lambda$).  c) Dispersive zero thickness sheet model using IE-GSTC-SD but with discrete sources matching the number of wires (i.e., 1 div/cell). d) Fictitious non-dispersive zero thickness sheet with constant tangential surface susceptibilities. Simulation parameters: $N= 51$~wires of radius $d_0 = 0.2$~mm, and length $\ell = 2.5$~mm, separated by $\Lambda = 2.15$~mm, operating frequency $60$~GHz and angle of incidence $30^\circ$ measured from $x-$axis. }
		\label{Fig:WireSemi}
\end{figure*}

The final set of simulations (Fig.~\ref{Fig:WireSemi}) are of a semicircular structure with a radius of $r = 0.0342$~m and consisting of 51 wire unit cells. The incident plane wave travels left to right at $30^\circ$ from horizontal striking the front of the open surface. Fig. \ref{Fig:WireSemi}(a) shows the HFSS simulation, which presents a complex transmitted field pattern. Besides, a complex spatial dispersion effect is expected since the plane wave is illuminating a curved structure with the angle of incidence varying along the surface. Due to the angular dependence of transmission, we see significant nulls in the field patterns internal to the curved surface, which also produce interference patterns due to multiple reflections in its interior. There is also, once again, evidence of the discrete nature of the resonators on the surface. Fig. \ref{Fig:WireSemi}(b) presents the IE-GSTC-SD model results. It should be noted that the susceptibility was extracted in a periodically infinite flat surface model ~\cite{Part_1_Nizer_SD}, implying that some errors would eventually show up when applied to a curved surface. However, we see in Fig. \ref{Fig:WireSemi}(b) that the basic field structure is captured very well -- all of the primary features of the interference pattern are predicted accurately. As previously, the introduction of quantization of the surface currents, shown in Fig. \ref{Fig:WireSemi}(c), creates some additional interference features as seen in the HFSS simulation. Finally, a comparison with the non-spatially dispersive case in Fig. \ref{Fig:WireSemi}(d) shows a significant absence of the fine field structure, establishing the importance of capturing the spatial dispersion property of the unit cell.

\section{Conclusion}

An IE-GSTC field solver to compute the scattered fields from spatially dispersive metasurfaces has been proposed and has been numerically confirmed using variety of examples. The work is a continuation of Part-1 \cite{Part_1_Nizer_SD}, which proposed the basic methodology of representing spatially dispersive metasurface structures in the spatial frequency domain, $k_y$. By representing the angular dependence of the surface susceptibilities in $k_y$ as a ratio of two polynomials, standard GSTCs have been extended to include the spatial derivatives of both the difference and average fields around the metasurface. These extended boundary conditions were successfully integrated into the standard IE-GSTC solver, which led to a new IE-GSTC-SD framework. The proposed IE-GSTC-SD platform has been confirmed using the semi-analytical Fourier decomposition method applied to uniform metasurfaces before testing it against a practical short conducting wire unit cell for various cases of finite size flat and curvilinear surfaces. All the results for finite-sized metasurface structures composed of spatially dispersive wire unit cells have confirmed the successful implementation and integration of spatial dispersion into the IE-GSTC simulation framework. Due to its inherent structural symmetry and simplicity, the wire unit cell has been proven to be an excellent example, where the proposed IE-GSTC-SD framework is tested to predict both high and low field amplitude features simultaneously and accurately. Moreover, the framework successfully modeled surfaces with conformal and curvilinear geometries, demonstrating its versatile architecture. Finally, some fine and subtle field interference patterns observed in HFSS have been traced to large unit cell periodicities in practical metasurfaces, which revealed those intricate interference features when accounted for in terms of discrete current sources.

This work has so far focused on integrating tangential surface susceptibilities and spatially symmetrical structures to avoid cumbersome mathematical developments and implementation details. Moreover, the focus has been on uniform metasurface structures of relatively small unit cell periodicities for simplicity. A natural extension of this work is to incorporate normal surface susceptibilities and the bi-anisotropic tensor terms, which can model general non-uniform metasurface structures with arbitrary structural symmetries. In addition, a more in-depth analysis must be performed to rigorously explain the current quantization phenomenon observed in this work, which could be of greater importance for electrically large unit cell periodicities. This proposed work thus represents an important developmental step for fast and efficient simulation of practical metasurface structures which are not necessarily deeply sub-wavelength and exhibit fundamental spatial dispersion effects.

\section*{Acknowledgements}

The authors acknowledge funding from the Department of National Defence's Innovation for Defence Excellence and Security (IDEaS) Program in support of this work.

\balance
\bibliographystyle{ieeetran}
\bibliography{PREP_2021_MS_Spatial_Dispersion_TAP}  

\end{document}